\documentclass[12pt]{article}
\usepackage{amsmath,amssymb,epsfig}
\usepackage{graphicx,floatflt,subfigure}
\usepackage{cite}

\makeatletter

\newcommand{\anti}[1]{\mbox{$\overline{\rm #1}$}}

\newcommand{\zv}{{\bf z}}

\newcommand{\yt}{{\tilde y}}

\def\Im{\mathop{\mathrm{Im}}\nolimits}

\def\Re{\mathop{\mathrm{Re}}\nolimits}

\def\tr{\mathop{\mathrm{tr}}\nolimits}

\def\half{{\frac{1}{2}}}

\def\unit{{1\kern-.65ex {\rm l}}}
\def\1{{1\kern-.65ex {\rm l}}}
\def\ap{{\alpha'}}
\def\ls{{\ell_{\rm s}}}
\def\gs{{g_{\rm s}}}

\def\CA{{\cal A}}
\def\CB{{\cal B}}

\def\CR{{\cal R}}

\def\bbC{{\mathbb{C}}}

\def\bbR{{\mathbb{R}}}

\newcount\hour \newcount\minute
\hour=\time \divide \hour by 60
\minute=\time
\def\now{%
\ifnum \hour<13
  \ifnum \hour=0 \advance \hour by 12 \number\hour:\else \number\hour:\fi%
     \ifnum \minute<10 0\fi%
     \number\minute%
\ A.M.%
\else \advance \hour by -12 \number\hour:%
  \ifnum \minute<10 0\fi%
  \number\minute%
  \ P.M.%
\fi%
}

\makeatother

\begin{document}

% format
\baselineskip=18pt  % a la harvmac
\numberwithin{equation}{section}  % make eq labels (sec.num)
\allowdisplaybreaks  % allow page breaks in displayed eqs

% print date, time and filename 
%\pagestyle{myheadings}
%\markright{{\tt \jobname.tex} -- \today{} \now}

%%%%%%%%%%%%%%%%%%%%%%%%%%%%%%%%%%%%%%%%%%%
%%%        TITLE BEGINS HERE
%%%%%%%%%%%%%%%%%%%%%%%%%%%%%%%%%%%%%%%%%%%

%% ========== title (note version) begins here ==========
%
%\vspace*{-1cm}
%\begin{center}
% {\Large\bf Title of the Document}
%\end{center}
%\vspace*{-.5cm}
%
%% ========== title (note version) ends here ==========

%% ========== title (paper version, a la harvmac) begins here ==========

\thispagestyle{empty}

% Report number
\vspace*{-2cm} 
\begin{flushright}
{\tt arXiv:0705.0983}\\
CALT-68-2644\\
ITFA-2007-15
\end{flushright}

% title, authors, affiliation
\vspace*{0.4cm} 
\begin{center}
 {\LARGE Nonsupersymmetric Brane/Antibrane Configurations\\[1 ex] in Type IIA and $M$ Theory\\}
 \vspace*{1.5cm}
 Joseph Marsano$^1$, Kyriakos Papadodimas$^2$ and Masaki Shigemori$^1$\\
 \vspace*{1.0cm} 
 $^1$ California Institute of Technology 452-48, Pasadena, CA 91125, USA\\[1ex]
 $^2$ 
Institute for Theoretical Physics, University of Amsterdam\\
Valckenierstraat 65, 1018 XE Amsterdam, The Netherlands\\
 \vspace*{0.6cm} 
  % can this prevent spam?
 {\tt ma{}r{}sa{}no@t{}h{{}}{e{ory}}.c{}a{}l{{}}te{{}}ch.e{{}}d{}u},
 {\tt kp{}a{}p{ado}@s{}ci{}en{}ce.{uva}.nl}, \\
 {\tt sh{}i{}ge{}@t{}h{{}}{e{ory}}.c{}a{}l{{}}te{{}}ch.e{{}}d{}u}
\end{center}
\vspace*{0.8cm}

% abstract
\noindent
We study metastable nonsupersymmetric configurations in type IIA string
theory, obtained by suspending D4-branes and \anti{D4}-branes between
holomorphically curved NS5's, which are related to those of
\texttt{hep-th/0610249} by $T$-duality.  When the numbers of branes and
antibranes are the same, we are able to obtain an exact $M$ theory lift
which can be used to reliably describe the vacuum configuration as a
curved NS5 with dissolved RR flux for $g_s \ll 1$ and as a curved M5 for
$g_s \gg 1$.  When our weakly coupled description is reliable, it is
related by $T$-duality to the deformed IIB geometry with flux of
\texttt{hep-th/0610249} with moduli exactly minimizing the potential
derived therein using special geometry.  Moreover, we can use a direct
analysis of the action to argue that this agreement must also hold for
the more general brane/antibrane configurations of
\texttt{hep-th/0610249}.  On the other hand, when our strongly coupled
description is reliable, the M5 wraps a nonholomorphic minimal area
curve that can exhibit quite different properties, suggesting that the
residual structure remaining after spontaneous breaking of supersymmetry
at tree level can be further broken by the effects of string
interactions.  Finally, we discuss the boundary condition issues raised
in \texttt{hep-th/0608157} for nonsupersymmetric IIA configurations,
their implications for our setup, and their realization on the type IIB
side.

\newpage
\setcounter{page}{1} % don't number title page

%% ========== title (paper version, a la harvmac) ends here ==========

%%%%%%%%%%%%%%%%%%%%%%%%%%%%%%%%%%%%%%%%%%%
%%%           TITLE ENDS HERE
%%%%%%%%%%%%%%%%%%%%%%%%%%%%%%%%%%%%%%%%%%%

\tableofcontents
%\printindex

%%%%%%%%%%%%%%%%%%%%%%%%%%%%%%%%%%%%%%%%%%%
%%%        MAIN TEXT BEGINS HERE
%%%%%%%%%%%%%%%%%%%%%%%%%%%%%%%%%%%%%%%%%%%

\section{Introduction and Summary}

Following the publication of \cite{Intriligator:2006dd}, the past year
has seen a great deal of interest in the study of metastable
supersymmetry-breaking vacua in supersymmetric gauge theories
% Gauge theory metastable refs %
\cite{Ooguri:2006pj, Forste:2006zc, Amariti:2006vk, Shih:2007av,
Intriligator:2007py,Ooguri:2007iu},
and string theory
% String theory metastable refs %
\cite{Franco:2006es,Garcia-Etxebarria:2006rw,Ooguri:2006bg,Franco:2006ht,Bena:2006rg,Ahn:2006gn,Ahn:2006tg,Argurio:2006ny,Aganagic:2006ex,Tatar:2006dm,Lebedev:2006qc,Ahn:2007eh,Ahn:2007uu,Heckman:2007wk,Ahn:2007er,Giveon:2007fk,Argurio:2007qk,Murthy:2007qm,Hirano:2007cj,Ahn:2007ve,Garcia-Etxebarria:2007vh,Kawano:2007ru,Angelantonj:2007ts,Douglas:2007tu,Ahn:2007ym}.
Because such configurations do not correspond to true vacua, many of the difficulties associated with the construction of realistic models of supersymmetry-breaking can be avoided.  As such, the ideas of \cite{Intriligator:2006dd} have already found wide phenomenological application
% Pheno applications refs
\cite{Abel:2006cr,Fischler:2006xh,Abe:2006xp,Abel:2006my,Dine:2006xt,Kitano:2006xg,Csaki:2006wi,Abel:2007uq,Anguelova:2007at}.

One particularly interesting system, proposed in \cite{Aganagic:2006ex},
realizes metastability by wrapping branes and antibranes on vanishing
2-cycles of a Calabi-Yau threefold.  The geometry can be engineered so
that these 2-cycles are homologous but, nevertheless, attain a finite
size away from the singular points, providing a barrier to
brane/antibrane annihilation.  Unlike previous examples, this setup is
inherently stringy in that the decay process cannot be described by a
quantum field theory with a finite number of degrees of
freedom{\footnote{ If one attempts to decouple stringy modes to get a
gauge theory description, one needs to take $\alpha'\rightarrow 0$, which
in turn means that one has to simultaneously scale the distance between
branes and antibranes to infinity in order to render the open string
tachyon massive.  Conversely, keeping the branes and antibranes a finite
distance apart, as was considered in \cite{Aganagic:2006ex}, 
implies that $\alpha'$ must also be finite and, indeed, is sufficiently
large that a field theory description is available only for the deep IR
physics near either the brane or the antibrane stack.}}.

A particularly nice feature of this system is that it appears to be
under fairly good calculational control.  When the numbers of branes and
antibranes are large, the authors of \cite{Aganagic:2006ex} suggest that
one can apply the large $N$ duality story of \cite{Vafa:2000wi,
Cachazo:2001jy, Cachazo:2001gh, Cachazo:2001sg, Cachazo:2002pr} even in
this nonsupersymmetric setting.  In other words, we can effectively
replace the branes by a deformed geometry with fluxes.
Moreover, it is argued that the ${\cal{N}}=2$ supersymmetry on the
deformed Calabi-Yau is actually spontaneously broken by the
opposite-sign fluxes, implying that both the superpotential and K\"ahler
potential continue to be determined by special geometry.  This makes it
possible to perform controlled computations and study, for instance, the
stabilization of the compact complex structure moduli.

\subsection{Brane/antibrane configurations in type IIA}
 
It has been well known for many years that geometrically engineered
systems of this sort are $T$-dual to Hanany-Witten type NS5/D4
configurations in type IIA theory \cite{Ooguri:1995wj, Kutasov:1995te,
Karch:1998yv}.
In this description, one studies the vacuum configuration by noting that the NS5/D4 system can also be described by a single M5-brane extended along a potentially complicated 6-dimensional hypersurface.  If one tunes the parameters appropriately, the M5 worldvolume theory is reliably approximated by the Nambu-Goto action and hence the vacuum configuration corresponds to an M5 extended along a minimal area surface.  

What can one hope to gain from such a description?  In supersymmetric
examples, BPS arguments indicate that the (holomorphic) physics should
not depend on $g_s$ so one expects that the vacuum configuration at all
values of the coupling is simply that obtained by applying $T$-duality
to the IIB picture.  We can verify this directly from the $M$ theory
point of view by noting that the minimal area M5 reduces, at small
$g_s$, to a curved NS5-brane with flux which is $T$-dual to the IIB
deformed geometry in the supersymmetric vacuum \cite{Dasgupta:2001um,
Oh:2001bf}.  This provides a nice alternative way of understanding the
large $N$ duality story of type IIB but does not teach us anything
fundamentally new about the physics.

In the nonsupersymmetric case at hand, though, we do not know \emph{a
priori} whether the vacuum configuration is protected as one moves to
different parameter regimes or not.  This has two important
consequences.  First, it means that we must take care to understand the
specific choices of parameters for which a description based on minimal
area M5's is valid.  Second, it indicates that the physics in one
regime, say strong coupling, need not resemble that in another, say weak
coupling, and hence there may be something new to be learned from a
IIA/M description depending on precisely when we can perform reliable
computations there.

\begin{figure}
\begin{center}
\subfigure[NS5/D4/\anti{D4} configuration $T$-dual to the brane/antibrane system of \cite{Aganagic:2006ex}]
{\epsfig{file=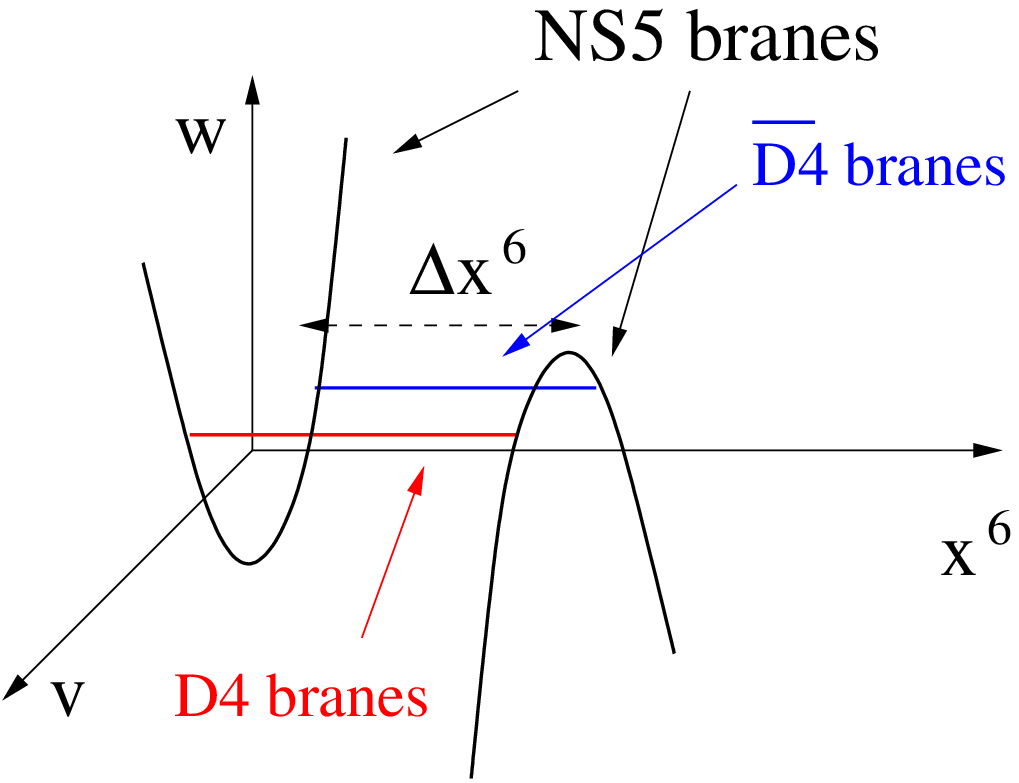,width=0.4\textwidth}\label{setupintro}}
\hspace{0.1\textwidth}
\subfigure[Tilting of D4's and \anti{D4}'s that gives rise to logarithmic bending along the $w$-direction]
{\epsfig{file=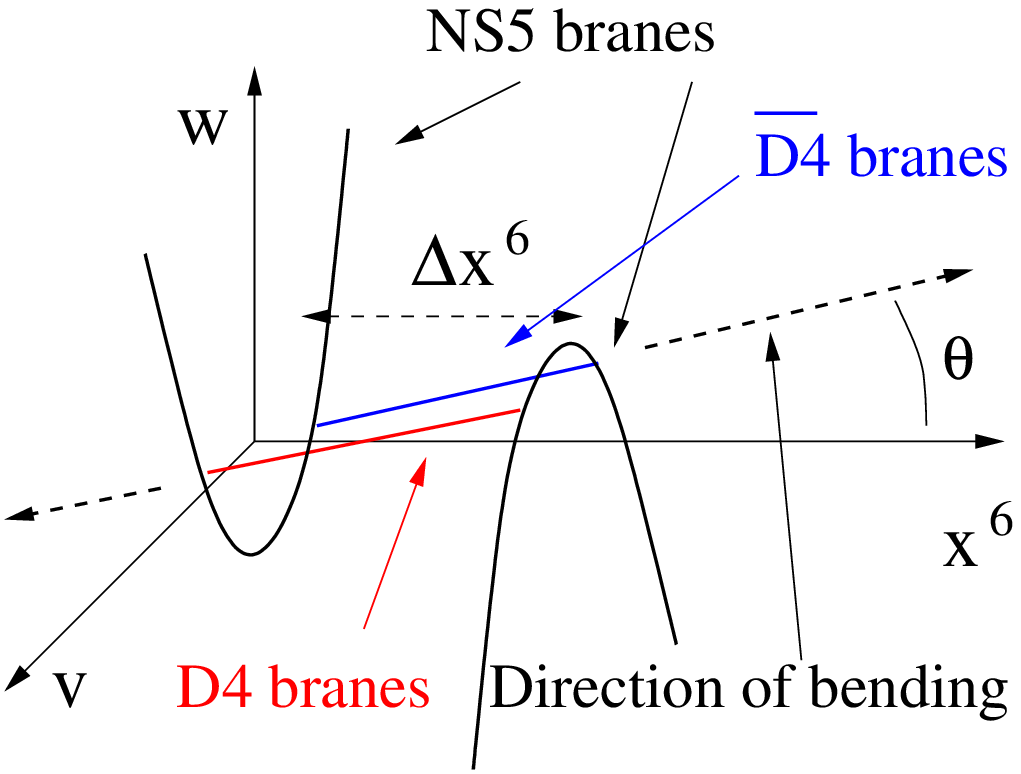,width=0.4\textwidth}\label{wbendingintro}}
\caption{The NS5/D4/\anti{D4} configuration under consideration}
\label{oursetup}
\end{center}
\end{figure}

In this paper, we thus endeavor to use techniques of $M$-theory to study
the nonsupersymmetric brane/antibrane configurations that are obtained
by applying $T$-duality to the system of \cite{Aganagic:2006ex}, namely
that with D5's and \anti{D5}'s placed at conifold singularities in a
local Calabi-Yau in type IIB{\footnote{Some comments on the IIA
description have been made previously in
\cite{Tatar:2006dm}.}}$^,${\footnote{Nonsupersymmetric brane
configurations in IIA have been studied before, for example by the
authors of \cite{Mukhi:2000te, Mukhi:2000dn}.  Unlike ours, the setups
studied there are stable.}}.  Specifically, the setup on which we focus
most of our attention is that illustrated in figure \ref{setupintro} and
consists of a pair of quadratically curved NS5-branes with stacks of
D4's and \anti{D4}'s suspended between them.  From the point of view of
$M$ theory, this system is described by a single M5-brane which, for
parameter regimes in which the Nambu-Goto piece of the worldvolume
theory is reliable, simply wraps a minimal area surface.  There are two
distinct parameter regimes for which this is the case and an analysis
based on minimal area surfaces is justified.  One lies at strong
coupling, where the $x^{10}$ radius is large and the M5 curvature small
in 11-dimensional Planck units.  The other, which does not seem to
receive as much attention in the literature\footnote{The reason perhaps
is that one typically is interested in the gauge theory limit where the
scales of the system become substringy and the Nambu-Goto action ceases
to be meaningful.}, lies at weak coupling.  There, the M5 is more
appropriately viewed as an NS5 with dissolved RR flux.  The worldvolume
action of the NS5 is simply the dimensional reduction of that of the M5
and the reduced Nambu-Goto term is reliable provided the NS5 is weakly
curved and a few other conditions, about which we will have more to say
in section \ref{subsubsec:reliability} and Appendix \ref{app:validity},
are met.  To summarize, once we find a minimal area M5 curve with the
right properties, we can use it to reliably describe our system as a
curved M5 at strong coupling or a curved NS5 with flux at weak coupling
provided we make appropriate choices of parameters.

When the numbers of branes and antibranes are equal, we are able to find
an \emph{exact} solution to the minimal area equations which has a
number of interesting properties.  
First and foremost, if we consider the regime in which the solution reliably describes our system at weak coupling as a curved NS5 with flux, it simplifies significantly to a configuration that is indeed $T$-dual to a deformed Calabi-Yau geometry with flux in type IIB\@.  Moreover, the moduli of the Calabi-Yau as determined by the minimal area condition in IIA \emph{exactly} solve the equations of motion which follow from the IIB potential derived in \cite{Aganagic:2006ex} using large $N$ duality and special geometry.  Consequently, we are able to understand large $N$ duality, even in this nonsupersymmetric context, from the IIA point of view as the replacement of the configuration of figure \ref{setupintro} by a curved NS5-brane with flux.  

Furthermore, we can extend this agreement to more general situations by studying the NS5 worldvolume action directly.  In the regime at weak coupling where 
our analysis is reliable,
we are able to explicitly demonstrate for arbitrary numbers of branes and antibranes that solving the equations of motion of this system is mathematically equivalent to starting with the deformed Calabi-Yau of type IIB and minimizing the potential obtained from special geometry.

That we find such agreement between the IIB and IIA pictures is slightly nontrivial and, for reasons that we now explain, further supports the idea that these brane/antibrane setups exhibit a degree of protection, at least at
small string coupling.
In particular, our IIA analysis implicitly assumes that the circle on
which we perform $T$-duality is large in string units while reliability
of the computations of \cite{Aganagic:2006ex} requires instead that the
dual circle on the IIB side be large.  Consequently, the two
descriptions we are comparing correspond to quite different parameter
regimes.  For supersymmetric situations, one does not worry about this
so much because the usual BPS arguments suggest that the system is
protected as one varies this radius.  When supersymmetry is broken,
though, this is not expected to be the case unless there is some
additional structure present.  As alluded to before, the authors of
\cite{Aganagic:2006ex} have argued that the brane/antibrane systems
under consideration still maintain some residual structure from
supersymmetry because, at least at string tree level, it is broken
spontaneously via FI terms.  It is this fact that must be responsible
for our ability to successfully relate the IIB and IIA stories at weak
coupling.

Given this success, it is natural to ask whether or not our system
remains protected, in any sense, as we move to the strong coupling
regime.  From our exact solution for the lift of figure \ref{setupintro}
with equal numbers of D4's and \anti{D4}'s, though, it is easy to see
that this does not seem to be the case in general.
The reason for this is that our solution exhibits new nonholomorphic
features which are small when the weak coupling interpretation is
reliable but which can become large when the strong coupling
interpretation is reliable.  The most obvious of these can be understood
by noting that it is favorable for the D4 and \anti{D4} stacks to tilt
slightly as depicted in figure \ref{wbendingintro}.  This is because the
energy cost associated with increasing their length is balanced by the
decrease in energy achieved when the branes and antibranes move closer
together.  Such tilting significantly impacts the entire geometry of the
resulting M5 curve because the D4's and \anti{D4}'s pull on and
``dimple'' the NS5's in a nonholomorphic way \cite{Witten:1997sc}.  In supersymmetric
setups, the direction of this ``dimpling'' is transverse to the NS5's
and gets combined with the RR gauge potential, or $x^{10}$ coordinate in
the $M$-theory language, to form a holomorphic quantity.  In the case at
hand, though, the ``dimpling'' is no longer transverse to the NS5's and
consequently nonholomorphicity is introduced throughout the curve, even
at infinity{\footnote{This is not the only nonholomorphic deformation of
the geometry that arises but it is the simplest to see without
discussing any details of the solution.}}$^,${\footnote{As we shall
explicitly demonstrate, the tilting described here becomes
parametrically small when the minimal area surface reliably describes
the system as an NS5 in IIA\@.  When the minimal area surface reliably
describes the system as an M5 in $M$-theory, this need not be the case
as we can choose it to be small or large.}}.

Having an exact solution to the minimal area equations in hand permits
us to not only see this nonholomorphic features explicitly but also to
demonstrate that they are controlled, at least when our description can
be trusted, by two parameters involving $g_s N$ and various
characteristic length scales of the geometry.  It is important to note
that, unlike at weak coupling, we can take these parameters to be large
at strong coupling while maintaining reliability of our description.
Hence, these features are truly present in at least some part of the
parameter space and are not simply an artifact of our formalism breaking
down.  However, the nice structure of our solution{\footnote{In
particular, the solution factorizes into a holomorphic piece, roughly
coming from the D4's, and an antiholomorphic piece, roughly coming from
the \anti{D4}'s, when both parameters are small.  Because each piece is
separately holomorphic with respect to a different complex structure,
each is individually supersymmetric but with respect to different sets
of supercharges.  In the absence of further backreaction which breaks
this factorization, the system thus seems to exhibit spontaneous
breaking of supersymmetry.  Of course, our solution is not reliable
everywhere so strictly speaking we only know for sure that this nice
structure is exhibited in the parameter regimes discussed in Appendix
\ref{app:validity}.}} seems to suggest that one can go further and
conjecture that the parameters we find are the only ones relevant for
determining the ``severity'' of supersymmetry breaking, meaning that the
system remains ``protected'' whenever both are small.

To summarize, we find that using the intuition of $M$-theory to view the configuration of figure \ref{setupintro} as a single object, namely an NS5 with flux at weak coupling or an M5 at strong coupling, allows us to not only obtain an alternative understanding of the large $N$ duality of \cite{Aganagic:2006ex} but also to probe the brane/antibrane system at strong coupling.  From this we learn that various features of this system seem to be protected as the radius of the $T$-dual circle is varied but that this protection does not persist throughout the full parameter space.  In particular, there exists a regime at strong coupling where our description is reliable and new nonholomorphic features become important.  In retrospect, perhaps it is not surprising that stringy interactions can remove the residual structure of the supersymmetry that is spontaneously broken at tree level.  It is gratifying to see this explicitly, though, and to learn something about the physics of metastable nonsupersymmetric configurations in string theory at strong coupling.

\subsection{Metastability of our configurations}

Finally, because this system admits IIA and IIB descriptions that we
understand well, it provides a nice example in which to study the
subtleties pointed out in \cite{Bena:2006rg} and how they arise in the geometric
engineering context.  A main point of emphasis in \cite{Bena:2006rg} is that
nonsupersymmetric configurations engineered in type IIA from NS5's and
D4's of the type we consider here have different boundary conditions,
once quantum effects are taken into account, from the supersymmetric
configurations into which they can decay.  We can see this quite easily
by studying the configuration of figure \ref{setupintro}, taking the numbers of
branes and antibranes to be equal for simplicity.  In the
nonsupersymmetric configuration, D4's and \anti{D4}'s pull on
the NS5's and dimple them as mentioned above in a manner that extends
out toward infinity.  The supersymmetric configuration that remains
after the branes annihilate, though, has no such bending because there
are no longer any D4's or \anti{D4}'s to cause it.  From this,
it is clear that the configurations have dramatically different boundary
conditions at infinity and hence should be viewed as states in different
quantum theories.

Does this mean that the nonsupersymmetric configurations we consider are
quantum mechanically stable?  It appears to us that the answer is no.
By annihilating the fluxes on the curved NS5-brane, the system can
indeed lower its energy and consequently it is favorable to do so via a
tunneling process.  Once the fluxes are gone, though, one can no longer
support the nontrivial curvature of the NS5-brane and it begins to
straighten.  This is much like following the ``quasikink'' solution of
\cite{Bena:2006rg} as the kink moves toward infinity{\footnote{More
precisely, the ``quasikink'' of \cite{Bena:2006rg} was actually the
opposite of what we discuss here with a supersymmetry-breaking
configuration in the interior glued to supersymmetric boundary
conditions.  The idea is the same, however.}}.  Because the kink never
actually reaches infinity in finite time, the system exhibits a runaway
behavior.  The decay wants to end, but the final state at which it can
end has moved off to infinity and hence left the theory entirely.  This
sort of picture has also been recently advocated in
\cite{Giveon:2007fk}.

As a result, our configurations are not in the spirit of \cite{Intriligator:2006dd} in that they are not metastable supersymmetry breaking configurations in a supersymmetric theory.  They are indeed metastable but, because of the boundary conditions, the theory in which they live is not supersymmetric.  Instead, supersymmetry is broken by a runaway potential in a manner that seems to be qualitatively similar to \cite{Affleck:1983mk}.

What does all of this mean from the type IIB point of view?  There, the bending of NS5's corresponds to turning on nontrivial NS 2-form $B^{NS}$ while energy stored in the NS5 tension is identified with the energy of NS 3-form flux $H^{NS}$.  When the RR-fluxes and ``anti''-RR-fluxes annihilate one another, the nontrivial $H^{NS}$ can no longer be supported and relaxes just as the NS5's did on the IIA side.  The picture of the decay process we had before thus carries over entirely, complete with runaway behavior.

There is a distinct difference, however, between the philosophy behind
typical NS5/D4 constructions in the literature and studies of the local
Calabi-Yau configurations to which they are related by $T$-duality which
affects how one interprets these results.  When one engineers gauge
theories and other perhaps nonsupersymmetric setups using extended NS5's
and D4's in type IIA, it is usually assumed from the outset that the
full theory under consideration is truly 10-dimensional type IIA with fully noncompact branes.
This means, for instance, that gauge theories realized in this manner
are provided with a specific UV completion from the outset, namely
MQCD{\footnote{In particular, this is precisely the situation studied by
the authors of \cite{Bena:2006rg}.}}.  That is not to say that a local
interpretation of these configurations is not possible but rather that
it does not seem as natural in the IIA context.

On the other hand, the philosophy behind local type IIB constructions is
quite different.  One imagines that the local Calabi-Yau is capturing
the physics in a particular region of a larger, compact Calabi-Yau.  The
situation is quite similar to effective field theory in that one imposes
a cutoff scale in order to perform computations and specifies the values
of noncompact moduli, which play the role of ``coupling constants'', at
that scale.  Of course, one can take the full noncompact Calabi-Yau
seriously by taking the cutoff scale to infinity{\footnote{One also
tries to keep IR quantities fixed in this limit.  Quantities for which
this is possible are well-described by effective field theory.}}.  This
would correspond to UV-completing the effective local description into
the $T$-dual of MQCD\@.  Typically, though, this is not our interest as
compact situations are more realistic for practical applications.

Nevertheless, we learn something very important about these local IIB
constructions from the subtleties that arose in type IIA\@.  While we
can study the tunneling process by which the fluxes and ``anti''-fluxes
annihilate in a local context, the eventual endpoint of the decay is
highly dependent on how we UV-complete the local configuration into a
compact Calabi-Yau.  This is already evident from the computations in
\cite{Aganagic:2006ex}, where the energy difference between the
supersymmetric and nonsupersymmetric configurations computed in a
regularization scheme with finite cutoff exhibits an explicit dependence
on that cutoff which cannot be removed.  The lesson here seems to be
that, while local constructions are useful for studying some aspects of
metastable systems in string theory, one must be careful of the inherent
limitations of such descriptions and take care to ask questions which
they are well-suited to answer.

\subsection{Outline of the paper}

The organization of this paper is as follows.  In section
\ref{sec:prelim}, we introduce the type IIB and IIA constructions that
shall be the focal point of our work and review their relation to one
another via $T$-duality.  In section \ref{sec:2susyex}, we review the
relation between these approaches in two supersymmetric examples.  In
the first, we consider branes at a single conifold singularity in type
IIB and their $T$-dual description in terms of a pair of NS5's with a
single stack of D4's suspended between them.  This will allow us to
review the basic philosophy behind the construction of parametric M5
curves.  In the second example, we consider branes at a pair of conifold
singularities and their $T$-dual description in terms of quadratically
bent NS5's with two stacks of D4's suspended between them.  This will
allow us to introduce a formalism for constructing genus-1 M5 curves
parametrically which will be required when considering nonsupersymmetric
configurations.  In section \ref{sec:br/anti}, we turn to the
brane/antibrane system of \cite{Aganagic:2006ex}, its $T$-dual
description in type IIA, and the exact $M$-theory lift.  In section
\ref{sec:disc}, we discuss various things we might learn from our
solution regarding the ``severity'' of supersymmetry breaking, address
issues related to boundary conditions and decays in greater detail, and
finally mention a few possible future directions.  The appendices
include various technical details.

\section{Preliminaries}
\label{sec:prelim}

In this section, we discuss the local Calabi-Yau geometries which play
an essential role in the IIB constructions of interest
\cite{Cachazo:2001jy,Cachazo:2001gh,Cachazo:2001sg} as well as the type
IIA brane setups \cite{Witten:1997sc,Witten:1998jd} to which they are
related by $T$-duality \cite{Ooguri:1995wj, Kutasov:1995te,
Karch:1998yv}.  Both of these are well-known, but we review them here
for completeness and in order to make precise the specific $T$-duality
dictionary we shall be using.

\subsection{IIB geometric constructions}

\begin{figure}
\begin{center}
\epsfig{file=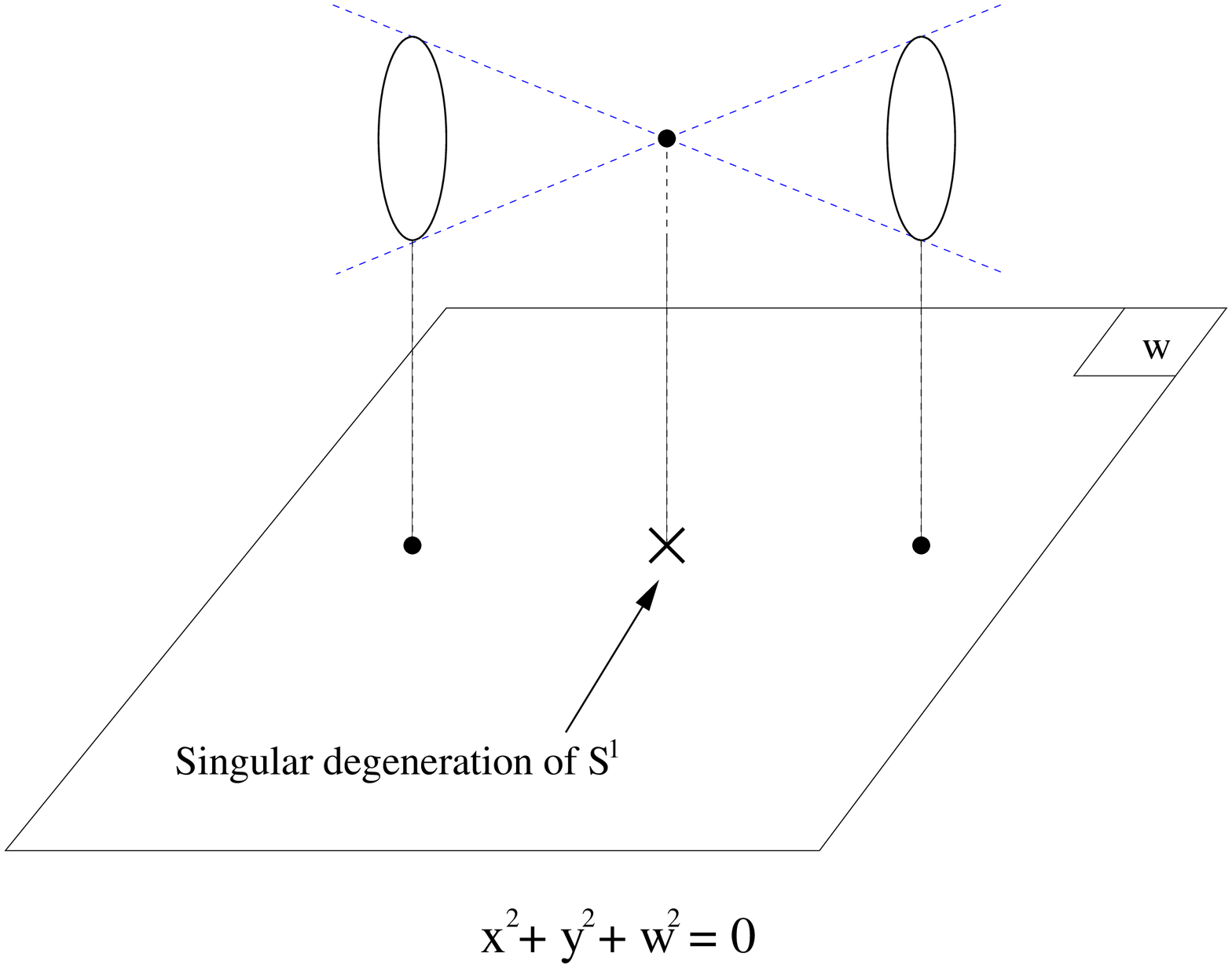,width=0.4\textwidth}
\hspace{0.1\textwidth}
\epsfig{file=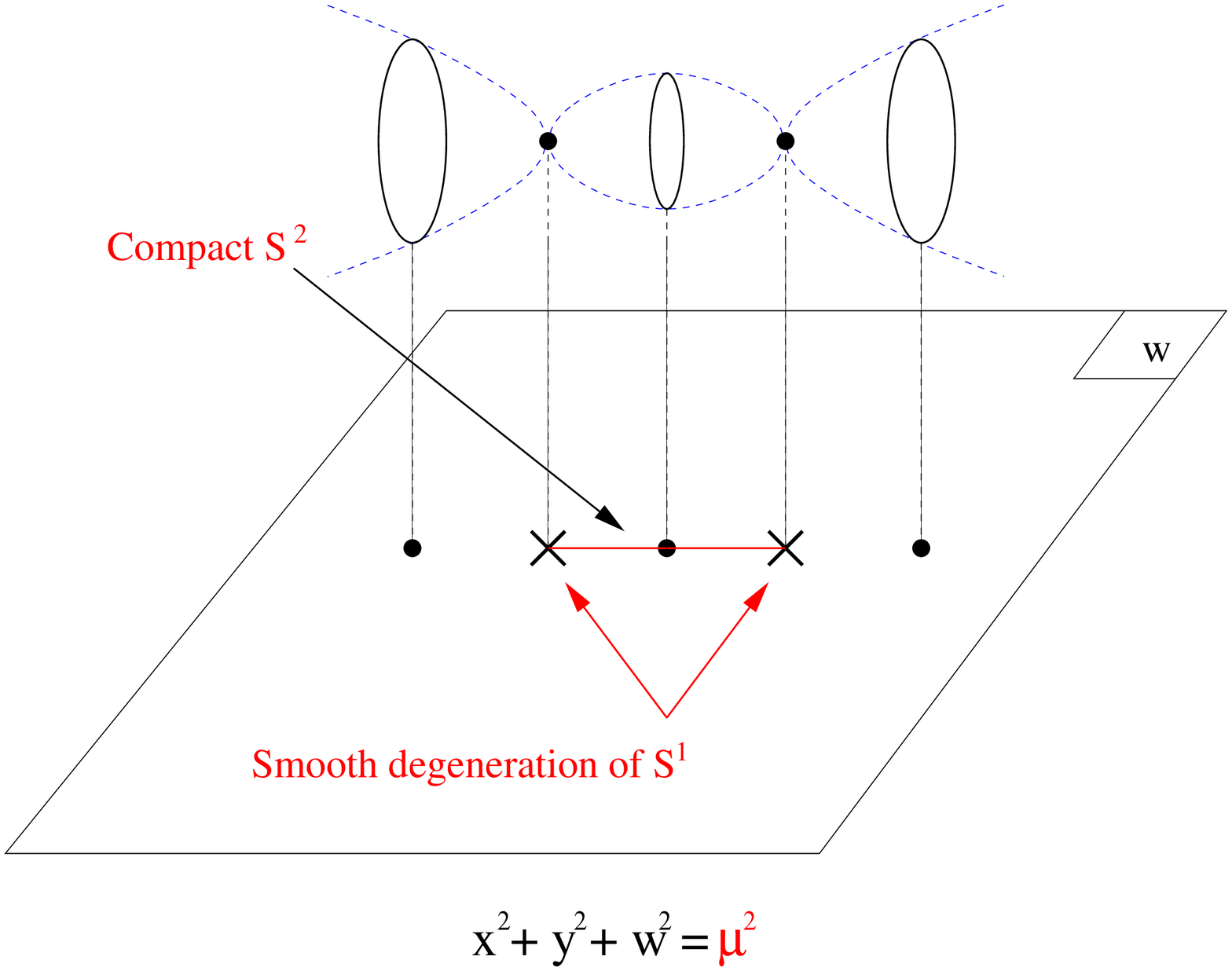,width=0.4\textwidth}
\caption{Singular and deformed $A_1$ singularities}
\label{A1sing}
\end{center}
\end{figure}

\begin{figure}
\begin{center}
\epsfig{file=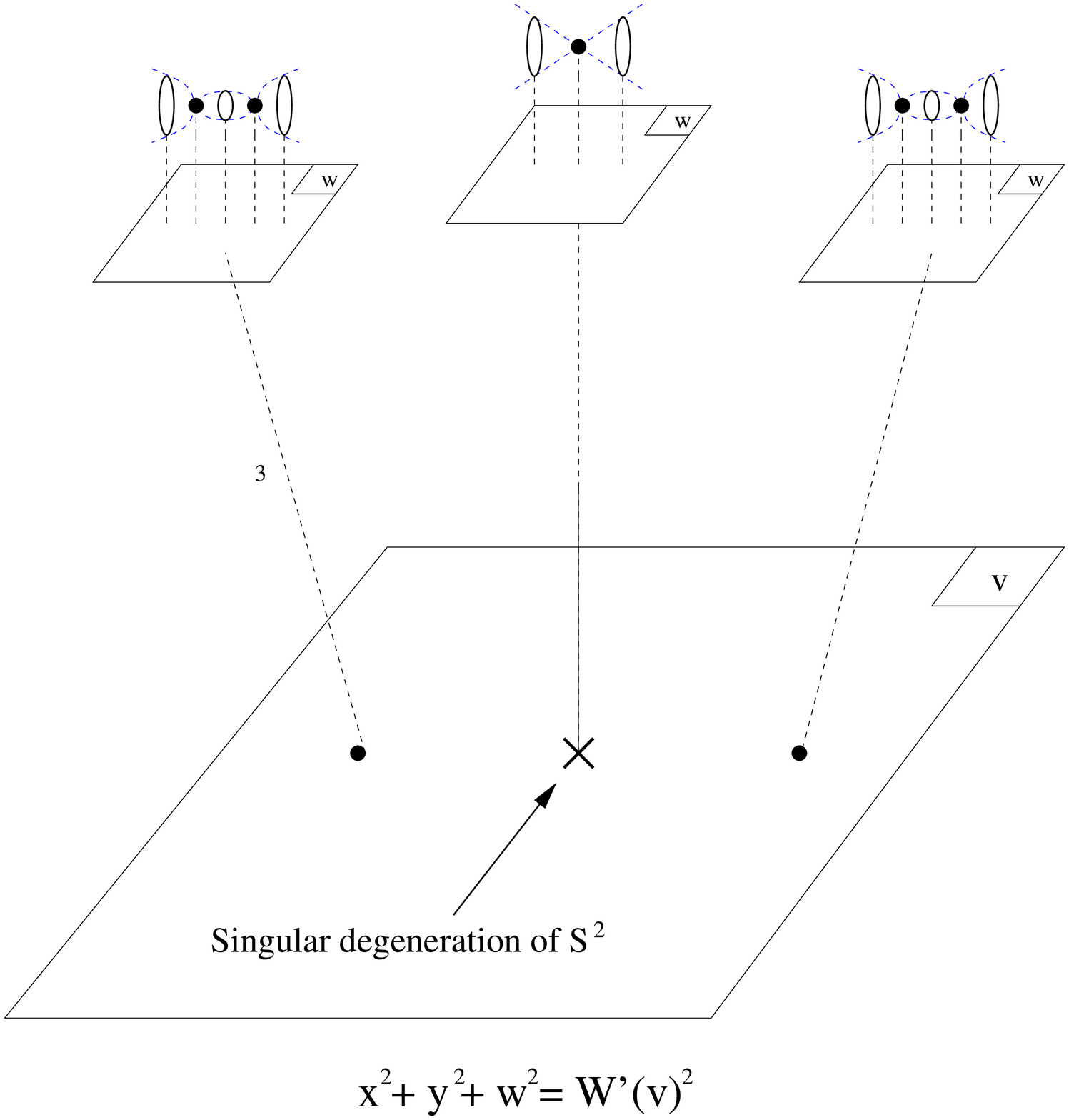,width=0.4\textwidth}
\hspace{0.1\textwidth}
\epsfig{file=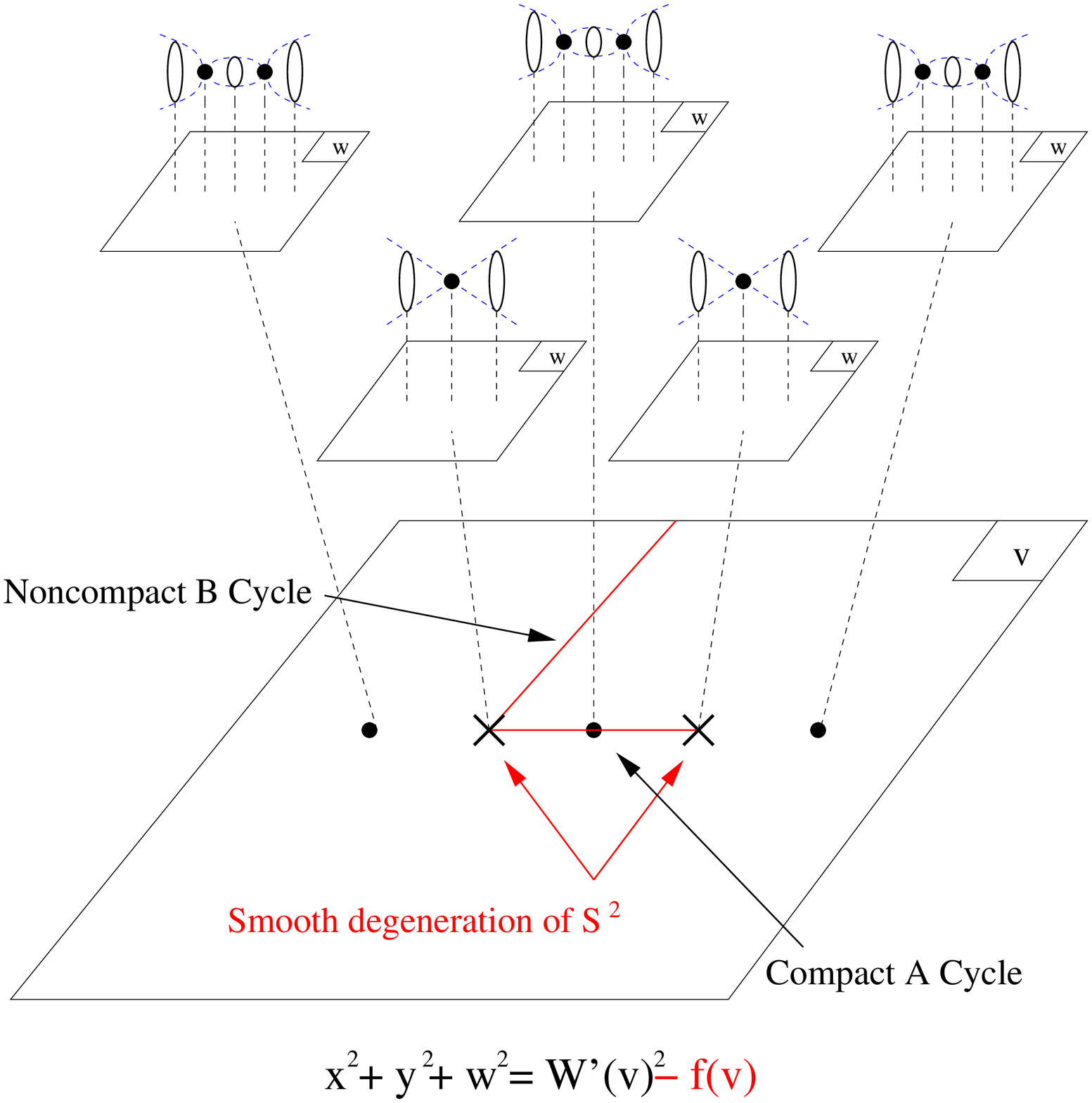,width=0.4\textwidth}
\caption{Singular and deformed $A_1$ fibration}
\label{A1fib}
\end{center}
\end{figure}

We begin by considering the $A_1$ ALE space, which can be realized by
the complex equation
\begin{equation}
x^2+y^2+w^2=0,\qquad x,y,z\in\bbC.
\end{equation}
One can view this as a $\mathbb{C}^{\ast}$ fibration over the $w$ plane as in figure \ref{A1sing}.
The singularity at $w=0$ corresponds to a singular double-degeneration of the nontrivial $S^1$ in $\mathbb{C}^{\ast}$ and can be removed by introducing a complex deformation $\mu$ as follows
\begin{equation}
x^2+y^2+w^2=\mu^2.\label{defale}
\end{equation}
The singular degeneration of $S^1$ at $w=0$ has now been replaced by
smooth degenerations at $w=\pm \mu$.  Fibering the $S^1$ over this
interval yields an $S^2$ which has grown in place of the singularity as
depicted in figure \ref{A1sing}.
We can now construct a local Calabi-Yau threefold by fibering this
deformed $A_1$ surface over a plane parametrized by a fourth complex
parameter, $v\in\bbC$, as in figure \ref{A1fib}.  This is easily accomplished
in \eqref{defale} by replacing the constant $\mu$ with a holomorphic
function $W'(v)$
\begin{equation}
x^2+y^2+w^2=W'(v)^2.\label{fibgeom}
\end{equation}
For generic values of $v$, the $S^2$ at the tip of the cone has finite
volume but it degenerates at the zeros of $W'(v)$.  Moreover, because
the roots of $W'(v)^2=0$ necessarily have multiplicity at least two,
this degeneration is singular.  To deal with this, one can proceed by
analogy to what we did for $A_1$ itself, namely introduce a complex
deformation that breaks the double-degeneracy
\begin{equation}
x^2+y^2+w^2=W'(v)^2-f(v).\label{defgeom}
\end{equation}
For generic nontrivial $f(v)$, each singular degeneration point of the
$S^2$ will split into two
points where the degeneration is smooth.  Fibering the $S^2$ over an
interval connecting these points then reveals that a compact $S^3$ has grown in place of the singularity.  We will refer to these 3-cycles as the ${\cal{A}}$ cycles of the geometry.  The dual ${\cal{B}}$-cycles of the geometry are noncompact and can be obtained by fibering the $S^2$ over an interval beginning at one point of a given pair and extending to infinity along $v$.  An illustration of this can be found in figure \ref{A1fib}.  

Let us now consider ``compactifying'' type IIB on the undeformed local
Calabi-Yau \eqref{fibgeom} and wrapping D5 branes on the degenerating $S^2$.  In order to prevent the effective 4d gauge coupling constant on the brane world-volume from diverging we must turn on a nontrivial NS-NS two-form field, $B^{NS}$, along the shrinking $S^2$.  A nontrivial $\theta$-angle can be introduced by turning on the $RR$ two-form $C_2^{RR}$ as well.  For a trivial fibration $W'(v)=0$, the world-volume theory is simply ${\cal{N}}=2$ SYM\@.  A simple expansion of the DBI action reveals that the $B$ and $C_2$ fields determine the effective 4d complexified coupling via
\begin{equation}
\frac{\theta}{2\pi}+\frac{4\pi i}{g_{YM}^2}=c+\frac{i b}{g_s}\label{potcoupling}
\end{equation}
where
\begin{equation}
b=\frac{1}{4\pi^2}\oint_{S^2\,\text{fiber}}B^{NS},\qquad\qquad c=\frac{1}{4\pi^2}\oint_{S^2\,\text{fiber}}C^{RR}_2.
\end{equation}
Note that, as we shall continue to do throughout all that follows, we have set
\begin{equation}
\alpha'=1.
\end{equation}
The fields of this theory include an adjoint scalar $\Phi$ which
parametrizes the location of the branes along $v$.  Nontrivially
fibering the deformed $A_1$ over the $v$-plane restricts the branes to
sit at the critical points of $W'(v)$, where the $S^2$'s they wrap
degenerate.  From the gauge theory point of view, this nontrivial
fibration corresponds to introducing a superpotential $W(\Phi)$ for the
adjoint superfield{\footnote{As the notation suggests, the function
$W'(v)$ that appears in the local Calabi-Yau \eqref{fibgeom} is nothing
more than the derivative of this superpotential \cite{Cachazo:2001gh}.}}
\cite{Cachazo:2001jy,Cachazo:2001gh}.  In what follows, we shall
consider superpotentials $W_n$ that are polynomials of degree $n+1$ and
restrict to deformations $f_{n-1}(v)$ that are polynomials of degree
$n-1${\footnote{In other words, we restrict to normalizable and
log-normalizable complex structure deformations
\cite{Cachazo:2001jy}.}}.  In these theories, the gauge group is
Higgs'ed according to $U(N)\rightarrow \prod_{i=1}^n U(N_i)$ with $N_i$
denoting the number of branes sitting at the $i$th critical point of
$W_n(v)$.

While this geometric construction provides a nice visualization for the
various Higgs branches that are present in the gauge theory, a direct
analysis of the quantum dynamics is not immediately obvious because it
requires going beyond the classical probe approximation for the
D5-branes.  It is by now well known that, in order to deal with this,
one can use large $N$ duality \cite{Cachazo:2001jy} to replace D5's at the singular points of the
geometry \eqref{fibgeom} with RR 3-form flux $H^{RR}$ wrapping $S^3$'s in the deformed
geometry \eqref{defgeom}.  In addition, we must also introduce some 3-form flux on the noncompact ${\cal{B}}$-cycles of \eqref{defgeom} in accordance with the $B^{NS}$ and $C_2$ that threaded the vanishing $S^2$'s of \eqref{fibgeom}.  The degrees of freedom of this system include $n$ chiral
superfields associated to the sizes of the $S^3$'s and $n$ Abelian vector
superfields obtained by reducing the $RR$ 4-form potential.  The former are identified with glueball superfields associated to the confined $SU(N_i)$ factors while the
latter correspond to the $n$ ``spectator'' $U(1)$'s.

Once we have replaced our brane configuration by deformed geometry with
fluxes, the quantum dynamics becomes easy to study because it is
captured by the well-known Gukov-Vafa-Witten (GVW) superpotential
\cite{Gukov:1999ya}
\begin{equation}
W=\int\,H\wedge\Omega,\label{gvw}
\end{equation}
where $\Omega$ is the holomorphic 3-form and
$H$ is a combination of the NS-NS and RR 3-forms
\begin{equation}
H=H^{RR}+\frac{i}{g_s}H^{NS}.
\end{equation}

Right away, however, we notice that noncompactness of the
${\cal{B}}$-cycles leads to problems because they have infinite
holomorphic volume.  In other words, the ${\cal{B}}$-periods of
$\Omega$, which appear directly in \eqref{gvw}, are divergent.  In order
to make \eqref{gvw} meaningful in practice, then, we must impose an
arbitrary cutoff $v_0$ on integrals over the base.  Concurrently, it is
also necessary to specify the boundary conditions of the system at this
cutoff scale.  We can accomplish this by fixing the integrals of the
2-form potential over the $S^2$ fiber at $v_0$
\begin{equation}
\int_{{\cal{B}}_i}^{v_0}H=\int_{S^2\text{ fiber at
}v_0}\left(C_2^{RR}+\frac{i}{g_s}B^{NS}\right)=4\pi^2\left[c(v_0)+\frac{i}{g_s}b(v_0)\right].
\end{equation}
which is equivalent to specifying the regulated $H$-flux along the
noncompact ${\cal{B}}$-cycles.  It is possible to incorporate
$v_0$-dependence in the boundary conditions $c(v_0),b(v_0)$ in such a
manner that explicit cutoff-dependence is removed from the
superpotential \eqref{gvw}.  From the gauge theory point of view, this
entire procedure is well-known.  We have simply introduced a UV cutoff
scale and specified the values of our ``coupling constants'' at that
scale.  Changes in cutoff scale must be accompanied by shifts in the
couplings consistent with RG flow if we wish to preserve the IR physics.

\begin{floatingfigure}{0.5\textwidth}
\begin{center}
%\vspace{0.5cm}
\epsfig{file=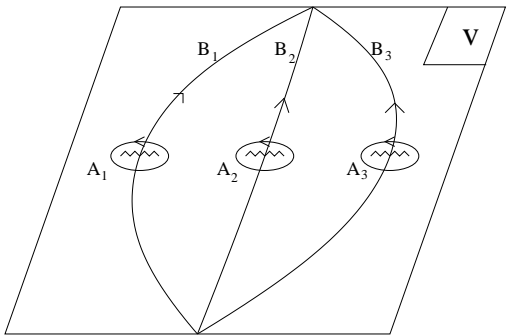,width=0.4\textwidth}
\caption{Sample hyperelliptic curve for $n=3$ with $A$ and $B$ cycles indicated.}
%\vspace{1cm}
\label{A1fibred}
\end{center}
\end{floatingfigure}

Finally, we note that it is often useful in practice to integrate $H$
and $\Omega$ over the $S^2$ fiber in order to reduce the problem to one
involving 1-forms defined on a Riemann surface.  We can do this
explicitly for $\Omega\,${\footnote{We insert the factor of
$\frac{1}{2}$ for convenience in order to absorb the factor of 2
difference between the ${\cal{A}}$ and ${\cal{B}}$ cycles of the local
Calabi-Yau, which pass along each cut once, and the $A$ and $B$ cycles
of the reduced hyperelliptic curve, which encircle each cut, effectively
passing along it twice (in equations,
$ \oint_\CA=\half\oint_A\oint _{S^2}$, $ \oint_\CB=\half\oint_B\oint _{S^2}$).}
\begin{equation}
\omega = \frac{1}{2}\oint_{S^2}\Omega=\frac{dv}{2}\sqrt{W'_n(v)^2-f_{n-1}(v)}\label{omegadef}.
\end{equation}
This 1-form is well-defined on the hyperelliptic curve
\begin{equation}
w^2=W_n'(v)^2-f_{n-1}(v),
\label{hyperdef}
\end{equation}
which can be visualized as a double-cover of the $v$-plane with cuts
connecting the various $S^2$ degeneration points.  The ${\cal{A}}$ and
${\cal{B}}$ cycles of the local Calabi-Yau \eqref{defconifold} now
descend to $A$ and $B$ cycles on this curve{\footnote{This is true up to
the usual factors of 2, which we have absorbed into the definitions of
$h$ and $\omega$.}}.  A convenient basis for these cycles is depicted in
figure \ref{A1fibred}.

On the other hand, we do not have an explicit expression for $H$ or its
corresponding reduced 1-form{\footnote{The minus sign is inserted for
convenience only.}}
\begin{equation}
h\equiv -\frac{1}{2}\oint_{S^2}H\label{hdef}
\end{equation}
but knowledge of the fluxes determines its periods along nontrivial
cycles of \eqref{hyperdef} as follows{\footnote{The factor of $4\pi^2$
here is simply the fundamental unit of 3-form flux,
$2\kappa_{10}^2\mu_5$, sourced by a single D5
brane.}}$^,${\footnote{Note that we use upper indices $N^i$ for the
fluxes as opposed to the lower indices $N_i$ denoting the number of
branes.  For branes, these are the same but for antibranes the sign is
opposite.}}
\begin{align}
\oint_{{\cal A}_i}H =4\pi^2 N^i\hspace{4.5ex}
  &\implies
 \frac{1}{2\pi i}\oint_{A_i}h=2\pi i N^i
\label{haper}
 \\
\oint_{{\cal B}_i}^{v_0}H=-4\pi^2\alpha(v_0)
 &\implies
 \frac{1}{2\pi i}\oint_{B_i}h=-2\pi i\alpha(v_0)
\label{hbper}
\end{align}
where we have defined
\begin{equation}
 -\alpha(v_0) \equiv c(v_0)+\frac{ib(v_0)}{g_s} = \frac{4\pi i}{g_{YM}^2}+\frac{\theta}{2\pi}.
\label{alpha_def}
\end{equation}

\subsection{IIA brane constructions}
\label{sec:IIAprelim}

Gauge theories that can be engineered using the type IIB constructions
reviewed in the previous section can also be obtained from brane
configurations in type IIA involving NS5's and D4's \cite{Witten:1997sc,
Witten:1998jd,Giveon:1998sr}, as we now briefly review.  In the
following, we shall consider configurations with two NS5-branes and $N$
D4-branes extended along the 0123 directions.  The NS5's also extend
along holomorphic curves of the form $w=w(v)$ where
\begin{equation}
v=x^4+ix^5,\qquad\qquad w=x^7+ix^8
\end{equation}
and are separated along $x^6$ by a distance $L$.  The D4's are then
suspended between the NS5's along $x^6$.

\begin{floatingfigure}{0.5\textwidth}
\begin{center}
\epsfig{file=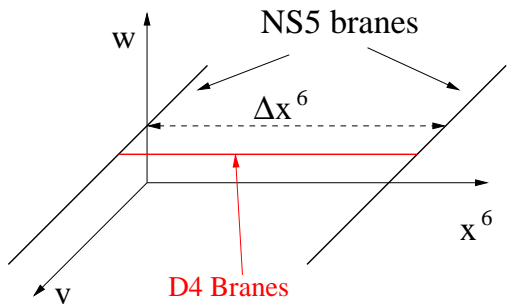,width=0.4\textwidth} \caption{A sample
NS5/D4-brane configuration which realizes ${\cal{N}}=2$ SYM on the D4
worldvolume}\label{HWex}
 \vspace*{4ex}
\end{center}
\end{floatingfigure}

Let us begin by considering the case of parallel NS5-branes wrapping the curves $w=0$ in $wv$ space.  This is the situation depicted in figure \ref{HWex}.  If we scale the length $L$ of the D4-branes along $x^6$ to zero, the theory on their worldvolume becomes effectively four-dimensional with gauge coupling constant given by
\begin{equation}
\frac{8\pi^2}{g_{YM}^2}=\frac{L}{g_s\sqrt{\alpha'}}.
\label{iibgym}
\end{equation}
The brane configuration preserves 8 supercharges so this theory has ${\cal{N}}=2$ supersymmetry in four dimensions.  The ${\cal{N}}=2$ vector superfield consists of an ${\cal{N}}=1$ vector superfield as well as an adjoint chiral superfield which parametrizes the location of the D4's along $v$.  

We can now introduce a superpotential $W(\Phi)$ for the adjoint
superfield by extending the NS5-branes instead along the nontrivial
curves \cite{Witten:1998jd, Hori:1997ab, deBoer:1997ap, Giveon:1998sr}
\begin{equation}
w(v)=\pm W'(v).\label{wvbdry}
\end{equation}
The various ways of distributing D4-branes at the critical points of $W(v)$ correspond to the different Higgs branches of the theory. 

While the classical brane picture is useful for visualizing the Higgs structure it is easy to see that, like the wrapped D5 brane configuration of the previous section, it obscures the quantum dynamics.  The reason for this is that the setup does not describe a key element of the quantum system, namely backreaction of the D4's on the NS5's.  In particular, it is well known that the NS5 throat is a region of large string coupling so it is difficult to analyze the NS5/D4 intersection, which plays a dominant role particularly when $L\ll \sqrt{\alpha'}$, in a IIA context.

To deal with this, Witten noted that NS5's and D4's are two different manifestations of the same object, namely the $M5$ brane, and hence their intersection could be smoothed out by looking at this system from the point of view of $M$-theory \cite{Witten:1997sc}.  There, our NS5/D4 configuration is thought of instead as a single $M5$ brane extended along a possibly complicated 6-dimensional hypersurface.  At large $g_s$ and small 11-dimensional Planck length, $\ell_{11}$, the worldvolume theory of the $M5$ is effectively described by the Nambu-Goto action so its embedding into target space  is one of minimal area.  At small $g_s$, on the other hand, this $M5$ is better thought of as a curved NS5-brane with dissolved RR flux whose worldvolume theory is obtained by dimensional reduction.  When the descendant of the Nambu-Goto term gives a reliable description of physics in the IR, this NS5 configuration can be obtained directly from the $M5$ one at large $g_s$ by reducing along
  the $M$-circle{\footnote{For supersymmetric configurations, one typically performs this IIA reduction without a second thought as the M5 is protected in such cases from corrections that arise as $g_s$ is decreased.  Because we are eventually interested in nonsupersymmetric configurations, we shall always be careful to specify when using the Nambu-Goto term at small $g_s$ is reliable.}}.  In practice, this simply means that we reinterpret the coordinate $x^{10}$ in our M5 solutions as the appropriate RR gauge potential.

In the situation at hand, the $M5$ brane extends along the 0123
directions and, due to supersymmetry, wraps a holomorphic curve $\Sigma$
in the remaining directions.  Because each D4 stack in the IIA
configuration fattens into a tube upon lifting to $M$ theory, the genus
of $\Sigma$ is related to the number $n$ of such stacks by $g=n-1$.  It
is convenient to use one of the complex coordinates, say $v$, to
parametrize this curve, permitting us to think of it as a double cover
of the $v$-plane with $n$ cuts{\footnote{Of course, this relies on the
fact that $\Sigma$ is actually hyperelliptic \cite{Witten:1997sc,
Witten:1998jd}.}}.

To determine the correct lift, we must impose the boundary conditions
\eqref{wvbdry} along the $w$ and $v$ directions as well as require that
the wrapping along $x^{10}$ be consistent with the D4 distribution in
IIA\@.  Turning first to $w(v)$, a hyperelliptic curve satisfying
$w(v)\sim \pm W'(v)$ as $v\to \infty$ can be written in the form
\begin{equation}
w(v)=\sqrt{W'(v)^2-f_{n-1}(v)}.
\label{wvmlift}
\end{equation}
Note that this is precisely the curve on which the 1-form $\omega$
\eqref{omegadef} was defined in the previous section.  As such, we will
continue to the basis of $A$ and $B$ cycles indicated in figure
\ref{A1fibred}.

To deal with the $x^{10}$ constraint, we first combine $x^{10}$ with
$x^6$ into a holomorphic variable \cite{Witten:1997sc}
\begin{equation}
s=R^{-1}(x^6+ix^{10}),
\label{s_def}
\end{equation}
where $R=g_s\sqrt{\ap}$ is the radius of the $M$-circle.  This is
crucial because in the end we are looking for a holomorphic curve.  With
this definition, the condition that $x^{10}$ wrappings be consistent
with the distribution of D4's in the IIA reduction can be expressed as
a constraint on the $A$-periods of the 1-form $ds$
\begin{equation}
\oint_{A_i}\,ds = 2\pi i N_i.\label{dsaper}
\end{equation}
The $B$-periods of $ds$, on the other hand, are related to the
separation between the NS5's along $x^6$ (and potentially along $x^{10}$
as well).  Generically, $x^6$ will vary logarithmically with $v$ due to
the fact that the D4's pull on and ``dimple'' the NS5
\cite{Witten:1997sc}.  As a result, integrals of $ds$ along the
noncompact $B$-cycle will in general diverge, forcing us to introduce a
cutoff on the $v$-integration at an arbitrary point $v_0$.  This
corresponds to introducing a UV cutoff in the gauge theory on the D4
worldvolume.  We identify this separation with the (complexified) 4d
gauge coupling at this scale
\begin{equation}
\oint_{B_i}^{v_0}\,ds =-2\pi i\alpha(v_0) =  -\frac{8\pi^2}{g_{YM}^2(v_0)}+i\theta_i(v_0)
\label{dsbper}
\end{equation}
and hence the dependence on $v_0$ simply corresponds to RG flow.  Once we have determined the holomorphic curve $\Sigma$, we can reduce to IIA by simply reinterpreting $x^{10}$, thus obtaining the curved NS5-brane with flux that results from backreaction of the D4's.

The $M$-theory description, as we have reviewed it so far, is purely
on-shell so that, unlike the case of IIB with fluxes, we have no analog
of the GVW superpotential that allows one to have an off-shell
understanding of the system.  Later, we will see one sense in which this
description can be extended to capture some off-shell information.
Another way this can be accomplished was suggested by Witten, who
conjectured a form for the superpotential which captures off-shell
physics of an $M5$ wrapping $\Sigma$ \cite{Witten:1998jd}.  In
particular, if we let $\Sigma_0$ denote a reference surface homologous
to $\Sigma$ and $\tilde{B}$ a 3-chain interpolating between the two,
this superpotential he wrote is
\begin{equation}
W(\Sigma)-W(\Sigma_0)=\frac{1}{2\pi i}\int_{\tilde{B}}\Omega.
\end{equation}
Later, de Boer and de Haro \cite{deBoer:2004he} noted that this can be rewritten in the form
\begin{equation}
W\sim \int_{\Sigma}ds\wedge w\,dv
\label{wdbdh}
\end{equation}
which is quite suggestive given the similarities between the 1-forms
$w\,dv$ and $ds$ here and the 1-forms $\omega$ \eqref{omegadef} and $h$
\eqref{hdef} of the previous section.  Of course, as we now review, this
similarity is not an accident.

\subsection{T-duality between IIA and IIB constructions}
\label{subsec:T-duality}

The above realizations of gauge theories in type IIA and IIB string
theories are related by $T$-duality \cite{Ooguri:1995wj, Kutasov:1995te,
Karch:1998yv, Dasgupta:2001um,Oh:2001bf}.  Here we briefly review this
relation and establish the mapping between quantities on both sides.

As we have seen, the type IIB construction is based on fibration of
deformed $A_1$ ALE space \eqref{defale} over the complex $v$-plane.
Consequently, if we understand the $T$-duality between the ALE fiber
over a point $v$ and the type IIA brane configuration at $v$,
the $T$-duality relation between the whole systems will follow by applying it fiberwise.
Therefore, we focus here on the $T$-duality between ALE space and NS5-branes.

The deformed $A_1$ ALE space whose complex structure is displayed in
\eqref{defale} can be realized as a two-center Taub-NUT space (see e.g.\
\cite{Witten:1997sc}, section 3).\footnote{Strictly speaking, a Taub-NUT
space is ALF and we must take $R_{\rm IIB}\to \infty$ to make it ALE\@.}
The metric of the Euclidean $k$-center Taub-NUT space is
\begin{equation}
\begin{split}
 ds^2_{\rm IIB}&=H^{-1}(d\yt+\omega)^2 +H\, d\zv^2+ds_\perp^2,\qquad 
 e^{2\Phi_{\rm IIB}}=1,\qquad
 0\le \yt\le 2\pi R_{\rm IIB},\\
 H(\zv)&=1+\sum_{p=1}^{k} H_p,\qquad
 H_p(\zv)={R_{\rm IIB}\over 2|\zv-\zv_p|},\qquad
 d\omega=*_3 dH.
\end{split}\label{k-TN_metric}
\end{equation}
Here $\zv_p$, $p=1,\dots, k$, is the position of the $p$-th center in
the base $K=\bbR^3$ parametrized by $\zv=(x^7,x^8,x^9)=(\Re w,\Im w,x^9)$,
$\omega(\zv)$ is a 1-form in $K$, and $ds^2_\perp$
is the metric for the remaining six directions 012345.
If we $T$-dualize the Taub-NUT metric \eqref{k-TN_metric} along $\yt$
using the standard Buscher rule \cite{Buscher:1987qj}, we obtain the
IIA metric
\begin{align}
  ds^2_{\rm IIA}&= H(d y^2+d\zv^2)+ds^2_\perp,\qquad
 e^{2\Phi_{\rm IIA}}=H,\qquad
 B_{\rm IIA}=\omega\wedge d\yt.
 \label{NS5_metric}
\end{align}
Here $y$ is the $T$-dual of $\yt$ whose periodicity is
\begin{align}
 y&\cong y +2\pi R_{\rm IIA},\qquad
 R_{\rm IIA}={\ap\over R_{\rm IIB}},\label{y_period}
\end{align}
and corresponds to $x^6$ in the last subsection.  The metric
\eqref{NS5_metric} is nothing but the geometry produced by $k$
NS5-branes located at $\zv=\zv_p$ in flat space.  In particular, if we
set $k=2$ and $\zv_{1,2}=\pm (\Re\mu,\Im\mu,0)$ (\textit{i.e.}, $w=\pm
\mu$), this shows that the deformed $A_1$ ALE space \eqref{defale} is
$T$-dual to two NS5-branes at $w=\pm \mu$.  Fibering this $T$-duality
over the $v$-plane, we see that the local CY space \eqref{fibgeom} is
$T$-dual to two NS5-branes placed along the $wv$ curve \eqref{wvbdry} in
a flat space.

In the metric \eqref{NS5_metric}, though, the NS5-branes are delocalized
(smeared) in the $y=x^6$ direction.  However, in string theory, the
NS5-branes are expected to become localized; indeed, it is known that
the $y=x^6$ position of the IIA NS5-brane is dual to $B$-field through
certain 2-cycles in the IIB Taub-NUT geometry \cite{Sen:1997zb,
Sen:1997js}.  Although one could study this localization of NS5-branes
using worldsheet CFT techniques \cite{Tong:2002rq, Harvey:2005ab}, in
Appendix \ref{app:T-dual_of_B} we have presented an alternative approach
to determining the position of NS5-branes, which, to our knowledge, is
new.

From \eqref{CB_Delta-s}, the 2-form fields in IIB are related to the
distance between two NS5-branes in IIA in the following manner:
\begin{align}
\int \left(C_2^{RR}+{i\over g_s^{\rm IIB}}B_2^{NS}\right)
 =4\pi^2\left(c+{i\over g_s^{\rm IIB}}b\right)=-4\pi^2\alpha(v_0)
 = -2\pi i \,\Delta s,
 \label{s_and_2-forms}
\end{align}
where $\alpha$ was defined in \eqref{alpha_def} and $s$ in
\eqref{s_def}.  The gauge theory couplings derived in IIB and IIA (Eqs.\
\eqref{potcoupling} and \eqref{iibgym}) can be shown to be identical
using this relation, as they should be.  From \eqref{s_and_2-forms}
immediately follows also the correspondence between the following
1-forms in IIB and IIA:
\begin{align}
 \frac{h}{2\pi i}~~\longleftrightarrow~~ ds.
\label{shident}
\end{align}
From this relation, it is clear that the periods of $h$ in IIB,
\eqref{haper} and \eqref{hbper}, are mapped into the periods of $ds$ in IIA,
\eqref{dsaper} and \eqref{dsbper}.  If we further note the equivalence
between the following 1-forms in IIB and IIA (Eq.\ \eqref{omegadef} and
\eqref{wvmlift}):
\begin{align}
 \omega&=\frac{1}{2}\oint_{S^2} \Omega
 ~~\longleftrightarrow ~~
 {1\over 2}w\,dv,
\end{align}
we can see that the IIB superpotential \eqref{gvw} is equivalent to the
IIA superpotential \eqref{wdbdh}:
\begin{align}
 W_{\rm IIB}\sim \int H\wedge \Omega \sim \int h\wedge \omega
 ~~\longleftrightarrow ~~
 W_{\rm IIA}\sim \int ds\wedge w\,dv.
\end{align}

Summarizing the discussion so far, local CY geometries in type IIB are
$T$-dual to NS5-brane configurations in IIA, and the realization of
gauge theories based on them are equivalent.  
There is one important issue that we have glossed over, though.  The $y$
and $\yt$ circles are compact with radii $R_{\rm IIA}$ and $R_{\rm
IIB}$, respectively, and are related to each other by
\eqref{y_period}.  In the IIB and IIA/M constructions in the previous
sections, we treated these circles as if they were noncompact, by
putting D5's in a noncompact (local) CY in IIB and putting NS5's and
D4's in noncompact $\bbR^6$ in IIA (or an M5 in $\bbR^6\times
S^1_{10}$).  However, the validity of such ``noncompact'' description is
not obvious, because if $R_{\rm IIA}$ is large then $R_{\rm IIB}$ is
small, and \textit{vice versa.}
In the supersymmetric case, if we take the gauge theory limit
(decoupling limit) where the scales of the system becomes
substringy,\footnote{In the gauge theory limit in type IIA (IIB), we
take the length $L$ of D4 (the size of the $S^2$ on which D5 is
wrapping) and the distance $\delta$ between different D4 stacks (D5
stacks) to be substringy, such that $L\sim g_s \sqrt{\ap}$ and
$\delta\sim\ap E$, where $E$ is the energy scale we are looking at.}
such a noncompact description can indeed be justified because the circle
direction becomes much larger than the system size if we take $R_{\rm
IIA}\sim R_{\rm IIB}\sim \sqrt{\ap}$.
What if we do not take the gauge theory limit?  Even then, as long as we
focus on holomorphic quantities such as the curve \eqref{hyperdef},
\eqref{wvmlift}, we can still use the noncompact description.  This is
because these holomorphic quantities are protected by supersymmetry and
do not depend on the scales of the system such as $R_{\rm IIA}, R_{\rm
IIB}$.  Namely, we are free to take them to be infinite.  In this sense,
the noncompact IIB and IIA/M constructions in previous sections are
$T$-dual to each other if supersymmetry is preserved.

In the nonsupersymmetric case that we shall study later, things can be
more subtle.  In order for the fundamental string stretching between
D-branes and anti-D-branes to be free of tachyonic modes, we must keep
the distance $\delta$ between them to be at least of the order of the
string length: $\delta\sim \sqrt{\ap}$.  However, because of the
relation \eqref{y_period}, it is impossible to make both $R_{\rm IIA}$
and $R_{\rm IIB}$ much larger than $\delta\sim \sqrt{\ap}$ at the same
time.  So, the {\em full\/} physics of the noncompact IIB and IIA/M
constructions is not going to be the same.  So, in a strict sense, by
studying noncompact IIA/M system we will be exploring the
nonsupersymmetric physics of a {\em new\/} system which is {\em
different\/} from the IIB system studied in \cite{Aganagic:2006ex,
Heckman:2007wk}.

However, even in the nonsupersymmetric case, it is possible that certain
quantities are still protected, if the supersymmetry breaking is soft
\cite{Lawrence:2004zk}.  For such quantities, scale parameters $R_{\rm
IIA,IIB}$ are again irrelevant.  Therefore, as far as such data are
concerned, we can still say that the noncompact IIB and IIA/M
constructions are in fact $T$-dual to each other and describing the {\em
same\/} physics.  Indeed, we will see that certain quantities computed in
IIA/M are the same as ones computed in IIB \cite{Aganagic:2006ex,
Heckman:2007wk}, although supersymmetry is broken.

\section{Two Supersymmetric Examples}
\label{sec:2susyex}

We now proceed to elaborate upon the connection between the type IIB and
IIA constructions reviewed in the previous section by looking at a pair
of simple examples.  Among other things, this will permit us to review
the parametric representation of genus zero $M5$ curves
\cite{Witten:1998jd} and introduce the formalism for extending this sort
of description to genus one situations in a more friendly,
supersymmetric setting.

\subsection{$A_1$ theory with quadratic superpotential --- IIB}
\label{subsec:A1quadIIB}

We begin with perhaps the simplest possible example, namely that of D5
branes at a conifold singularity \cite{Vafa:2000wi, Cachazo:2001jy}.
The relevant geometry is \eqref{fibgeom} with a quadratic superpotential
of the form $W(v)=m v^2/2$.  After implementing large $N$ duality, we
obtain the deformed geometry
\begin{equation}
x^2+y^2+w^2=m^2 v^2+f_0
\label{defconifold}
\end{equation}
with 3-form fluxes on the compact and noncompact cycles
\begin{equation}
\frac{1}{2\pi i}\oint_{\cal{A}} H = 4\pi^2 N,
\qquad\qquad \frac{1}{2\pi i}\oint_{\cal{B}} H=-4\pi^2\alpha.
\label{simpexfluxes}
\end{equation} 
This geometry has one modulus, $f_0$, whose value is fixed dynamically.
To study this further, we introduce ${\cal{A}}$ and ${\cal{B}}$ periods
of the holomorphic form $\Omega$ as usual
\begin{equation}
S\equiv\frac{1}{2\pi i}\oint_{\cal{A}}\Omega,
 \qquad\qquad
 \Pi\equiv\frac{1}{2\pi i}\oint_{\cal{B}}\Omega = \frac{\partial{\cal{F}}}{\partial S},
\label{SPidef}\end{equation}
where ${\cal{F}}$ is the ${\cal{N}}=2$ prepotential.  The field $S$ serves as an alternate means of parametrizing the (one-dimensional) moduli space of complex structures and, in fact, its relation to the modulus $f_0$ is easy to work out in this simple example 
\begin{equation}
S = -\frac{f_0}{4m}.
\end{equation}
The expectation value of $S$, and hence of $f_0$, can now be obtained by minimizing the GVW superpotential \eqref{gvw}.  Using the Riemann bilinear relations, one can rewrite $W_{GVW}$ as
\begin{equation}
W_{GVW}=\int\,H\wedge\Omega \sim N\Pi+\alpha S
\end{equation}
and immediately obtain the supersymmetric vacuum condition
\begin{equation}
\alpha + \hat{\tau}N=0
\label{susyvacuum}
\end{equation}
where $\hat{\tau}$ is the period ``matrix'' of the Calabi-Yau%
{\footnote{We use a rather unconventional notation here, with
$\hat{\tau}$ denoting the period matrix as opposed to $\tau$.  The
reason for this is to avoid confusion later when $\tau$ is used as the
complex structure modulus of an auxiliary torus.}}:
\begin{equation}
\hat{\tau}\equiv\frac{\partial\Pi}{\partial S}=\frac{\partial^2{\cal{F}}}{\partial S^2}.
\end{equation}

Though the solution $\hat{\tau}=-\alpha/N$ to \eqref{susyvacuum} provides a perfectly good description of the complex structure of \eqref{defconifold} at the supersymmetric vacuum, it is often desirable to translate this into a statement about the expectation value for $S$.  
This requires us to compute ${\hat{\tau}}(S)$, a task that is easily achieved in this simple example.
 Imposing a cutoff $v_0$ on $B$-periods as discussed in the previous section, we find
\begin{equation}
\hat{\tau}=\frac{1}{2\pi i}\ln\left(\frac{S}{mv_0^2}\right)+\ldots,
\end{equation}
where terms that vanish as $v_0\rightarrow\infty$ have been dropped.  Inverting this and applying \eqref{susyvacuum} the expectation value immediately follows
\begin{equation}
S^N=(mv_0^2)^Ne^{-2\pi i\alpha(v_0)}=m^N\Lambda_{n=1}^{2N}.
\label{svev}
\end{equation}
In this expression, we have exhibited the explicit $v_0$-dependence of
the ``coupling constant'' $\alpha$ that is needed to render $W_{GVW}$
cutoff-independent as well as introduced an ``RG-invariant'' scale
$\Lambda_{n=1}\,${\footnote{The subscript $n=1$ refers to the fact that, in the theory we are studying, $W'(v)$ has degree $n=1$.  This is to distinguish $\Lambda_{n=1}$ from the analogous quantity introduced later for theories having $W'(v)$ of degree 2.}}
\begin{equation}\Lambda_{n=1}^{2N}\equiv v_0^{2N}e^{-2\pi i\alpha(v_0)}=v_0^{2N}\exp\left\{-\frac{8\pi^2}{g_{YM}^2(v_0)}+i\theta(v_0)\right\}.\label{lambdan1def}\end{equation}
This completes our brief review of the system obtained by wrapping $N$ D5 branes at a conifold singularity.  We have seen that the supersymmetric vacuum is described by the deformed geometry \eqref{defconifold} with fluxes \eqref{simpexfluxes} and complex modulus \eqref{svev}.

\subsection{$A_1$ theory with quadratic superpotential --- IIA/M}
\label{subsec:A1quadIIA}

\begin{floatingfigure}{0.5\textwidth}
\begin{center}
\epsfig{file=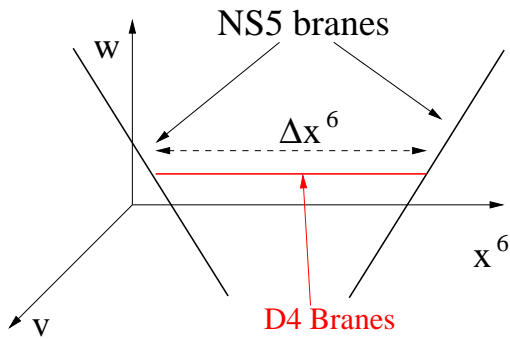,width=0.4\textwidth}
\caption{NS5/D4 configuration obtained by applying $T$-duality to collection of D5 branes at a conifold singularity.}
\label{A1quadTdualD4}
\end{center}
\end{floatingfigure}

We now proceed to study this system from the IIA/M perspective
\cite{Witten:1998jd, Dasgupta:2001um, Oh:2001bf} .  Applying
$T$-duality, we obtain a brane configuration with two NS5's extended
along the curves $w=\pm mv$ and separated along $x^6$ with $N$ D4's
suspended between.  This configuration is depicted in figure
\ref{A1quadTdualD4}.  As discussed in section \ref{sec:IIAprelim}, in
order to describe this system away from $g_s=0$ we should view it
instead as an M5 extended along a genus zero holomorphic curve with
boundary conditions
\begin{equation}
w\sim\pm mv\qquad\text{ as }v\rightarrow\infty
\label{wvlinbdry}
\end{equation}
and embedding coordinate $s$, defined in \eqref{s_def} to describe the
wrapping along $x^{10}$, satisfying
\begin{align}
\oint_{A}\,ds &= 2\pi i N,\label{sacond}\\
\oint_{B}\,ds &= -2\pi i\alpha(v_0).\label{sbcond}
\end{align}
Though an explicit representation of this curve is well-known
\cite{Witten:1998jd}, we shall review the parametric one here because it
will more easily generalize to the nonsupersymmetric curves of interest
later.

The curve we seek to study has genus zero so it can be parametrized by a single copy of the complex plane with a pair of marked points, corresponding to the preimages of $\infty$ on each of the two NS5-branes.  For definiteness, we refer to the complex parameter as $\lambda$ and place the marked points at $\lambda=0$ and $\lambda=\infty$.  At these points, the holomorphic functions $w(\lambda)$ and $v(\lambda)$ must diverge and, moreover, because the embedding is 1-1 near $\infty$ these divergences must come in the form of first order poles.
Combined with the boundary conditions \eqref{wvlinbdry}, this is sufficient to fix their form up to an overall rescaling
\begin{equation}
\begin{split}
v(\lambda)&=\lambda+\frac{a}{\lambda},\\
w(\lambda)&=m\left(\lambda-\frac{a}{\lambda}\right).
\label{wvcurve}
\end{split}
\end{equation}
From this, it is easy to verify that $w$ and $v$ are related as in \eqref{wvmlift}
\begin{equation}
w^2=m^2\left(v^2-4a\right)
\end{equation}
and hence that \eqref{wvcurve} provides a parametric description of the hyperelliptic curve depicted in figure \ref{quadsupex}.  The $A$ and $B$ cycles shown there appear on the $\lambda$ plane as illustrated in figure \ref{lambdaplane}.

\begin{figure}
\begin{center}
\subfigure[Hyperelliptic curve \eqref{wvcurve} with $A$ and $B$ cycles indicated]
{\epsfig{file=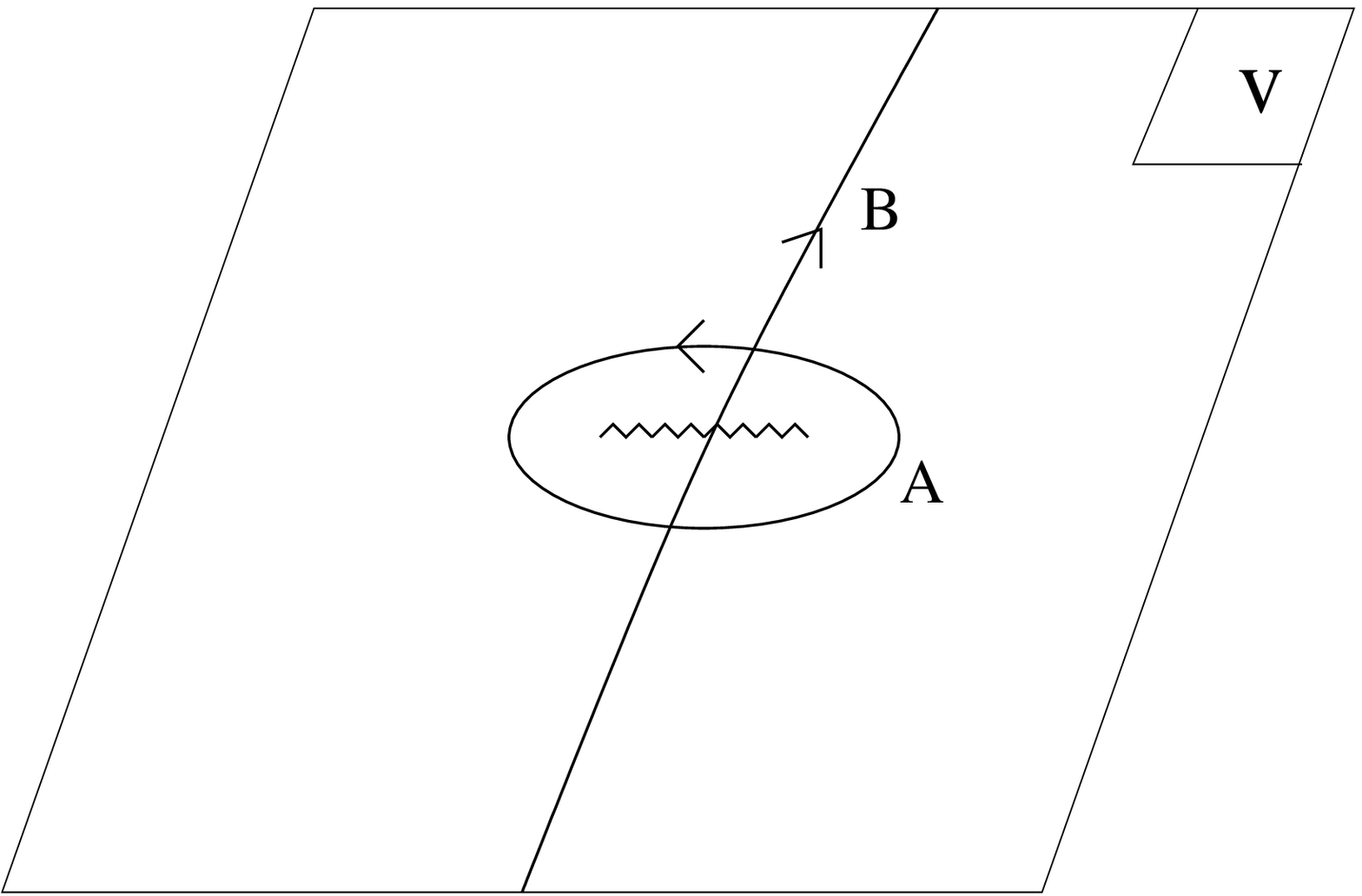,width=0.4\textwidth}\label{quadsupex}}
\subfigure[$\lambda$ plane used to parametrize the hyperelliptic curve \eqref{wvcurve} with $A$ and $B$ cycles indicated]
{\epsfig{file=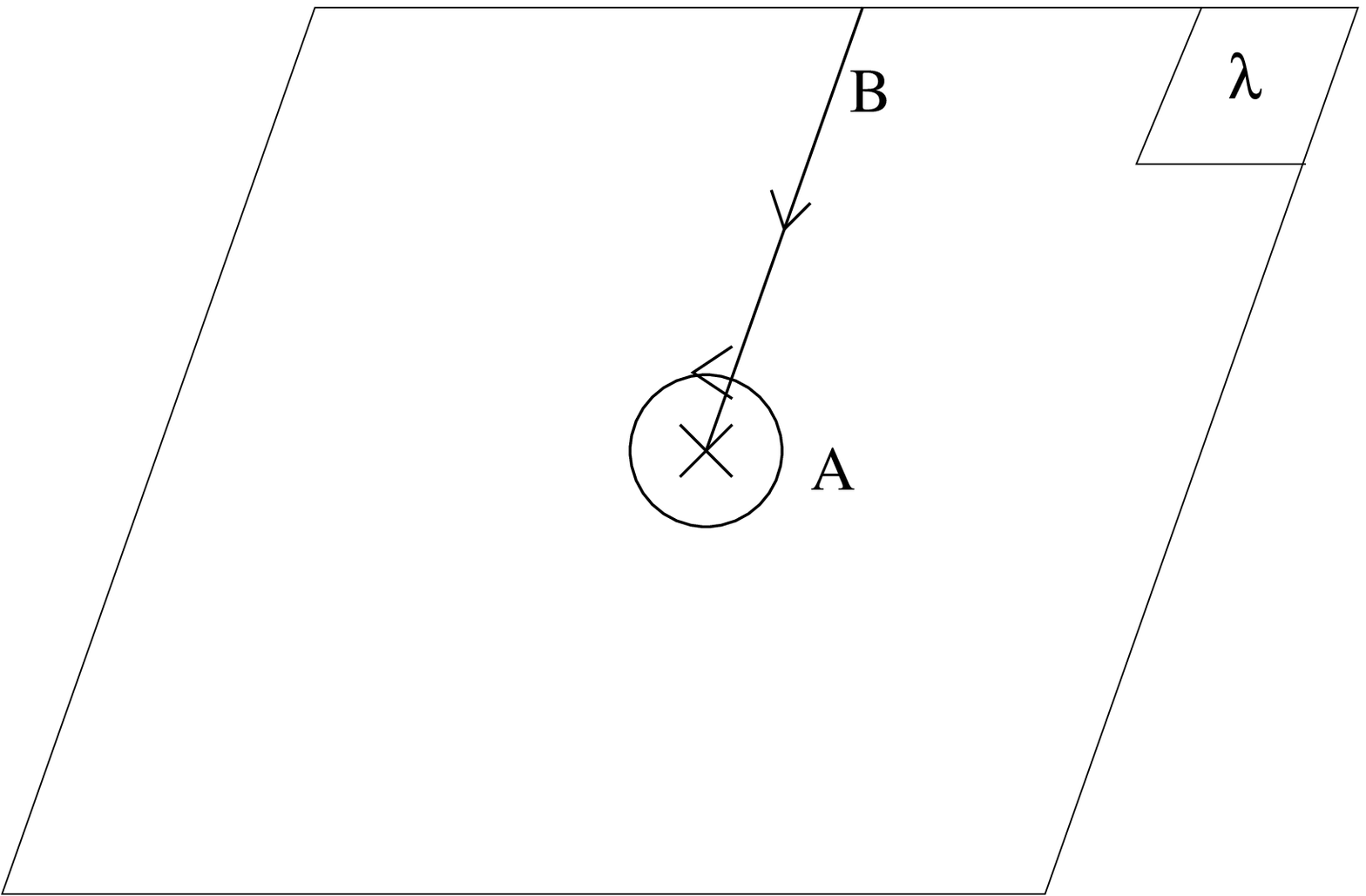,width=0.4\textwidth}\label{lambdaplane}}
\caption{Parametrizations of the hyperelliptic curve \eqref{wvcurve}}
\label{wvparams}
\end{center}
\end{figure}

We now turn to the embedding coordinate $s$, which characterizes wrapping along $x^{10}$.  A holomorphic $s(\lambda)$ with $A$-period \eqref{sacond} is easily seen to be
\begin{equation}
s=N\ln\lambda.
\label{scurve}
\end{equation}
The logarithmic behavior seen here, which we alluded to in the previous
section, necessitates the introduction of a cutoff $v_0$ in order to
study the $B$-period constraint \eqref{sbcond}
\begin{equation}
-2\pi i\alpha=\oint_{B}^{v_0}\,ds=-N\ln\left(\frac{v_0^2}{a}\right)+\ldots
\end{equation}
From this, we see that a holomorphic 1-form $ds$ with the properties \eqref{sacond} and \eqref{sbcond} exists on the hyperelliptic curve \eqref{wvcurve} provided the complex parameter $a$ satisfies
\begin{equation}
a^N=v_0^{2N}e^{-2\pi i\alpha(v_0)}=\Lambda_{n=1}^{2N},
\label{afix}
\end{equation}
where $\Lambda_{n=1}$ is as in \eqref{lambdan1def}.  This completely
fixes the complex structure of the hyperelliptic curve
\eqref{wvcurve}{\footnote{Modulo $N$ choices parametrized by $N$th roots
of unity.}}.

\subsubsection{Comparison with IIB}

If we interpret \eqref{wvcurve} and \eqref{scurve} as describing a curved NS5-brane with flux, the configuration at hand is $T$-dual to a deformed geometry of the type \eqref{defconifold} with flux \eqref{simpexfluxes}.  In fact, we can even verify that the modulus of our IIA configuration is identical to that determined in IIB by dynamics of the GVW superpotential \eqref{gvw}.  One way to do this, for instance, is to check the vacuum equation \eqref{susyvacuum} directly.
This can be done because the period ``matrix'' $\hat{\tau}$ of the curve \eqref{wvcurve} is easy to compute in terms of $a$
\begin{equation}
\hat{\tau}=\frac{1}{2\pi i}\ln\left(\frac{a}{v_0^2}\right)+\ldots
\end{equation}
Using \eqref{afix}, it immediately follows that \eqref{susyvacuum} is satisfied.  Alternatively, we can make the comparison by computing $S$ directly.  The $T$-duality dictionary provides us with an expression for $S$ that is easy to evaluate on the IIA side
\begin{equation}
S=\frac{1}{2\pi i}\oint_A\,{1\over 2}w\,dv=ma=m\Lambda_{n=1}^2.
\label{Scomp}
\end{equation}
Using \eqref{afix}, we see that the expectation value \eqref{svev} is reproduced.  Consequently, we see that going from the NS5/D4 configuration of figure \ref{A1quadTdualD4} to the curved NS5 with flux described by \eqref{wvcurve} and \eqref{scurve} is exactly $T$-dual to the large $N$ duality in type IIB \cite{Dasgupta:2001um,Oh:2001bf}!

\subsubsection{Reliability of the M5 and NS5 descriptions}
\label{subsubsec:reliability}

Now that we have completed our description of the M5 configuration which
reduces to that of figure \ref{A1quadTdualD4} at $g_s=0$, we must
address the question of when this analysis is reliable.  What we have
found is a minimal area surface or, in other words, a solution to the
equations of motion which follow from the Nambu-Goto contribution to the
worldvolume action.  Viewing our setup as a curved M5-brane, this is
justified provided a number of conditions are met.  First, we require
that the curvature of the M5 be everywhere small in 11-dimensional
Planck units and the radius of the $x^{10}$ circle large.  Furthermore,
we must avoid letting $N$ become too large in order to prevent the
density of windings along $x^{10}$ from growing to the point that the M5
comes within a Planck length of intersecting itself.  As discussed in
Appendix \ref{app:validity}, one can demonstrate that all of these
conditions are satisfied provided $g_s\gg 1$ and the conditions
\eqref{validityM5} involving $N$ are satisfied.

On the other hand, we can attempt to provide a direct IIA interpretation
of our setup as a curved NS5-brane with flux.  Because the NS5
worldvolume action is simply the dimensional reduction of the M5 one
\cite{Bandos:2000az}, our analysis is valid in the IIA regime, where
$x^{10}$ is identified with the appropriate RR gauge potential, provided
we can restrict attention only to the descendant of the Nambu-Goto term.
This, in turn, requires that the curvature of the NS5 be small in string
units and that the flux $N$ not be too large{\footnote{The reason for
this is to prevent excited string states from becoming light.  In
particular, M2 branes ending on the M5 and wrapping $x^{10}$ in the
$M$-theory picture descend in IIA to strings with tension that can be
made arbitrarily small unless $g_sN$ is sufficiently small. See Appendix
\ref{app:validity}.}}.  We show in Appendix \ref{app:validity} that this
leads to the constraints
\begin{equation}
 \begin{split}
  g_s&\ll 1,\\
  g_sN&\ll \min\left(\Lambda_{n=1},\frac{\Lambda_{n=1}}{m}\right)=\min\left(\sqrt{\frac{S}{m}},\sqrt{Sm}\right),\\ 
  1& \ll\min\left(m^2\Lambda_{n=1},\frac{\Lambda_{n-1}}{m}\right)=\min\left(\sqrt{m^3S},\sqrt{\frac{S}{m^3}}\right).
\label{IIAregimeg0}
 \end{split}
\end{equation}
where $\Lambda_{n=1}$ is as in \eqref{lambdan1def} and $S$ is given by \eqref{Scomp}.  Physically, these come about because when $m>1$, for example, $S/m^3$ sets the scale of the ``tubes'' into which the D4's blow up in the lift while $S/m$ determines the flux density.  The symmetry under $m\rightarrow m^{-1}$ simply reflects our ability to interchange $w\leftrightarrow v$.  

Outside of the regimes \eqref{validityM5} and \eqref{IIAregimeg0}, our analysis based on the Nambu-Goto term of the worldvolume action is no longer reliable.  One still expects the system to be described by an M5 along the curve described by \eqref{wvcurve} and \eqref{scurve} but this is dependent on a BPS argument that relies on supersymmetry.

\subsection{$A_1$ theory with cubic superpotential --- IIB}

We now move on to the second example, namely that of D5 branes at the conifold singularities of the $A_1$ fibration \eqref{fibgeom} with 
\begin{equation}
W(v)=g\left(\frac{v^3}{3}-\frac{\Delta^2 v}{4}\right).
\label{cubWv}
\end{equation}
In particular, we have two singularities at $v=\pm\Delta/2$ at which we place $N_1$ and $N_2$ D5's, respectively.  After the geometric transition, we are left with the deformed geometry \cite{Cachazo:2001jy}
\begin{equation}
x^2+y^2+w^2= g^2\left(v^2-\frac{\Delta^2}{4}\right)^2-f_1v-f_0
\label{cubdefgeom}
\end{equation}
and 3-form fluxes
\begin{equation}
 \oint_{{\cal{A}}_i}H=4\pi^2 N^i,\qquad \oint_{{\cal{B}}_i}H=-4\pi^2\alpha_i,
\end{equation}
where $\alpha_1=\alpha_2=\alpha\,$.{\footnote{We could actually choose
$\alpha_1$ and $\alpha_2$ to differ by an integer.  This corresponds to
shifting the $\theta$ angle associated with one stack of branes by
$2\pi$ relative to the other stack.  For simplicity, we take the
$\theta$ angles identical so $\alpha_1=\alpha_2$.\label{ftnt:IIBalpha}}}
This system has two complex moduli corresponding to the holomorphic
volumes of the compact $S^3$'s
\begin{equation}
S_i=\frac{1}{2\pi i}\oint_{{\cal{A}}_i}\Omega,
\end{equation}
which can in turn be related to the complex deformation parameters $f_0$
and $f_1$, though we do not do it explicitly here.  Introducing the
${\cal{B}}$-period $\Pi_i$ and period matrix $\hat{\tau}_{ij}$ as usual
\begin{equation}
\Pi_i\equiv\frac{1}{2\pi i}\oint_{{\cal B}_i}\Omega=\frac{\partial{\cal{F}}}{\partial S_i},\qquad\qquad\hat{\tau}_{ij}\equiv\frac{\partial\Pi_i}{\partial S_j}=\frac{\partial{\cal{F}}}{\partial S_i\partial S_j},
\end{equation}
we can write the GVW superpotential as
\begin{equation}
W_{GVW}\sim N_i\Pi_i+\alpha_iS_i
\end{equation}
and obtain the supersymmetric vacuum condition
\begin{equation}
\alpha_i+\hat{\tau}_{ij}N_j=0.
\label{susyvac}
\end{equation}
This specifies $\hat{\tau}_{ij}$ in terms of the fluxes $\alpha,N_i$ and hence completely fixes the complex moduli.  To translate this into a statement about the $S_i$, it is necessary to determine the dependence of $\hat{\tau}_{ij}$ on the $S_i$.  This can be done using the results of \cite{Cachazo:2001jy}, who compute the prepotential ${\cal{F}}$ as an expansion in the variables
\begin{equation}
t_i = \frac{S_i}{g\Delta^3}.
\label{tidef}
\end{equation}
Applying their result, we find that $\hat{\tau}_{ij}$ is given by the following to leading order in the $t_k$
\begin{equation}
\hat{\tau}_{ij}=\frac{1}{2\pi i}\left[\begin{pmatrix}\ln t_1 & 0\\ 0 & \ln t_2\end{pmatrix}-\ln\left(\frac{v_0}{\Delta}\right)^2\begin{pmatrix}1 & 1\\ 1 & 1\end{pmatrix}\right]+\ldots
\end{equation}
This subsequently leads to the expectation values
\begin{equation}
t_1^{N_1}=t_2^{N_2}=\left(\frac{\Lambda_{n=2}}{\Delta}\right)^{2N},
\label{iibti}
\end{equation}
where $\Lambda_{n=2}$ is the RG-invariant combination of $\alpha(v_0)$ and the cutoff $v_0$
\begin{equation}\Lambda_{n=2}^{2N}\equiv v_0^{2N}e^{-2\pi i\alpha(v_0)}.\label{lambdan2def}\end{equation}
We see \emph{a posteriori} that this result is valid in the regime $\Lambda_{n=2}/\Delta \ll 1$ and further corrections amount to an expansion in this parameter.

\subsection{$A_1$ theory with cubic superpotential --- IIA/M}
\label{A1cubIIAM}

\begin{floatingfigure}{0.45\textwidth}
%\begin{figure}
\begin{center}
 \vspace*{2ex}
\epsfig{file=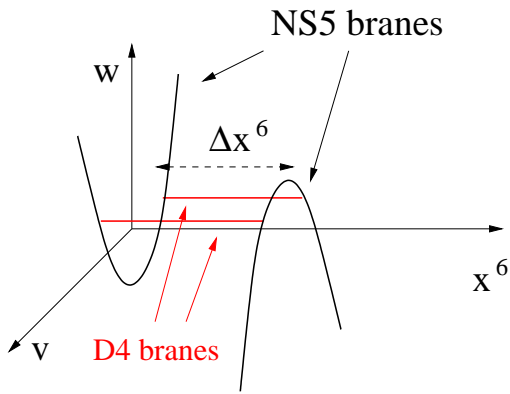,width=0.4\textwidth}
\caption{NS5/D4 configuration obtained by applying $T$-duality to the geometry \eqref{fibgeom} with cubic $W(v)$ \eqref{cubWv} and D5 branes wrapping the conifold singularities at $v=\pm\Delta/2$.
}
\label{A1cubTdualD4}
% \vspace*{2ex}
\end{center}
%\end{figure}
\end{floatingfigure}

We now turn to the description of this system in the IIA/M picture.  The relevant brane configuration here consists of two NS5's extended along the quadratic curves
\begin{equation}
w=\pm g\left(v^2-\frac{\Delta^2}{4}\right)
\end{equation}
and separated along $x^6$ with $N_1$ ($N_2$) D4-branes suspended in
between at $v=\Delta/2$ $(-\Delta/2)$.  This configuration is depicted
in figure \ref{A1cubTdualD4}.  The corresponding $M$-theory lift is
described by a genus one holomorphic curve with boundary conditions
\begin{equation}
w\sim \pm g\left(v^2-\frac{\Delta^2}{4}\right)\qquad\text{as }v\rightarrow\infty
\label{wvbdrygone}
\end{equation}
and embedding coordinate $s$ satisfying{\footnote{As in IIB (see
footnote \ref{ftnt:IIBalpha}), we could choose $\alpha$ to be different
by an integer for $B_1$ and $B_2$ cycles.  This would correspond to
nontrivial wrapping of M5 along $x^{10}$ when one goes around the
compact cycle $B_1-B_2$.}}
\begin{align}
\oint_{A_i} ds &= 2\pi i N^i,\label{ssacond}\\
\oint_{B_i} ds &= -2\pi i\alpha(v_0).\label{ssbcond}
\end{align}
As in the genus zero case, an explicit representation is well-known but we shall seek a parametric description here.  Such an approach may be unfamiliar, so we shall discuss it at length.

Because the desired curve has genus one, we shall parametrize it by a
single complex variable $z$ subject to the identifications $z\sim z+1$
and $z\sim z+\tau$ where $\tau$ is the modulus of the torus.  In all
that follows, we will focus only on the fundamental parallelogram, which
is depicted in figure \ref{zplane}.  As in the genus zero case, we must
specify two marked points $a_1$ and $a_2$ on the parallelogram as the
preimages of the points at infinity on the NS5-branes.  Unlike the
previous example, though, the embedding is no longer one-to-one at
infinity.  This is due to the quadratic curving of the NS5's and leads
to the constraint that, while $v(z)$ has single poles at $a_1$ and
$a_2$, $w(z)$ necessarily has double poles at these points.  The
embedding coordinate $s(z)$, on the other hand, is multivalued on the
$z$-plane with a cut connecting $a_1$ and $a_2$ and monodromies
consistent with \eqref{ssacond} and \eqref{ssbcond}.

With the analytic structure of $v(z)$, $w(z)$, and $s(z)$ at hand we can
in principle proceed to write them down.  To do this in practice,
though, we need the analog of the functions $\lambda^{-1}$ and
$\ln\lambda$ which allowed us to introduce poles and cuts in the
previous example.  A convenient choice of building blocks for
constructing genus one curves is based on the function{\footnote{This
collection of building blocks was recently used by \cite{Janik:2003hk}
for essentially the same purpose as ours.}}
\begin{equation}
F(z)=\ln\theta(z-\tilde{\tau}),
\label{Fdef}
\end{equation}
where
\begin{equation}
\theta(z)=\sum_{n=-\infty}^{\infty}e^{i\pi n^2\tau+2\pi inz},
\qquad\qquad
{\tau}\equiv \frac{1}{2}(\tau+1).
\label{thetadef}
\end{equation}
Because $\theta(z)$ has a simple zero at $z=-\tilde{\tau}$, we see that $F(z)\sim \ln z$ near $z=0$.  This means that we can use $F(z)$ to introduce branch points and derivatives of $F(z)$ to introduce poles.  In what follows, we shall adopt the notation
\begin{equation}
F_i^{(n)}=\left(\frac{\partial}{\partial z}\right)^{n}F(z-a_i).
\label{Fderivs}
\end{equation}
Detailed properties of these functions and their relation to Weierstrass
elliptic functions can be found in Appendix \ref{app:ellipfcns}\@.  The
most important feature to keep in mind is that $F_i^{(n)}$ introduces an
$n$th order pole at the point $a_i$.  It is also worth noting here that
the $F_i^{(n)}$ are elliptic for $n>1$ and have the following
monodromies for $n=0,1$
\begin{equation}\begin{split}
F_i(z+1)&=F_i(z),\\
F_i(z+\tau)&=F_i(z)+i\pi - 2\pi i(z-a_i),\\
F_i^{(1)}(z+1)&=F_i^{(1)}(z),\\
 F_i^{(1)}(z+\tau)&=F_i^{(1)}(z)-2\pi i.
\label{Fmonods}
\end{split}\end{equation}

With our building blocks handy, we are now ready to begin writing
general expressions for the embedding functions.  We start with $v(z)$
and $w(z)$, whose analytic structure was described above.  If we add the
requirement that $w\propto \pm v^2$ near $a_1$ and $a_2$, the most
general possibility, up to constant shifts, is given by{\footnote{The
relative sign of $F_1^{(1)}$ and $F_2^{(1)}$ in $v$ is fixed by
requiring $v$ to be elliptic while the relative sign of $F_1^{(2)}$ and
$F_2^{(2)}$ in $w$ is obtained by requiring $w\propto \pm v^2$ at $a_1$
and $a_2$.}}
\begin{equation}
\begin{split}
v&= X\left(F_1^{(1)}-F_2^{(1)}-\left[F^{(1)}(a)-i\pi\right]\right),\\
w&=C\left(F_1^{(2)}-F_2^{(2)}\right),
\label{vwparamrep}
\end{split}\end{equation}
where we have inserted the constant shift $F^{(1)}(a)-i\pi$ so that
$w(v)$ behaves like \eqref{wvbdrygone} and $a$ is the separation between
marked points
\begin{equation}
a\equiv a_2-a_1.
\label{adef}
\end{equation}
As described in Appendix \ref{app:ellipfcns}, it is possible to work out
the polynomial relation between $w$ and $v$ explicitly.  It takes the
form of a hyperelliptic curve
\begin{equation}
w^2=P_2(v)^2-f_1v-f_0
\label{hypergone}
\end{equation}
with $P_2(v)$ a particular quadratic polynomial in $v$ from which we can
read off relations between the curve parameters $X$ and $C$ of
\eqref{vwparamrep} and the ``physical'' parameters $g$ and $\Delta$ of
\eqref{wvbdrygone}
\begin{equation}
g=\frac{C}{X^2},\qquad\qquad \Delta^2 = 12X^2\wp(a).
\label{physparams}
\end{equation}
The function $\wp(z)$ appearing here is the Weierstrass $\wp$-function.

We have thus seen that a holomorphic curve with the desired analytic properties along $w$ and $v$ corresponds to a hyperelliptic curve of the form \eqref{hypergone} and admits the parametric representation \eqref{vwparamrep}.  We have also found a convenient way to parametrize the moduli space of such curves, as they depend on two complex parameters, $\tau$ and $a$.  These are analogous to the quantities $S_i$ on the IIB side as they encode essentially the same information.

Let us finally turn our attention to the embedding coordinate $s$, which
is a multivalued function of $z$ satisfying \eqref{ssacond} and
\eqref{ssbcond}.  To determine its form, we must identify those cycles
on the $z$-plane that correspond to our $A$ and $B$ cycles.
Illustrations of both representations of the hyperelliptic curve
\eqref{hypergone} which identify all the relevant cycles can be found in
figure \ref{wvparamsgone}.  Imposing the $A$-periods \eqref{ssacond}
uniquely fixes the form of $s$ up to an integration constant, for which
we make a convenient choice{\footnote{Our choice of integration constant
simplifies the limit used to obtain the local geometry near one of the
D4 stacks.  It also renders our curve invariant under a particular
$\mathbb{Z}_2$ symmetry of the parametrization $z\rightarrow \tau-z$,
$a_1\rightarrow\tau-a_2$, $a_2\rightarrow\tau-a_1$,
$s\rightarrow -s$.}}$^,${\footnote{The relative coefficient of $F_1$ and
$F_2$ is fixed by requiring $ds$ to be elliptic.}}
\begin{equation}
s=(N_1+N_2)\left(F_1-F_2-i\pi a\right)+2\pi i N_1\left(z-A\right),
\label{susys}
\end{equation}
where we defined
\begin{align}
 A&\equiv {a_1+a_2\over 2}.
\end{align}

\begin{figure}
\begin{center}
\subfigure[Hyperelliptic curve \eqref{hypergone} with $A$ and $B$ cycles indicated]
{\epsfig{file=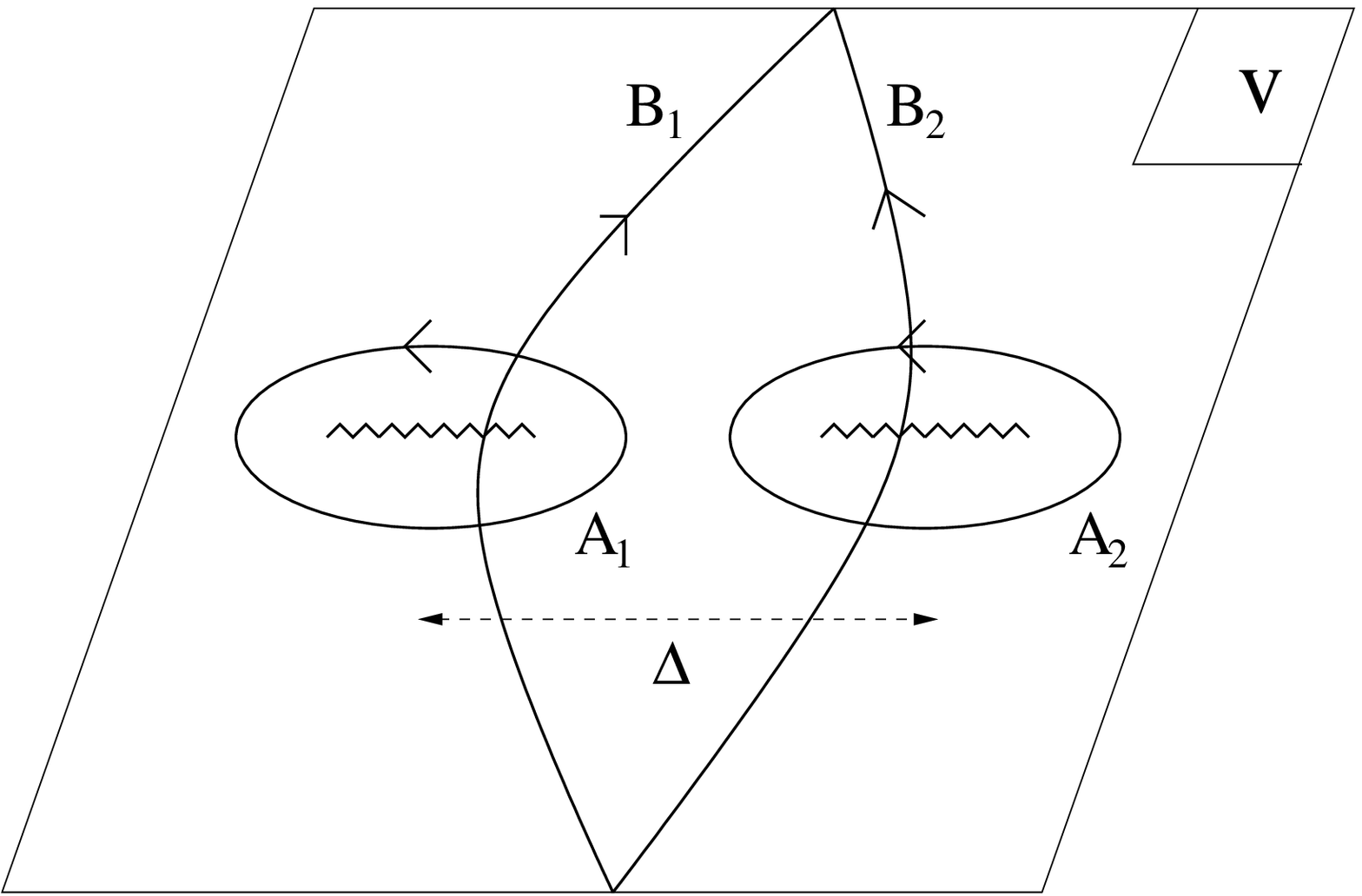,width=0.4\textwidth}\label{cubsupex}}
\subfigure[Fundamental parallelogram on $z$ plane used to parametrize the curve \eqref{hypergone} with $A$ and $B$ cycles indicated]
{\epsfig{file=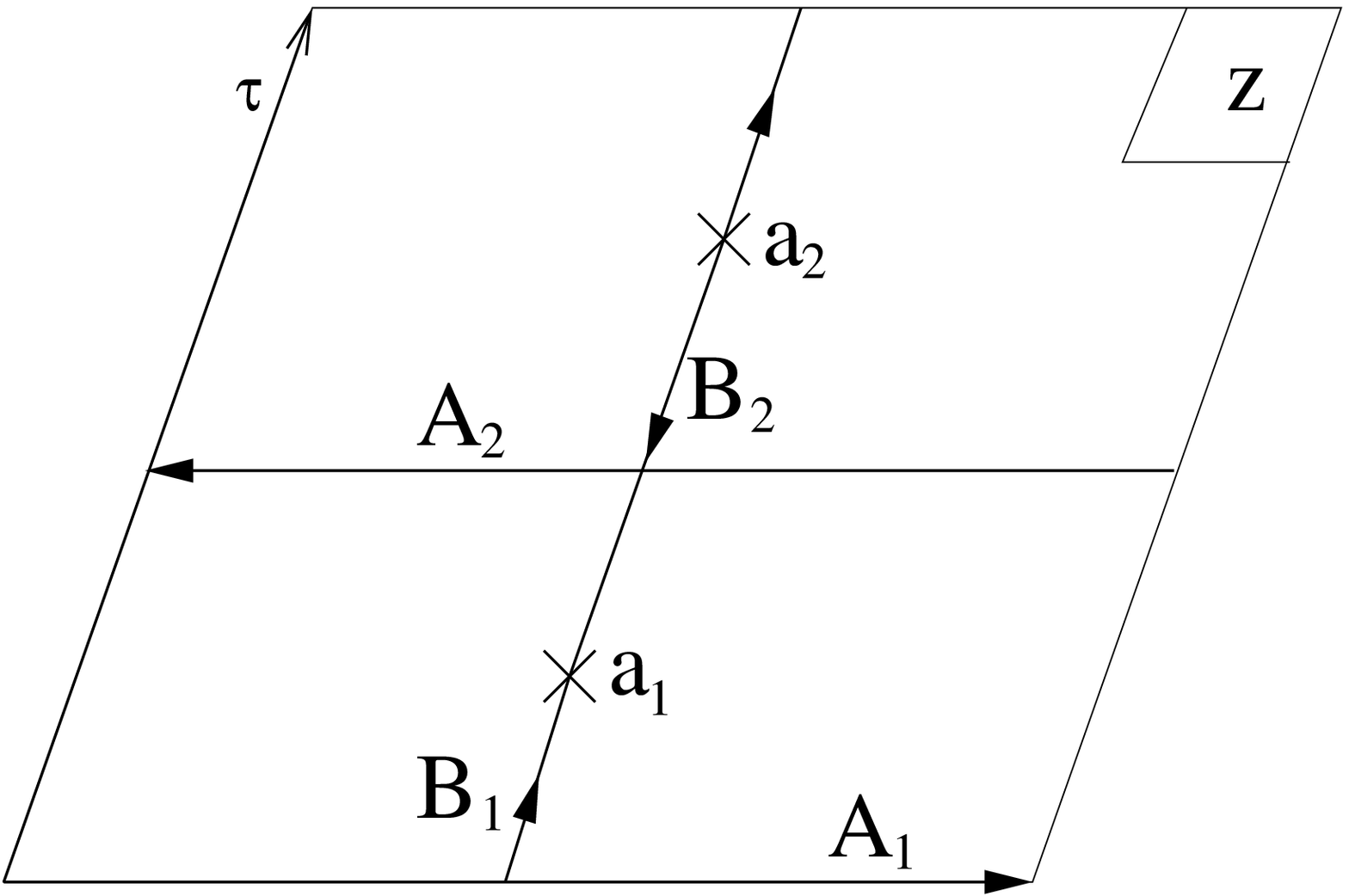,width=0.4\textwidth}\label{zplane}}
\caption{Parametrizations of the hyperelliptic curve \eqref{hypergone}}
\label{wvparamsgone}
\end{center}
\end{figure}

We can now fix the moduli $\tau$ and $a$ by imposing the $B$-period constraints \eqref{ssbcond}.  First note that equivalence of the two noncompact $B$-periods implies that
\begin{equation}
\oint_{B_2-B_1}ds=0,
\end{equation}
or in other words
\begin{equation}
s(z+\tau)=s(z).
\end{equation}
This leads immediately to a relation between $a$ and $\tau$
\begin{equation}
a=\frac{N_1\tau}{N}.
\label{ataurel}
\end{equation}
Evaluating the noncompact $B$-periods then leads to the further condition
\begin{equation}
-2\pi i\alpha(v_0)=\oint_{B_1}ds = -N \left[L(a,\tau)+\ln\left(\frac{v_0^2}{\Delta^2}\right)\right]+\frac{2\pi i N_1N_2\tau}{N},
\label{alphataurel}
\end{equation}
where
\begin{equation}
L(a,\tau)\equiv\ln\left(\frac{12\wp(a)\theta(a-\tilde{\tau})^2}{\theta'(\tilde{\tau})^2}\right).
\label{Ldef}
\end{equation}
The conditions \eqref{ataurel} and \eqref{alphataurel} are our final
result for the constraints on moduli of the M5 curve.  

\subsubsection{Comparison with IIB}

As in our genus zero example, there are several ways to compare with IIB\@.
The most direct is to compute the period matrix $\hat{\tau}_{ij}$ of the
elliptic curve \eqref{vwparamrep} explicitly in terms of the parameters
$a$ and $\tau$.  For this, we find
\begin{equation}
\hat{\tau}_{ij}=-\frac{1}{2\pi i}\left[L(a,\tau)+\ln\left(\frac{v_0^2}{\Delta^2}\right)\right]\begin{pmatrix}1&1\\ 1&1\end{pmatrix}+\begin{pmatrix}\tau-a & 0 \\ 0 & a\end{pmatrix}.
\label{taumat}
\end{equation}
It is now easy to verify that the condition \eqref{susyvac} on the moduli for supersymmetric vacua in IIB is solved exactly when \eqref{ataurel} and \eqref{alphataurel} are satisfied.

If we are interested in computing the $S_i$, though, these can also be
determined by direct integration as discussed in Appendix
\ref{app:ellipfcns}\@.  In principle, they can be computed exactly as
functions of $a$ and $\tau$, though the expressions are somewhat
complicated and hence not very enlightening.  A natural limit to study
is that of $\Im\tau\gg 1$ which, based on the identification of cycles
in figure \ref{zplane}, we naively expect to be identified with a limit
of large separation $\Delta$.  Indeed, we can justify this expectation
by noting that, in this regime, the ratio $\Lambda_{n=2}/\Delta$ becomes
\begin{equation}
\left(\frac{\Lambda_{n=2}}{\Delta}\right)^{2N}\sim e^{2\pi i N_1N_2\tau/N}+\ldots
\label{ldtau}
\end{equation}
where we have suppressed terms that are further exponentially suppressed at large $\Im\tau$.  Expanding the $S_i$ in this limit as well, we find
\begin{equation}
t_1^{N_1}=t_2^{N_2}=\left(\frac{\Lambda_{n=2}}{\Delta}\right)^{2N}+\ldots
\label{iiati}
\end{equation}
where the $t_i$ are as in \eqref{tidef}.  This agrees with the IIB
result \eqref{iibti} that was obtained in the same regime.  Because the
moduli of the $M5$ curve exactly solve the vacuum equation
\eqref{susyvac}, this agreement will persist to all orders in the
parameter $\Lambda_{n=2}/\Delta$.  Though it was not our main objective,
it is nice that the elliptic function formalism leads to exact results
for the moduli $S_i$ as obtaining them from the IIB side requires a full
solution of the Dijkgraaf-Vafa matrix model \cite{Dijkgraaf:2002fc,
Dijkgraaf:2002vw, Dijkgraaf:2002dh} (see also \cite{Janik:2003hk}).

\subsubsection{Reliability of the M5 and NS5 descriptions}
\label{subsubsec:g1reliability}

Finally, let us address the question of when our M5 and NS5
interpretations of the curve of \eqref{vwparamrep} and \eqref{susys} are
reliable without the use of BPS arguments.  For this, we can borrow many
of the results of section \ref{subsubsec:reliability} and Appendix
\ref{app:validity} provided $\Lambda_{n=2}/\Delta$ is sufficiently
large.  In this case, we note that the effective size $S$ of each tube
is captured by the corresponding $S_i$ while the effective mass,
obtained by expanding the superpotential \eqref{cubWv} near one of its
critical points, is given by $m=g\Delta$.  As usual, the conditions for
a reliable M5 interpretation at strong coupling are straightforward but
a bit complicated.  These can be easily worked out from equation
\eqref{validityM5} in Appendix \ref{app:validity}\@.

The conditions for a reliable NS5 interpretation at weak coupling, on the other hand, take a fairly simple form
\begin{equation}
 g_s\ll 1,
 \qquad 
 g_s N \ll \min\left(\sqrt{\frac{S_i}{g\Delta}},\sqrt{g\Delta S_i}\right),
 \qquad
 1\ll \min\left(\sqrt{\frac{S_i}{g^3\Delta^3}},\sqrt{g^3\Delta^3S_i}\right).
\label{ns5conds}
\end{equation}
It is not difficult to see that these conditions can indeed be
satisfied.  Consider, for instance, the behavior of \eqref{ns5conds} at
leading order in $\Lambda_{n=2}/\Delta$
\begin{equation}
 g_s\ll 1,
  \qquad
  \frac{g_sN}{\Delta}\ll \left(\frac{\Lambda_{n=2}}{\Delta}\right)^{N/N_i}\min(1,g\Delta),
  \qquad 
  1\ll \frac{1}{g}\left(\frac{\Lambda_{n=2}}{\Delta}\right)^{N/N_i}\min\left(1,g^3\Delta^3\right).
\label{IIAregime}\end{equation}
These are easily seen to hold for a wide range of parameters.  
This condition will be of interest to us later when studying configurations with D4's and \anti{D4}'s.

\section{The Brane/Antibrane System}
\label{sec:br/anti}

We now proceed with our study of IIA/M configurations obtained by
applying $T$-duality to the brane/antibrane system of
\cite{Aganagic:2006ex}.  After first reviewing the IIB construction of
\cite{Aganagic:2006ex}, we will discuss the NS5/D4 configuration and its
$M$ theory lift.

\subsection{Branes and antibranes on local CY in IIB}

In the recent paper \cite{Aganagic:2006ex}, it was suggested that interesting SUSY-breaking configurations could be constructed by wrapping D5's and \anti{D5}'s at singular points of local Calabi-Yau.  In particular, if the singular $S^2$'s wrapped by branes and antibranes are homologous, the absence of a conservation law preventing their eventual annihilation suggests that this system is metastable and will eventually decay.

Because supersymmetry is broken, it might seem that obtaining any quantitative information about the system is out of the question.  However, it was suggested in \cite{Aganagic:2006ex} that the breaking is sufficiently soft that one retains a significant amount of computational control.  In particular, they conjectured that large $N$ duality continues to hold, permitting one to replace the branes and antibranes with fluxes on a deformed geometry.  Then, it was further argued that the ${\cal{N}}=2$ SUSY of IIB strings on this deformed geometry is broken spontaneously by the fluxes, leading one to expect that essential aspects of the physics continue to be captured by the ${\cal{N}}=2$ prepotential which, in turn, is determined by special geometry.

The simplest example of such a system has been considered at length in \cite{Aganagic:2006ex,Heckman:2007wk} and consists of wrapping $N_1$ D5's and $N_2$ \anti{D5}'s at the singular points of the $A_1$ fibration \eqref{fibgeom} with cubic superpotential \eqref{cubWv}.  After the geometric transition, we are left with the deformed geometry \eqref{cubdefgeom}, repeated here for convenience
\begin{equation}
x^2+y^2+w^2=g^2\left(v^2-\frac{\Delta^2}{4}\right)^2-f_1v-f_0
\label{cubdefgeomm}
\end{equation}
and 3-form fluxes
\begin{equation}
 \oint_{{\cal{A}}_i}H=4\pi^2 N^i,\qquad \oint_{{\cal{B}}_i}H=-4\pi^2\alpha_i.
\label{sbfluxes}
\end{equation}
We will use $N^i$ to refer to flux numbers, which can be negative, and $N_i$ to the number of branes or antibranes so that
\begin{equation}
N_1=N^1,\qquad\qquad N_2=-N^2 >0.
\end{equation}
We also take $\alpha_1=\alpha_2=\alpha$ for simplicity as in the supersymmetric example of the previous section.

The presence of the fluxes \eqref{sbfluxes} leads to generation of the superpotential
\begin{equation}
W_{GVW}\sim N^i\Pi_i+\alpha_iS^i,
\end{equation}
where as usual we define
\begin{equation}
S^i\equiv{1\over 2\pi i}\oint_{{\cal{A}}_i}\Omega,
 \qquad 
 \Pi_i\equiv{1\over 2\pi i}\oint_{{\cal{B}}_i}H=\frac{\partial{\cal{F}}}{\partial S^i},
 \qquad
 \hat{\tau}_{ij}=\frac{\partial\Pi_i}{\partial S^j}=\frac{\partial^2{\cal{F}}}{\partial S^i\partial S^j}.
\end{equation}

The absence of solutions to the supersymmetric vacuum equation \eqref{susyvac} {\footnote{That no solutions exist follows from the fact that $\Im\hat{\tau}$ is positive definite.}}
\begin{equation}
\alpha+\hat{\tau}_{ij}N^j=0
\end{equation}
explicitly demonstrates that SUSY is broken and necessitates a computation of the scalar potential for the lowest components of $S^i$ to determine the actual vacuum.  Because the breaking is spontaneous, it is natural to conjecture, as the authors of \cite{Aganagic:2006ex} did, that the K\"ahler potential of the system continues to be determined by special geometry
\begin{equation}
{\cal{K}}_{ij}(S^k)=\Im\hat{\tau}_{ij}(S^k)
\label{kahlerpot}
\end{equation}
and hence that the scalar potential is given by
\begin{equation}
V=\left(\bar{\alpha}+\bar{\hat{\tau}}_{ij}N^j\right)\left(\Im\hat{\tau}\right)^{-1\,jk}\left(\alpha+\hat{\tau}_{kl}N^l\right).
\label{IIBnonsusypot}
\end{equation}
The moduli $S^i$ are now determined by minimizing this quantity.  This
was studied by the authors of \cite{Aganagic:2006ex} who found that, to
leading order in the $S^i$, the minimization equations can be written in
the relatively simple form
\begin{equation}
\begin{split}
\Re(\alpha)+\Re(\hat{\tau}_{ij})N^j&=0,\\
\Im(\alpha)+\Im(\hat{\tau}_{ij})|N^j|&=0
\label{leadingnsvac}
\end{split}\end{equation}
and lead to the expectation values
\begin{equation}
t_1^{N_1}=t_2^{N_2}
 =\left(\frac{v_0}{\Delta}\right)^{2N_1}\left(\frac{\bar{v}_0}{\bar{\Delta}}\right)^{2N_2}e^{-2\pi i\alpha(v_0)}
 =\tilde{\Lambda}^{2(N_1+N_2)}{\Delta}^{-2N_1}\bar{\Delta}^{-2N_2},
\end{equation}
where the $t_i$ are as in \eqref{tidef}.  In the above, we have
implicitly defined the ``RG-invariant'' scale $\tilde{\Lambda}$
\begin{equation}
\tilde{\Lambda}^{2(N_1+N_2)}=v_0^{2N_1}\bar{v}_0^{2N_2}e^{-2\pi i\alpha(v_0)}.
\label{Ltildaprx}
\end{equation}
As in the supersymmetric example of the previous section, we see that this approximation is valid for small $|\tilde{\Lambda}/\Delta|$.  Subsequent corrections can in principle be determined using the Dijkgraaf-Vafa matrix model technology \cite{Dijkgraaf:2002fc,Dijkgraaf:2002vw,Dijkgraaf:2002dh}.  This was recently done in \cite{Heckman:2007wk} and an interesting phase structure uncovered.  We will see later how this structure arises in the IIA/M analysis.

\subsection{Brane/antibrane configurations in type IIA and $M$ theory}
\label{braneantibraneIIAM}

\begin{floatingfigure}{0.5\textwidth}
\begin{center}
\epsfig{file=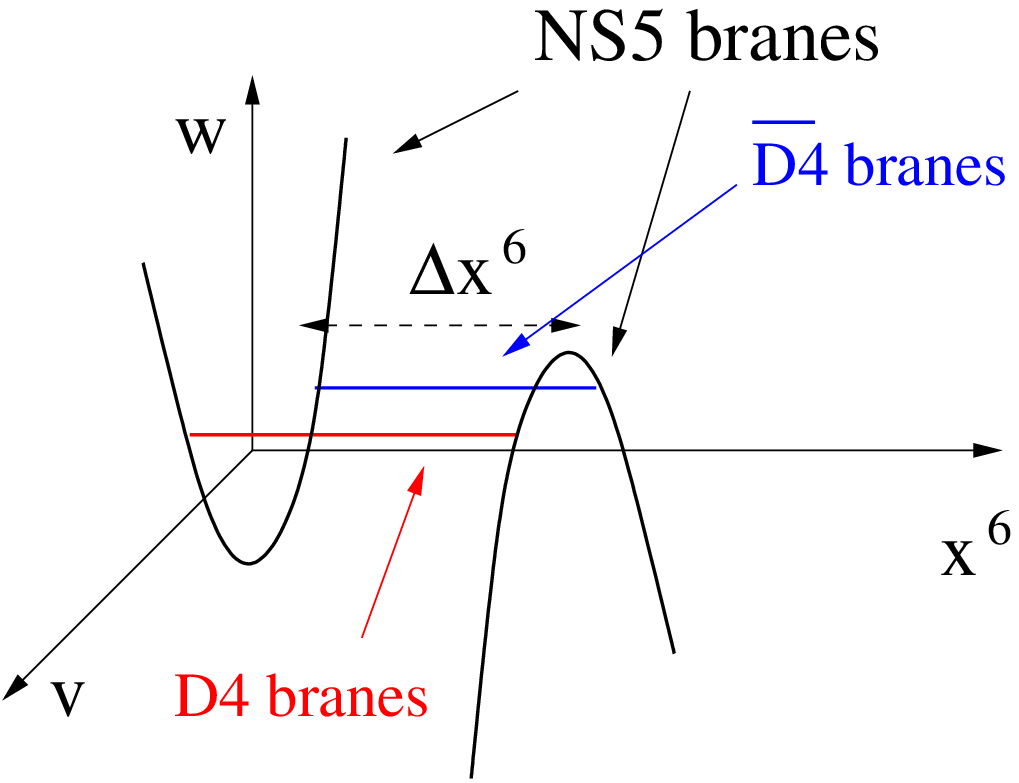,width=0.4\textwidth}
\caption{NS5/D4/\anti{D4} configuration $T$-dual to the brane/antibrane system of \cite{Aganagic:2006ex}}
\label{A1cubTdualD4D4bar}
\end{center}
\end{floatingfigure}

We now consider the brane/antibrane configuration of \cite{Aganagic:2006ex} from the IIA/M point of view.  Applying $T$-duality, we find a pair of quadratically curved NS5-branes extended along the curves
\begin{equation}
w=\pm g\left(v^2-\frac{\Delta^2}{4}\right)
\end{equation}
and separated along $x^6$.  There are also $N_1$ D4's suspended between
the NS5's at $v=\Delta/2$ and $N_2$ \anti{D4}'s at $v=-\Delta/2$.  This
configuration is depicted in figure \ref{A1cubTdualD4D4bar}.

This classical brane configuration fails to capture any quantum effects
of the system and, strictly speaking, is valid only for $g_s=0$.  We
expect that, just as in the supersymmetric examples discussed
previously, quantum corrections will smooth out the NS5/D4 and
NS5/\anti{D4} intersection points and also lead to a ``dimpling'' of the
NS5's.  To study this process, we propose to follow the procedure
adopted before and lift this configuration to $M$ theory.  At large
$g_s$, this system is described by an $M5$ brane wrapping a smooth
minimal area surface in $wvs$ space.  At small $g_s$, this $M5$ is more
appropriately viewed as a curved NS5-brane with flux.  

As usual, we shall work in a regime where the $M5$ (NS5) worldvolume
theory is reliably approximated by the Nambu-Goto action (or its
descendant)
\begin{equation}S_{NG}=\int\,\sqrt{g}\end{equation}
which, when reduced along the flat 0123 directions, looks similar to the ``worldsheet'' action of the bosonic string \cite{Witten:1998jd}.  Standard analysis of this action indicates that a given surface is an extremum provided the embedding functions are harmonic on the ``worldsheet''
\begin{equation}
\partial\bar{\partial}v(z)=\partial\bar{\partial}w(z)=\partial\bar{\partial}s(z)=0
\label{harmonic}
\end{equation}
and satisfy a ``Virasoro''-type constraint{\footnote{We have assumed
here a ``target-space'' metric of the form
\begin{equation}ds^2 = g_s^2|ds|^2+|dv|^2+|dw|^2+\ldots\end{equation}
where $\alpha'$ has been set to 1 as usual and the manifest $g_s$-dependence has been introduced in accordance with the factor of $R^{-1}$ in \eqref{s_def}.}}
\begin{equation}
g_s^2\,\partial s\,\partial\bar{s}+\partial v\,\partial\bar{v}+\partial w\,\partial\bar{w}=0.
\label{virasoro}
\end{equation}
Notice that holomorphic embeddings automatically satisfy both
conditions.  In our situation, though, we do not expect to find a
holomorphic curve because the presence of antibranes signals a breaking
of supersymmetry.

Before proceeding let us note a few technical points.  First, in order to prevent the open string tachyon from destabilizing our system, we need the separation $\Delta$ between D4 and \anti{D4} stacks to satisfy $\Delta>1$.  Second, in what follows we will also take $\frac{\tilde{\Lambda}}{\Delta}$ to be small when necessary, where $\tilde{\Lambda}$ is as in \eqref{Ltildaprx}.  Note that such a condition is not very stringent because, as we shall see later in section \ref{subsec:phase}, if we make the physical assumption that $|v_0| > \Delta$ then all solutions corresponding to true minima have this property.

\subsubsection{First attempts at an $M$-theory lift}

For illustrative purposes, let us begin by attempting to construct a holomorphic lift of the configuration in figure \ref{A1cubTdualD4D4bar} to see what goes wrong.
The first step along this direction is to address the boundary conditions along $w$ and $v$
\begin{equation}
w\sim \pm g\left(v^2-\frac{\Delta^2}{4}\right)\qquad\text{ as }v\rightarrow\infty.
\label{holobdryconds}
\end{equation}
We have already seen how to deal with these in the previous section.  In particular, we saw that the resulting $wv$ geometry must be a hyperelliptic curve which admits a parametric description of the form \eqref{vwparamrep}
\begin{equation}
\begin{split}
v&= X\left(F_1^{(1)}-F_2^{(1)}-\left[F^{(1)}(a)-i\pi\right]\right),\\
w&=C\left(F_1^{(2)}-F_2^{(2)}\right)
\label{wvparamcurve}
\end{split}\end{equation}
with $g$ and $\Delta$ determined as in \eqref{physparams}
\begin{equation}
g=\frac{C}{X^2},\qquad\qquad\Delta^2 = 12X^2\wp(a).
\end{equation}
Our only task, then, is to write a holomorphic 1-form $ds$ with the appropriate periods
\begin{equation}
\frac{1}{2\pi i}\oint_{A_i}ds = N^i,\qquad
 {1\over 2\pi i}\oint_{B_i}ds = -\alpha_i.
\label{nspers}
\end{equation}
The condition on the $A$-periods uniquely fixes{\footnote{As in the
supersymmetric example of section \ref{A1cubIIAM}, $s$ is fixed only up
to an integration constant for which we make a particularly convenient
choice.}}
\begin{equation}
s=(N_1-N_2)\left(F_1-F_2-i\pi a\right)+2\pi i N_1 \left(z-A\right).
\end{equation}
If we now try to impose the $B$-period constraints, though, we find a problem.  In particular, because $\alpha_1=\alpha_2=\alpha$ we must have
\begin{equation}
\oint_{B_2-B_1}ds=0
\end{equation}
which in turn implies that
\begin{equation}
a\equiv a_2-a_1 = \frac{N_1}{N_1-N_2}\tau > \tau.
\end{equation}
This is impossible because both $a_1$ and $a_2$ lie within the
fundamental parallelogram by assumption.  This result is not surprising.
It simply illustrates that the obstruction to finding a holomorphic lift
of the configuration in figure \ref{A1cubTdualD4D4bar} is the lack of a
well-defined embedding coordinate $s$ that yields the appropriate
wrappings along $x^{10}$.

This situation is easily improved, though, if we take away the
constraint of holomorphy and instead simply require that $s$ be
harmonic, continuing to satisfy one of the minimal area conditions
\eqref{harmonic}.  In this case, we can write it as the sum of a
holomorphic and an antiholomorphic function and it is easy to find a
two-parameter family of such with the right $A$-periods{\footnote{We
have again made a convenient choice for the integration constant in
$s$.}}
\begin{equation}
\begin{split}
s&=(N_1-N_2+\gamma)\left(F_1-F_2-i\pi a\right)+i\pi(N_1+\delta)\left(z-A\right)\\
&\qquad+\gamma\left(\bar{F}_1-\bar{F}_2+i\pi\bar{a}\right)+i\pi \left(N_1-\delta\right)\left(\bar{z}-\bar{A}\right).
\end{split}
\label{sfam}
\end{equation}
We now have enough freedom to fix the $B$-periods to whatever we like without saying anything about $a$ and $\tau$.  It is thus generically possible to write down a harmonic embedding $s(z,\bar{z})$ with all the desired properties for arbitrary values of the moduli{\footnote{More generally, a holomorphic 1-form on a genus $g$ curve of this type will have $g$ free parameters, which is enough freedom to fix the $A$-periods but further imposing constraints on $B$-periods fixes the moduli.  On the other hand, a harmonic 1-form on a genus $g$ curve of this type will have $2g$ free parameters, which is enough freedom to fix all periods for arbitrary values of the moduli.}}.

It may seem that, with the moduli unfixed, we have too much freedom and indeed this is the case.  With the introduction of nonholomorphic contributions to $s$, we are now faced with the daunting task of addressing the nonlinear constraint \eqref{virasoro}, which is no longer satisfied.  As we shall see, this will select a particular $s(z,\bar{z})$ from the two-parameter family \eqref{sfam} and, in so doing, combine with the $B$-period constraints to fix the moduli.  For now, however, let us be very naive and try to use physical reasoning to pick a particular $s$, postponing a further discussion of \eqref{virasoro} to the next subsection.

At large separation $\Delta$, which we argued in the previous section corresponds to large $\Im(\tau)$, we expect the curve to roughly ``factorize'' into a holomorphic piece, describing the local geometry near the D4's, and an antiholomorphic piece, describing the local geometry near the \anti{D4}'s.  A realistic expectation, then, is that if we write $s(z,\bar{z})$ as the sum of holomorphic and antiholomorphic parts
\begin{equation}
s(z,\bar{z})=s_H(z)+\overline{s_A(z)}
\end{equation}
then the periods of $ds_H$ will reflect only the contribution from the D4's and the periods of $d\overline{s_A}$ only the contribution from the \anti{D4}'s when $\Im(\tau)$ is large.  Imposing this condition picks out the choice $\gamma=N_2$, $\delta=N_1$ in \eqref{sfam}
\begin{equation}
s=N_1\left(F_1-F_2-i\pi a\right)+2\pi i N_1\left(z-A\right)+N_2\left(\bar{F}_1-\bar{F}_2+i\pi\bar{a}\right).
\label{sguess}
\end{equation}
Further imposing the $B$-period constraints for this particular $s$ then
leads to the following conditions on $a$ and $\tau$
\begin{equation}
\begin{split}
 N_1\tau&=N_1a-N_2\bar a,\\
2\pi i\alpha(v_0)&=N_1\left[L(a,\tau)+\ln\left(\frac{v_0^2}{\Delta^2}\right)\right]+N_2\left[\overline{L(a,\tau)}+\overline{\ln\left(\frac{v_0^2}{X^2}\right)}\right]+2\pi i N_2\bar{a}.
\label{nsvacguess}
\end{split}
\end{equation}
Note the rough similarity of these equations to the vacuum equations \eqref{leadingnsvac} obtained on the IIB side at leading order in $\tilde{\Lambda}/\Delta$.  We can make the comparison more explicit by using the expression \eqref{taumat} for the period matrix $\hat{\tau}_{ij}$ of the hyperelliptic curve \eqref{wvparamcurve} to write \eqref{leadingnsvac} as
\begin{equation}
\begin{split}
0&=\Re\left\{\left(\alpha - \frac{N_1-N_2}{2\pi i}\left[L(a,\tau)+\ln\left(\frac{v_0^2}{\Delta^2}\right)\right]\right)\begin{pmatrix}1\\ 1\end{pmatrix}+\begin{pmatrix}N_1(\tau-a)\\ -N_2a\end{pmatrix}\right\},\\
0&=\Im\left\{\left(\alpha-\frac{N_1+N_2}{2\pi i}\left[L(a,\tau)+\ln\left(\frac{v_0^2}{\Delta^2}\right)\right]\right)\begin{pmatrix}1\\ 1\end{pmatrix}+\begin{pmatrix}N_1(\tau-a)\\ N_2a\end{pmatrix}\right\}.
\label{nsvaciib}
\end{split}
\end{equation}
It is now easy to see that \eqref{nsvacguess} and \eqref{nsvaciib} are equivalent and hence that the curve \eqref{wvparamcurve} with embedding coordinate \eqref{sguess} satisfying \eqref{ssacond} and \eqref{ssbcond} has the same moduli as the nonsupersymmetric IIB vacuum at leading order in $\tilde{\Lambda}/\Delta$.

\subsubsection{A few problems}

Though our ``physically''-motivated curve \eqref{wvparamcurve},
\eqref{sguess} enjoys some success when comparing to IIB, two important
problems remain.  First, the connection with IIB is valid only at
leading order in $\tilde{\Lambda}/\Delta$ and fails to account for any
further corrections.  The second problem, which is not unrelated to the
first, is that our curve fails to satisfy the additional constraint
\eqref{virasoro} and hence is not a true minimal area surface.

In fact, from \eqref{virasoro} we see that the introduction of any
nonholomorphic dependence to $s$ necessitates further nonholomorphic
contributions to $v$ and $w$.  This further alters the geometry and, in
particular, makes it impossible to write down an $M$-theory lift of the
configuration in figure \ref{A1cubTdualD4D4bar} with a holomorphic
relation between $w$ and $v$.  As a result, we cannot hope to obtain a
geometry that is $T$-dual to the local CY \eqref{cubdefgeomm} with
fluxes except in an approximate sense.

How then can we explain the nice agreement with IIB produced by our
``physically''-motivated curve \eqref{wvparamcurve}, \eqref{sguess}?
Because it cannot be an exact solution, the best we can hope for is
that, as alluded to above, it is an approximate one.  To make this more precise, note that, at leading order in $\tilde{\Lambda}/\Delta$, the characteristic size of contributions to \eqref{virasoro} from the embedding coordinate $s$ is $g_s N$ while those from $v$ and $w$ are $\tilde{\Lambda}^2/\Delta$ and $g\tilde{\Lambda}^2$, respectively{\footnote{The easiest way to see this is by mapping the parameters $\tilde{\Lambda}$, $\Delta$, and $g$ to the effective quantities $\Lambda_{n=1}$ and $m$ which determine the geometry near either brane stack.  In particular, $\Lambda_{n=1}=\tilde{\Lambda}^2/\Delta$ and $m=g\Delta$.  We now use the fact that the genus 0 curve has characteristic scales $v\sim \Lambda_{n=1}$ and $w\sim m\Lambda_{n=1}$ to obtain the desired result.  One can also obtain the scaling by studying the elliptic functions in \eqref{vwparamrep} directly, but this is less transparent.}}.  This means that the nonholomorphic contributions $\delta v,\delta w$ to $v,w$ typically scale like
\begin{equation}
\delta v \sim\frac{(g_sN)^2\Delta}{\tilde{\Lambda}^2},\qquad \delta w\sim \frac{(g_sN)^2}{g\tilde{\Lambda}^2},
\end{equation}
and hence are suppressed relative to the holomorphic ones precisely when
\begin{equation}
\frac{g_sN}{\Delta}\ll \left(\frac{\tilde{\Lambda}}{\Delta}\right)^2 \min(1,g\Delta).\label{partialregime}
\end{equation}
Quite nicely, this is less stringent than the second of the conditions
\eqref{IIAregime} required for our analysis based on the Nambu-Goto
action to yield a reliable weakly-coupled IIA description of the
configuration as a curved NS5-brane with flux\footnote{The second
equation in \eqref{IIAregime} gives different conditions depending on
the values of $N_{1,2}$, but it is least stringent for $N_1=N_2$, for
which it is the same as \eqref{partialregime}.}.  In other words, when a
IIA interpretation is reliable the nonholomorphic corrections to $w$ and
$v$ are always suppressed!
%
% without this, somehow the page number overlaps with the footnotes..
\thispagestyle{empty}

This does not help at all with the problem of identifying the right $s$
from the 2-parameter family of possibilities \eqref{sfam}, though.  As
we shall see, there are a couple of ways to deal with this.  The most
direct is to actually find an exact solution to \eqref{virasoro} and
study it in the limit of equation \eqref{partialregime}.  This will be
done in the next subsection.  We could alternatively try to discern the
behavior of the curve in this regime directly from the action without
actually finding an exact solution.  This will be addressed in section
\ref{subsec:action}.

\subsection{An exact $M$-theory lift}

\begin{floatingfigure}{0.45\textwidth}
\begin{center}
 \vspace*{2ex}
\epsfig{file=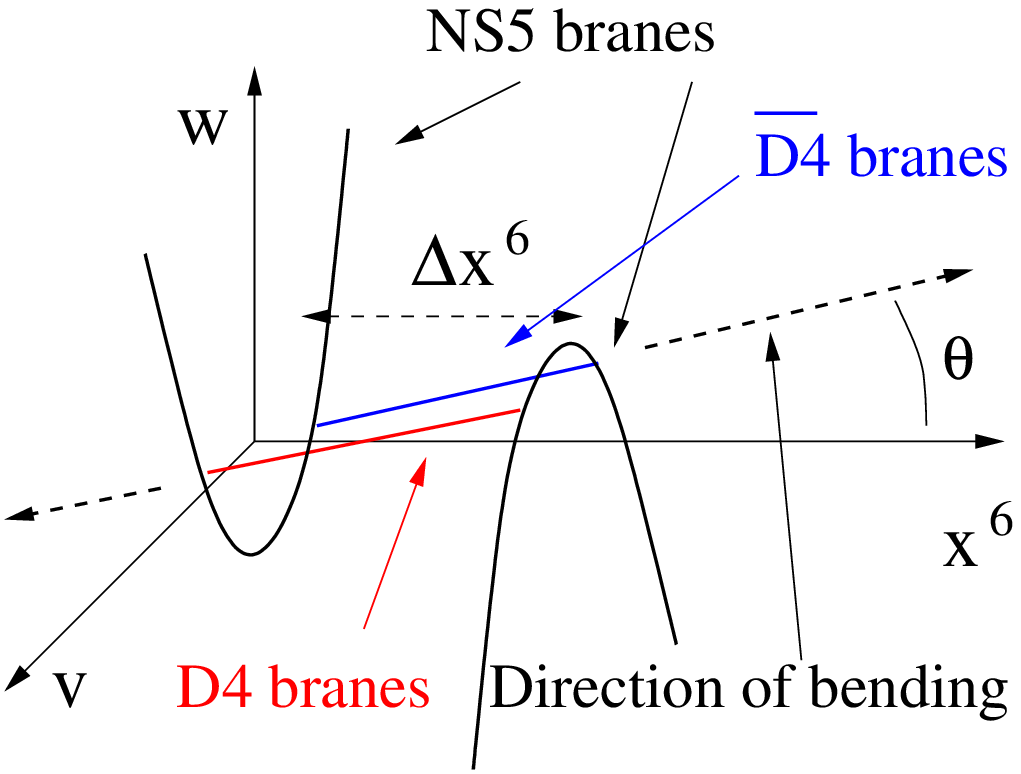,width=0.4\textwidth}
\caption{Tilting of D4's and \anti{D4}'s that gives rise to logarithmic bending along the $w$-direction in the $M$-theory lift \eqref{exactsol}.}
\label{wbending}
\end{center}
\end{floatingfigure}

Following the discussion above, we are led to conjecture that the configuration depicted in figure \ref{A1cubTdualD4D4bar} can be lifted to an $M5$ curve that, in the limit \eqref{partialregime}, is approximated by a holomorphic geometry of the form \eqref{wvparamcurve} with a suitable harmonic embedding coordinate $s$.  To test this conjecture, let us first search for an exact solution.
This is most easily accomplished if the number of D4-branes, $N_1$, is equal to the number of \anti{D4}-branes, $N_2$
\begin{equation}
N_1=N_2\equiv N
\end{equation}
because the additional symmetry leads to significant simplifications.  

As described in Appendix \ref{app:exactsol}, the resulting curve satisfies
\begin{equation}\Re(\tau)=0,\qquad\qquad\tau = 2a\end{equation}
which causes a number of elliptic quantities to vanish and permits us to write a relatively-simple exact solution
\begin{equation}
\begin{split}s&= Nr_0\cos\theta\left[\left(F_1-F_2-\frac{i\pi\tau}{2}+i\pi \left(z-A\right)\right)+\text{cc}\right]+i\pi N\left(z-A+\bar{z}-\bar{A}\right),\\
v&= X\left(F_1^{(1)}-F_2^{(1)}-\left[F^{(1)}(a)-i\pi\right]\right)+\frac{2\xi(g_sN)^2}{\bar{X}}\left(\bar{F}_1^{(1)}-\bar{F}_2^{(2)}-\left[\overline{F^{(1)}(a)}+i\pi\right]\right),\\
w&= g_sNr_0\sin\theta\left[\left(F_2-F_1-\frac{i\pi\tau}{2}-i\pi (z-A)\right)+\text{cc}\right]+\frac{g_sN\xi}{r_0\sin\theta}\left(F_1^{(2)}-F_2^{(2)}\right),
\label{exactsol}
\end{split}\end{equation}
where
\begin{equation}
r_0^2\equiv\frac{3\pi^2\wp(\tau/2)}{3\wp(\tau/2)^2-g_2},
 \qquad\qquad\xi\equiv\frac{\pi^2}{6\wp(\tau/2)^2-2g_2}
\end{equation}
and $g_2$ is one of the Weierstrass elliptic invariants as defined in
Appendix \ref{app:ellipfcns}\@.  This solution as written admits three
free parameters, $X$, $\theta$, and $\tau$.  The first two are analogous
to the quantities $X$ and $C$ in the supersymmetric curve
\eqref{vwparamrep} \eqref{susys} and encode the boundary conditions of
the curve along $w$ and $v$.  The third, $\tau$, is determined in terms
of the boundary data $\alpha$ by the noncompact $B$-period of $ds$.

There are a number of interesting features to this solution but let us focus for now on one in particular, namely that the functions $F_i$, which exhibit logarithmic behavior and have described the bending of NS5's along $x^6$ in our supersymmetric examples, make their appearance in $w$ {\footnote{Note that $F_1$, $F_2$, and their conjugates enter in a combination that is single-valued in a manner analogous to the function $\ln (\lambda/\bar{\lambda})$.}}.  The reason for this is quite simple to understand.  The branes and antibranes can lower their energy by moving closer together.
Because of the NS5's, though, this requires a rotation into the $w$
direction and carries an energy cost associated to the corresponding
increase in length.  In the equilibrium configuration, the D4's and
\anti{D4}'s are thus rotated a bit, changing the
direction along which they pull on and ``dimple'' the NS5's as
illustrated in figure \ref{wbending}.  This is directly reflected in the
exact solution \eqref{exactsol}, where $\theta$ indicates the rotation
angle.

\subsubsection{The IIA regime}

Let us now turn to the regime in which the configuration
\eqref{exactsol} can be reliably interpreted as a curved NS5-brane with
flux.  From the discussion of section \ref{subsubsec:g1reliability}, we
see that this corresponds to the regime \eqref{IIAregime}
\begin{equation}
 g_s\ll 1,\qquad \frac{g_s N}{\Delta}\ll \left(\frac{\tilde{\Lambda}}{\Delta}\right)^2\min(1,g\Delta),\qquad g\ll \left(\frac{\tilde{\Lambda}}{\Delta}\right)^2\min\left(1,(g\Delta)^3\right).\label{IIAconsts}
\end{equation}
We now look for solutions with approximately holomorphic boundary conditions along $w$ and $v$ of the form
\begin{equation}w(v)\sim g\left(v^2-\frac{\Delta^2}{4}\right)+\ldots\end{equation}
where ellipses are used to indicate the possibility of nonholomorphic terms which we have previously argued must be suppressed when the second condition of \eqref{IIAconsts} is satisfied. 
To do this, let us rewrite \eqref{exactsol} in the form of \eqref{wvparamcurve} with nonholomorphic corrections.  Using \eqref{physparams} to relate curve parameters to the physical ones $g$ and $\Delta$, we obtain
\begin{equation}
 \begin{split}
  s &= Nr_0\sqrt{1-\frac{4(g_sN r_0)^2}{g^2\Delta^4}}\left[\left(F_1-F_2-\frac{i\pi\tau}{2}+i\pi (z-A) + \text{cc}\right)\right]+i\pi N(z-A+\bar{z}-\bar{A}),\\
  v&= \frac{\Delta}{\sqrt{12\wp(\tau/2)}}
  \biggl\{\left(F_1^{(1)}-F_2^{(1)}-\left[F^{(1)}(a)-i\pi\right]\right)\\[-1ex]
  &\qquad\qquad\qquad\qquad\qquad
  +4\left(\frac{g_s N r_0}{|\Delta|}\right)^2\left(\bar{F}_1^{(1)}-\bar{F}_2^{(1)}-\left[\overline{F^{(1)}(a)}+i\pi\right]\right)\biggr\},\\
  w&= \frac{2}{g}\left(\frac{g_s N r_0}{\Delta}\right)^2\left[\left(F_2-F_1+\frac{i\pi\tau}{2}+i\pi (z-A)\right)+\text{cc}\right]+\frac{g\Delta^2}{12\wp(\tau/2)}\left(F_1^{(2)}-F_2^{(2)}\right).
\label{exactsolexp}
 \end{split}
\end{equation}

From this, it is quite easy to see that the nonholomorphic contribution to $v$ is negligible provided{\footnote{Note that we use the fact that $r_0\sim 1$, which is true when $\tilde{\Lambda}/\Delta$ is small as we are assuming.}}
\begin{equation}\frac{g_sN}{\Delta}\ll 1\end{equation}
To determine precisely when the nonholomorphic contribution to $w$ is negligible, though, one must look a bit more closely at the properties of the elliptic functions $F_i^{(n)}$.  The largest contributions to the ratio $\delta w/w$ come from the regions near near the D4 and \anti{D4} ``tubes'' and hence near the midpoints between $a_1$ and $a_2$.  There, it is not difficult to show that the $F_i^{(n)}$ all scale like $q^{1/2}=(\tilde{\Lambda}/\Delta)^2$.  This means that the first term of the expression for $w$ in \eqref{exactsolexp} scales like $g^{-1}(g_s N/\Delta)^2\,\,${\footnote{Note in particular that the contribution from the $F_i$ is suppressed relative to that of the constant term $(z-A)$.}} while the second term scales like $g\tilde{\Lambda}^2$.  This means that the nonholomorphic contribution to $w$ is suppressed when
\begin{equation}\frac{g_sN}{\Delta}\ll g\tilde{\Lambda}\end{equation}
and hence that our solution reduces to a hyperelliptic curve along $w$ and $v$ with harmonic embedding coordinate $s$ when
\begin{equation}\frac{g_sN}{\Delta}\ll \min(1,g\tilde{\Lambda})\label{redregime}\end{equation}
Note that this condition is less stringent than what we expected from general reasoning \eqref{partialregime}.  The reason for this is that while $s$ as a whole scales like $g_sN$, the division into holomorphic and antiholomorphic pieces is not symmetric.  In particular, the holomorphic part of $s$ scales like $g_sN$ but the antiholomorphic piece actually goes instead like $g_sN\tilde{\Lambda}^2/\Delta^2<g_sN$.  This means that $g_sN$ can actually be a bit larger than our previous reasoning would have suggested.  We expect this behavior to persist even for more general numbers of branes and antibranes so that nonholomorphic contributions to $w$ and $v$ can be neglected for a parameter range that is wider than what we need for the IIA interpretation to be reliable.

\subsubsection{The Reduced Curve}

Let us now focus on the regime \eqref{redregime} in which the nonholomorphic corrections to $w$ and $v$ in \eqref{exactsolexp} can be dropped and we are left with the approximate solution
\begin{equation}\begin{split}
s&= r_0 N\left[\left(F_1-F_2-\frac{i\pi\tau}{2}+i\pi (z-A)\right)+\text{cc}\right]+i\pi N(z-A+\bar{z}-\bar{A}),\\
v&=\frac{\Delta}{\sqrt{12\wp(\tau/2)}}\left(F_1^{(1)}-F_2^{(1)}-\left[F^{(1)}(a)-i\pi\right]\right),\\
w&=\frac{g\Delta^2}{12\wp(\tau/2)}\left(F_1^{(2)}-F_2^{(2)}\right).
\label{redcurve}
\end{split}\end{equation}
This geometry looks precisely like its supersymmetric counterpart \eqref{vwparamrep} with the holomorphic embedding along $s$ \eqref{susys} simply replaced by a harmonic one.  Note that a very specific expression for $s$ has emerged with an extra factor of $r_0$ that we were unable to produce without the exact solution \eqref{exactsol}.  As expected, this reduces to our guess \eqref{sguess} when $\Im(\tau)$ is large because in this regime $r_0$ becomes
\begin{equation}
r_0^2 = 1+40e^{i\pi\tau}+\ldots
\end{equation}
Without the solution \eqref{redcurve} in hand, though, we were
unable to say anything about possible subleading corrections away from the limit of infinite
$\Im(\tau)$.  Requiring the existence of an exact solution has now
completely fixed the ambiguity of \eqref{sfam} for all values of $\tau$.

This geometry is connected by $T$-duality to a local CY of the form
\eqref{cubdefgeomm} with fluxes just as the large $N$ duality conjecture
of \cite{Aganagic:2006ex} would suggest.  But what about the moduli?  We
saw in the previous section that they agree in the limit $\Im(\tau)\rightarrow\infty$
but, armed with the correction terms in $r_0$, can we go further?

Quite nicely, the answer to this question is a resounding yes.  The
complex structure of the curve \eqref{redcurve} is completely fixed by
the relations
\begin{equation}\Re(\tau)=0,\qquad\qquad \tau=2a,\label{nonsusyminima}\end{equation}
\begin{equation} 
2\pi i
\alpha(v_0)=r_0N\left(\left[L(\tau/2,\tau)+\ln\left(\frac{v_0^2}{\Delta^2}\right)\right]+\left[\overline{L(\tau/2,\tau)}+\ln\left(\frac{\bar{v}_0^2}{\bar{\Delta}^2}\right)\right]-i\pi
\tau\right)
\label{nonsusyminimaalph}
\end{equation}
where $L(a,\tau)$ is as in \eqref{Ldef} and the condition
\eqref{nonsusyminimaalph} arises from the $B$-period constraint in
\eqref{nspers}.

As discussed in Appendix \ref{app:exactsol}, it is now easy to verify,
using the expression \eqref{taumat} for the period matrix
$\hat{\tau}_{ij}$ of the geometry \eqref{redcurve}, that $a$ and $\tau$
satisfying \eqref{nonsusyminima} \emph{exactly} solve the equations of
motion which follow from the IIB potential \eqref{IIBnonsusypot}.  In
the regime where our exact solution \eqref{exactsol} provides a reliable
IIA description of the system in terms of a curved NS5-brane with flux,
it is thus exactly $T$-dual to the IIB system after the large $N$
transition to all orders in $\tilde{\Lambda}/\Delta$!

\subsection{Direct analysis of the NS5 action}
\label{subsec:action}

In order to better understand this agreement, we now turn to a direct
study of the M5/NS5 action in the IIA regime \eqref{IIAconsts} where the
M5-brane can be described as an NS5-brane with flux.\footnote{The
worldvolume action has previously been used to obtain K\"ahler
potentials in \cite{deBoer:1997zy}.}  We shall find that, in this limit,
the IIB potential \eqref{IIBnonsusypot} makes a natural appearance and
is responsible for fixing the moduli from the IIA/M point of view as
well.

In M theory the M5-brane was curved in the $v,w,s=R^{-1}(x^6+i x^{10})$
directions. Upon reduction to IIA we expect a single NS5-brane curved in
$v,w,x^6$.  The winding of the original M5-brane around the M-theory
circle $x^{10}$ is encoded in the nontrivial configuration of the
one-form field strength $F_m$ of the worldvolume theory of the NS5-brane
in IIA\@.  The relevant part of the NS5-brane Lagrangian is
\cite{Bandos:2000az, Bena:2006rg}:
\begin{equation}I = {1\over g_s ^2}\int \sqrt{\det(g_{mn} + g_s^2 F_m F_n)}.
\label{fullns}\end{equation}
The NS5-brane configuration is specified by imposing the boundary conditions 
\begin{equation}w(v)\sim \pm W_n'(v)\label{bcs}\end{equation}
at infinity, requiring the fluxes of $F_m$ to be consistent with the numbers of D4's
and \anti{D4}'s that we started with and fixing the logarithmic bending of $x^6$
at infinity. 

Because the boundary conditions \eqref{bcs} along $v$ and $w$ are
independent of $g_s$ and $N$, the corresponding embedding functions can
remain macroscopically large even regardless of the value of $g_s N$.
This means that, provided we take $g_sN$ to be sufficiently small, it is
possible to obtain a parametric separation of scales relevant for the
$w,v$ and $s$ parts of the geometry, respectively.  Indeed, the second
condition defining the IIA regime \eqref{IIAconsts} is essentially just
this because it is equivalent to imposing $|ds/dw|,|ds/dv|\ll
1\,${\footnote{Recall that, on general grounds, we expect $|ds|\sim
g_sN$, $|dv|\sim \tilde{\Lambda}^2/\Delta$, $|dw|\sim
g\tilde{\Lambda}$.}}.  As a result, it is natural to separate the
contribution to the induced metric coming from $x^6$ and write the NS5
worldvolume action as:
\begin{equation}{I = {1\over g_s ^2}\int \sqrt{\det(g'_{mn} + \partial_m x^6 \partial_n x^6 + g_s^2 F_m F_n)}}
\label{fullnsb}\end{equation}
where the induced metric $g'$  is that associated to the embedding coordinates $w$ and $v$, treating $x^6$ as a constant.
As explained above, the last two terms scale like $(g_s N)^2$.  So defining:

\begin{equation}{ds= {dx^6 \over g_s}+i F}\end{equation} and expanding to the first nontrivial order in $s$ we find
\begin{equation}
 I 
= \frac{1}{g_s^2}\int\,\sqrt{g'} + {1\over 2}\int\,ds\wedge *\overline{ds},\label{reducedns}\end{equation}
where $*$ is with respect to the metric $g'$.

Because we do not have to worry about the complications associated with
finding exact solutions to the equations of motion, let us now be
completely general here and try to minimize \eqref{reducedns}, starting with a configuration of NS5's wrapping
holomorphic curves
\begin{equation}w^2(v)=W_n'(v)^2\end{equation}
for $W'_n(v)$ a polynomial of degree $n$ and suspending arbitrary numbers of D4's and \anti{D4}'s between the them at the $n$ critical points of $W_n'(v)$.  These configurations are related by $T$-duality to the general ones of \cite{Aganagic:2006ex} with many conifold singularities. As we argued above we can  replace
the NS5's and D4/\anti{D4} branes with a single NS5-brane with fluxes turned on. As usual, we shall denote the RR fluxes associated to these branes by $N^i$ with negative fluxes corresponding to antibranes. The embedding of the NS5-brane has to satisfy the boundary conditions:

\begin{equation}w(v)\sim \pm W_n'(v)\label{bcsb}\end{equation}
at infinity and the periods of $ds$ must be consistent with the fluxes as in \eqref{dsaper} and \eqref{dsbper}
\begin{equation}
\oint_{A_i}ds = 2\pi iN^i,\qquad\qquad\oint_{B_j}ds = -2\pi i\alpha.  \label{abperrepeat}\end{equation} 

If we consider only the first term of \eqref{reducedns}, ignoring the second for a moment, we find that the embedding described by $w$ and $v$ alone must be of minimal area.  Since we impose holomorphic boundary conditions \eqref{bcsb} at infinity, the entire surface must be holomorphic and we are again led to the family of hyperelliptic curves \eqref{hyperdef}
\begin{equation}w^2=W'(v)^2-f_{n-1}(v).\label{hyperdefrepeat}\end{equation}
Note that the complex structure moduli of these curves are true flat directions of the first term in \eqref{reducedns}.

If we now consider only the second term in the action and minimize it with respect to $s$ for given and fixed embedding $w(v)$, the equations of motion imply that $ds$ is a harmonic 1-form with respect to the induced metric $g'$.  As we shall see explicitly in a moment, there is a unique harmonic 1-form $ds$ with specified $A$ and $B$ periods for
each value of the complex structure moduli of the curve \eqref{hyperdefrepeat}.

This gives us our embedding but what about the moduli?  Despite the fact that they correspond to flat directions of the first term in \eqref{reducedns}, this is not true of the second $s$-dependent term.  The moduli must therefore be chosen to minimize this quantity which, as we shall see, is nothing more than the IIB potential \eqref{IIBnonsusypot}{\footnote{Consequently, we see that the two terms in \eqref{reducedns} are, in a sense, not quite decoupled.}}.

\subsubsection{Construction of $ds$ and the effective potential}

To construct the harmonic form $ds$ with the desired periods we proceed
in the following way: we can parametrize the complex structure of the
surface \eqref{hyperdefrepeat} by the periods of $\half w\,dv$ on the $A$
cycles:
\begin{equation}
 {1\over 2\pi i}\oint_{A_i} \half w\, dv = S_i.
\end{equation}
The periods of $w\,dv$ on the $B$ cycles are determined by the
holomorphic prepotential ${\cal F}$ of the curve:
\begin{equation}
 {1\over 2\pi i}\oint_{B_i} \half w\, dv = {\partial  {\cal F} \over \partial S_i}.
\end{equation}
For each value of $S_i$ we want to construct a harmonic form $ds$ on the surface
with the desired periods. A harmonic form can be written as the sum of
holomorphic and antiholomorphic forms. It is convenient to pick the following
basis of holomorphic forms:
\begin{equation}{\omega_i = {1\over 4\pi i}{\partial  w  \over \partial S_i} dv}\end{equation}
as they have nice periods on the A cycles:
\begin{equation}{\oint_{A_i} \omega_j = \delta_{ij}}\end{equation}
and on the $B$ cycles:
\begin{equation}{\oint_{B_i} \omega_j = {\partial^2 {\cal F} \over
\partial S_i \partial S_j} = \hat{\tau}_{ij}}\end{equation}
where $\hat{\tau}_{ij}$ is the period matrix of the surface.
The harmonic form $ds$ is the sum of holomorphic and antiholomorphic forms:
\begin{equation}{ds = h_i \omega_i + \overline{\ell}_i \overline{\omega}_i}.
\end{equation}

The $2n$ constraints \eqref{abperrepeat} arising from imposing the specific $A$ and $B$ periods now determine the $2n$
coefficients $h_i$ and $\ell_j$ completely in terms of $\alpha$ and the
period matrix $\hat{\tau}$ in a manner that is straightforward to
determine.  Collecting $h_i,\ell_i,N^i,\alpha^i$ into vectors
$\vec{h},\vec{\ell},\vec{N},\vec{\alpha}$ and recalling that the
$\alpha_i$ are all identical, we can write the result in a compact form
\begin{equation}\begin{split}\vec{h}&=\vec{\bar{\ell}}+2\pi i \vec{N},\\
\vec{\bar{\ell}}&
 = -\pi\left(\Im\hat{\tau}\right)^{-1}\left(\vec{\alpha}+\hat{\tau}\vec{N}\right).
\label{hlcoeffs}
\end{split}
\end{equation}
To fix the moduli, we must now minimize \eqref{reducedns}.  We focus on
the second term because this is the only one that depends on the moduli.
Using \eqref{hlcoeffs}, we can write it directly in terms of $\alpha$
and $\hat{\tau}$
\begin{equation}\begin{split}V &= \frac{1}{(2\pi)^2}\int_{\Sigma}ds\wedge*\overline{ds}\\
&=-\frac{1}{2\pi^2}\Im\left(\vec{h}^T\overline{[\hat{\tau}\vec{h}]}+\ell^T\overline{[\hat{\tau}\ell]}\right)\\
&=\left(\vec{\alpha}+\hat{\tau}\vec{N}\right)\left(\Im\hat{\tau}\right)^{-1}\left(\bar{\alpha}+\bar{\hat{\tau}}\vec{N}\right)-2\Im\left(\vec{\alpha}^T\vec{N}\right).
\end{split}\end{equation}
This is precisely the effective potential on the IIB side obtained from
special geometry
\cite{Aganagic:2006ex} including the ``zero-point
shift'' introduced therein.  Consequently, we see that when we can reliably use the Nambu-Goto action to provide a reliable description of the system at weak coupling, the IIA picture quite
generally reproduces the IIB story complete
with some aspects of the off-shell physics.  Note that from this point
of view, the supersymmetric vacua also fall out nicely because any
holomorphic $ds$ that satisfies the constraints \eqref{abperrepeat}
automatically minimizes \eqref{reducedns}.

\subsubsection{More on the connection with IIB}

We can make the above agreement of the effective potentials computed in the IIB and IIA systems more transparent as follows.
In IIB after the geometric transition we have a noncompact Calabi-Yau with fluxes.  Without flux, the system would have flat directions parametrized by the complex structure moduli of the Calabi-Yau.  The fluxes then create a potential for these moduli which stabilizes them.  This potential \eqref{IIBnonsusypot} can be written suggestively as
\begin{equation}
V_{\rm IIB} = {\cal{K}}^{ij}\partial_iW\overline{\partial_jW}
\label{scalariib}
\end{equation}
where $W$ is the GVW superpotential \eqref{gvw} and ${\cal{K}}_{ij}$ the K\"ahler metric \eqref{kahlerpot}.  Using standard results from rigid special geometry, it is not difficult to show that the scalar potential \eqref{scalariib} is simply the ``electromagnetic'' energy of the $H$ flux of the system
\begin{equation}
V_{\rm IIB} = {\cal{K}}^{ij}\partial_iW\overline{\partial_jW}=\int_{CY}H\wedge \ast\overline{H}.
\end{equation}
Now the analogy with IIA should be clear.  There, we started with two NS5-branes on the curves $w\sim\pm W'(v)$ which get replaced by a single curved NS5 on a hyperelliptic curve of the form \eqref{hyperdef}
\begin{equation}w^2=W_n'(v)^2-f_{n-1}(v)\end{equation}
with 1-form flux $F_m$ on its worldvolume turned on and logarithmic
bending along $x^6$, both of which are combined into our complex
coordinate $s$.  Without the flux and accompanying bending, the system
has flat directions parametrized by the complex structure moduli of the
hyperelliptic curve.  The presence of nontrivial $ds$ induces an
effective potential for the moduli and lifts the degeneracy.  The form
of this potential is simply
\begin{equation}
V_{\rm IIA}\sim\int\,ds\wedge \ast\overline{ds}
\end{equation}
which, given the identification \eqref{shident}
\begin{equation}
ds\sim \oint_{S^2}H
\end{equation}
is nicely consistent with $T$-duality.  Indeed, the tension of our bent NS5 with flux in IIA is directly identified with the energy stored in the nontrivial $H$-flux in IIB\@.

\subsection{When are our solutions true minima?}
\label{subsec:phase}

Before closing this section, let us come back to a subtle point that
needs to be addressed.  While we have demonstrated that our curves solve
the equations of motion that follow from the IIB potential
\eqref{IIBnonsusypot}, we have not yet said anything about the nature of
such solutions or even demonstrated that they exist in the first place.
In particular, the moduli are fixed by \eqref{nonsusyminima} but
inverting this equation to actually determine $\tau$ is nontrivial.

Precisely this issue was addressed on the IIB side in the recent paper \cite{Heckman:2007wk}, which used the Dijkgraaf-Vafa matrix model technology to compute subleading corrections to the potential \eqref{IIBnonsusypot} at large $\Im(\tau)$ and study the structure of its critical points.  They found a fairly intricate phase structure that should make an appearance in our formalism as well.  In this section, we briefly comment further on this.  While far from our original goal, one nice benefit of our work is that we have exact expressions in hand for both the solutions as well as the IIB potential for a cubic superpotential and equal numbers of branes and antibranes.  This allows us to obtain exact results for many of the quantities which characterize the phase structure of \cite{Heckman:2007wk}.  It is important to point out, though, that the primary simplification arises not from a particular property of the IIA/M description itself but rather from the formalism we develop
 ed to study it{\footnote{In particular, it is the parametrization of the complex moduli space by $a$ and $\tau$ that makes things simple.  This does not come without a price, though, because the dictionary relating these quantities to the $S_i$ \eqref{Sexactt2a} is quite complicated.}}, which does not easily generalize to curves with higher genus.

Before proceeding, let us first revisit the manner in which moduli are
fixed from the IIA/M point of view.  As described in detail in Appendix
\ref{app:exactsol}, the conditions \eqref{nonsusyminima} follow directly
from requiring that $s$ be periodic along the compact $B$-cycle and
imposing a $\mathbb{Z}_2$ symmetry on the full solution.  The former is
a necessary condition for existence of a well-defined curve{\footnote{Of
course, we could have taken $s$ to be periodic along the compact
$B$-cycle up to an integer multiple of $2\pi R_{10}$, but we have chosen
this integer to be zero for simplicity.}} while the latter is required
because we imposed boundary conditions that preserve the obvious
$\mathbb{Z}_2$ of the brane/antibrane system.  Consequently, from the
IIA/M point of view, simply having a smooth curve with the right
boundary conditions already fixes all of the moduli except for the
imaginary part of $\tau$, which we hereafter refer to as $T$
\begin{equation}
T\equiv-i\tau.
\end{equation}
This quantity is finally fixed in terms of the boundary data $\alpha(v_0)$ by imposing the noncompact $B$-period constraint in \eqref{nspers}.  The result is \eqref{nonsusyminimaalph}
\begin{equation}
2\pi i\alpha(v_0)=r_0N\left(2L(iT/2,iT)+\pi T +\ln\left|\frac{v_0}{\Delta}\right|^4\right)
\label{nsa}
\end{equation}
where we have implicitly used the fact that $L(iT/2,iT)$ is real.  As we saw in the previous section, imposing this condition is equivalent to further extremizing the approximate action \eqref{reducedns} with respect to $T$.  The form of this action is equivalent to the IIB potential \eqref{IIBnonsusypot} and takes the form
\begin{equation}
V(T)=\frac{N^2}{\pi}\left(\frac{\left(2\pi i \alpha/N\right)^2}{2\left[L(iT/2,iT)+\pi T\right]+\ln\left|\frac{v_0}{\Delta}\right|^4}+\pi T\right).
\label{tpot}
\end{equation}
In the IIA/M, approach, we are thus naturally led to study the potential \eqref{tpot} as a function of a single real parameter $T$, which is easily seen to correspond essentially to the parameter $\delta$ in \cite{Heckman:2007wk}.  Critical points of this potential correspond to solutions of the full equations of motion.  Because the expression \eqref{tpot} is exact for all $T\ge 0$, we should in principle be able to study the nature of its critical points in quantitative detail even away from the regime $T\gg 1$.

As a first use of \eqref{tpot}, we can evaluate the potential at the degeneration point $T=0$.  In \cite{Heckman:2007wk}, it was argued that this quantity is independent of $N_1$ and indeed this can be verified explicitly
\begin{equation}
V(0)=\frac{2\pi(i\alpha)^2}{L(0,0)+\frac{1}{2}\ln\left|\frac{v_0}{\Delta}\right|^4} .
\label{Vzero}
\end{equation}
In fact, this is precisely the form of equation (103) of
\cite{Heckman:2007wk} with $B_t=L(0,0)$.

It is also fairly straightforward to study the qualitative structure of
the critical points of \eqref{tpot}, which correspond to solutions of
\eqref{nsa}.  To do so, we note that
\begin{equation}
2L(iT/2,iT)+\pi T
\label{LLt}
\end{equation}
is a monotonically increasing function of $T$ satisfying{\footnote{Because of these properties, we must have that $|v_0|>|\Delta|/2\sqrt{2}$ in order prevent $V(t)$ from diverging and becoming negative at small $t$.  The need for such a condition is clear as we do not expect to be able to take the cutoff smaller than the physical scale $|\Delta|$.  The factor of $2\sqrt{2}$ can be understood by noting that the curve \eqref{vwparamrep} becomes
\begin{equation}w^2=g^2v^2(v^2-\Delta^2/2)\end{equation}
at $t=0$.  The cuts extend from $\pm\Delta/\sqrt{2}$ to 0 and are centered at $\Delta/2\sqrt{2}$.  The above condition simply corresponds to the sensible requirement that the physical cutoff to be larger than this scale.}}  
\begin{equation}
2L(0,0) = 6\ln 2,\qquad\qquad \lim_{T\rightarrow\infty}\left(L(iT/2,iT)+\pi T\right)=\infty
\label{LLtprops}
\end{equation}
while $r_0$ is a monotonically decreasing function of $T$ satisfying
\begin{equation}
\lim_{T\rightarrow 0}r_0=\infty\qquad\qquad\lim_{T\rightarrow\infty}r_0=1.
\end{equation}
\begin{figure}
\begin{center}
\epsfig{file=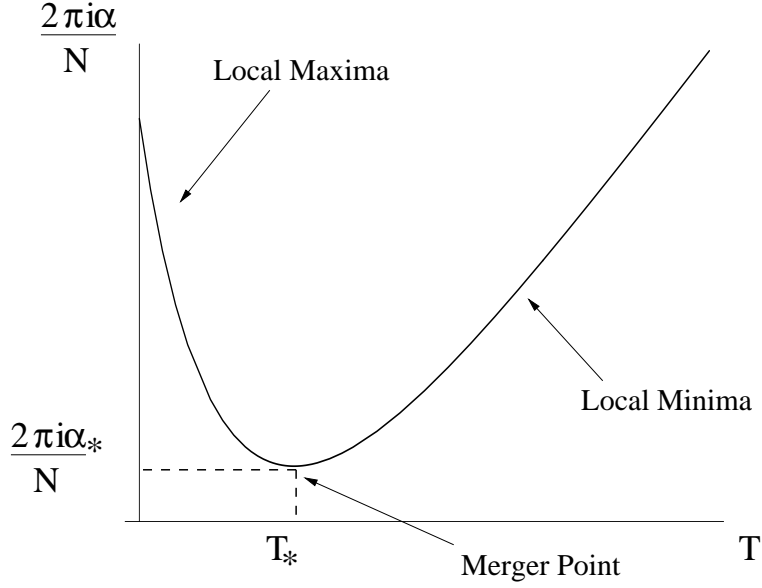,width=0.6\textwidth}
\caption{Solutions to the constraint \eqref{nsa}.  The $T_{\ast}>0$ branch corresponds to local minima of \eqref{tpot} while the $T_{\ast}<0$ branch corresponds to local maxima.}
\label{alphatplotfig}
\end{center}
\end{figure}
Consequently, if we fix $|v_0/\Delta|$ and view \eqref{nsa} as defining a function $\alpha$ as a function of $T$, the situation is as depicted in figure \ref{alphatplotfig} with $2\pi i\alpha$ attaining a minimum value $2\pi i\alpha_{\ast}$ at the point $T=T_{\ast}$.  What we see from this is that for $2\pi i\alpha>2\pi i\alpha_{\ast}$ there are two branches of solutions.  At $2\pi i\alpha_{\ast}$, these two branches merge so that for $2\pi i\alpha < 2\pi i\alpha_{\ast}$ there are no solutions at all.  From this structure, we can already conclude that one of the branches describes local minima and the other local maxima of \eqref{tpot}.  Physically, we expect that the $T>T_{\ast}$ branch yields the minima and indeed it is easy to see that this is the case by looking at the derivative of \eqref{tpot}, 
\begin{equation}
\partial_TV(T)=\frac{N^2}{\pi}\left(1-\frac{(2\pi i\alpha/N)^2}{r_0^2\left(2L(iT/2,iT)+\pi T+\ln\left|\frac{v_0}{\Delta}\right|^4\right)^2}\right)
\label{dtpot}
\end{equation}
which of course vanishes when \eqref{nsa} is satisfied.  Because $\partial_TV(T)$ is positive at both $T=0$ and $T=\infty$ it follows that, whenever $V(T)$ has two critical points, the one at larger (smaller) $T$ is a local minimum (maximum), in agreement our expectations and the results of \cite{Heckman:2007wk}.  Also note that, depending on the size of $2\pi i\alpha$, solutions along the $T>T_{\ast}$ branch may have higher or lower energy than the point $T=0$ \eqref{Vzero} and hence do not always correspond to global minima of the potential.  Sample plots of $V(T)$ that exhibit all of these features are presented in figure \ref{Vexs}.

\begin{figure}
\begin{center}
\subfigure[$2\pi i\alpha>2\pi i\alpha_{\ast}$ and sufficiently large that
the local minimum is also a global minimum of $V(T)$]
{\epsfig{file=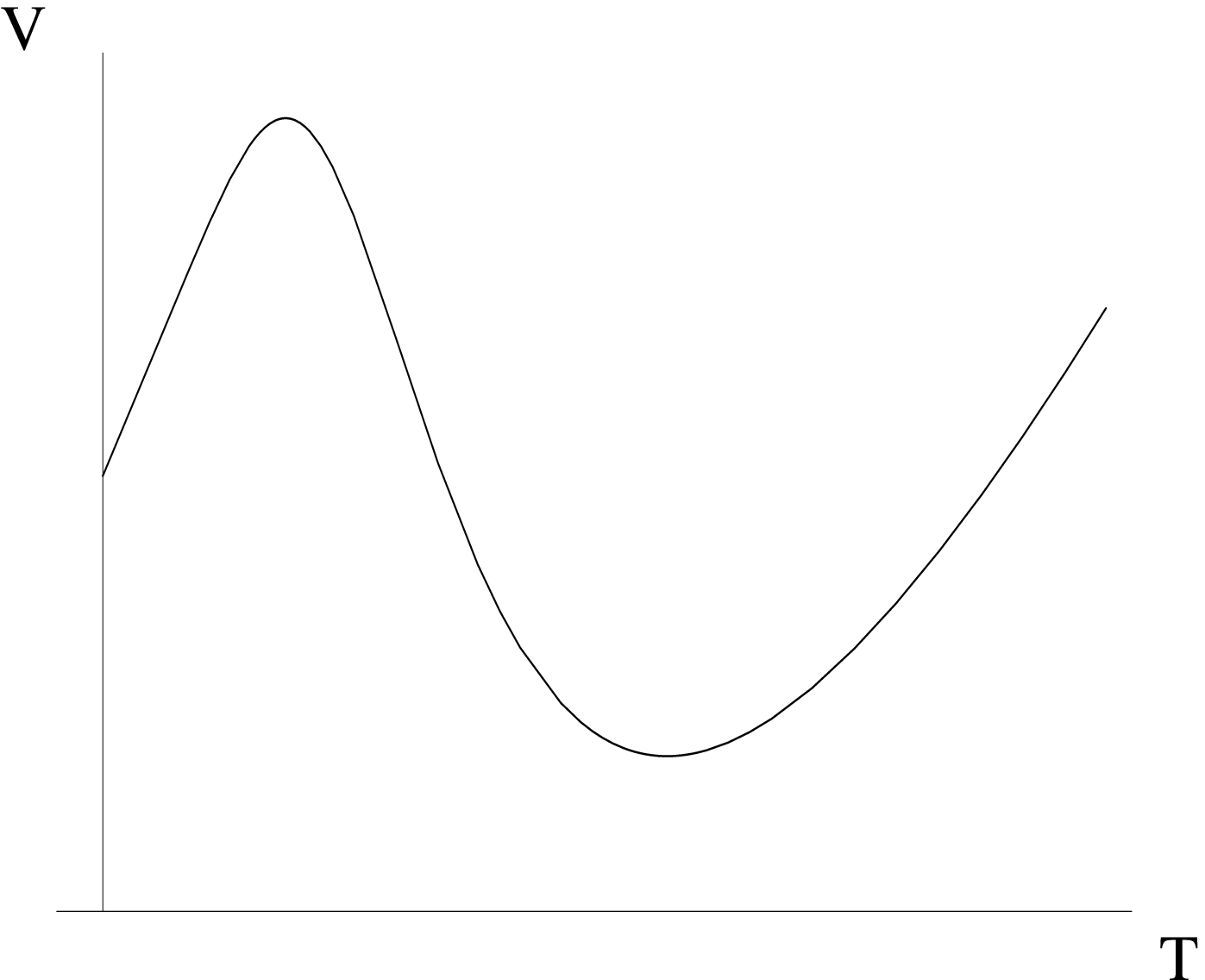,width=0.3\textwidth}\label{Vglobalmin}}
\subfigure[$2\pi i\alpha> 2\pi i\alpha_{\ast}$ but sufficiently small that
the local minimum is not a global minimum of $V(T)$]
{\epsfig{file=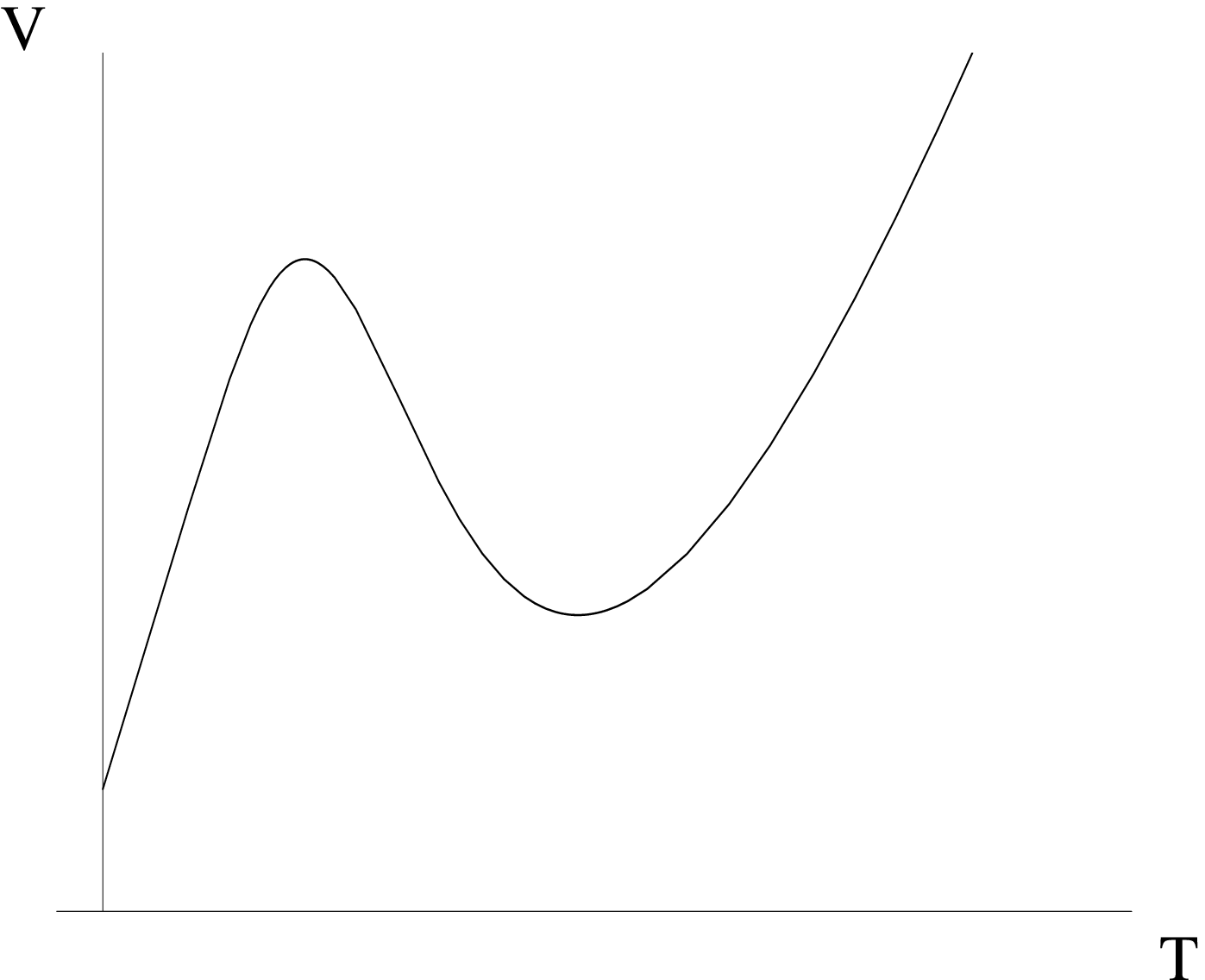,width=0.3\textwidth}\label{Vlocalmin}}
\subfigure[$2\pi i\alpha < 2\pi i\alpha_{\ast}$ so there is no local minimum and hence no solution to \eqref{nsa}.]
{\epsfig{file=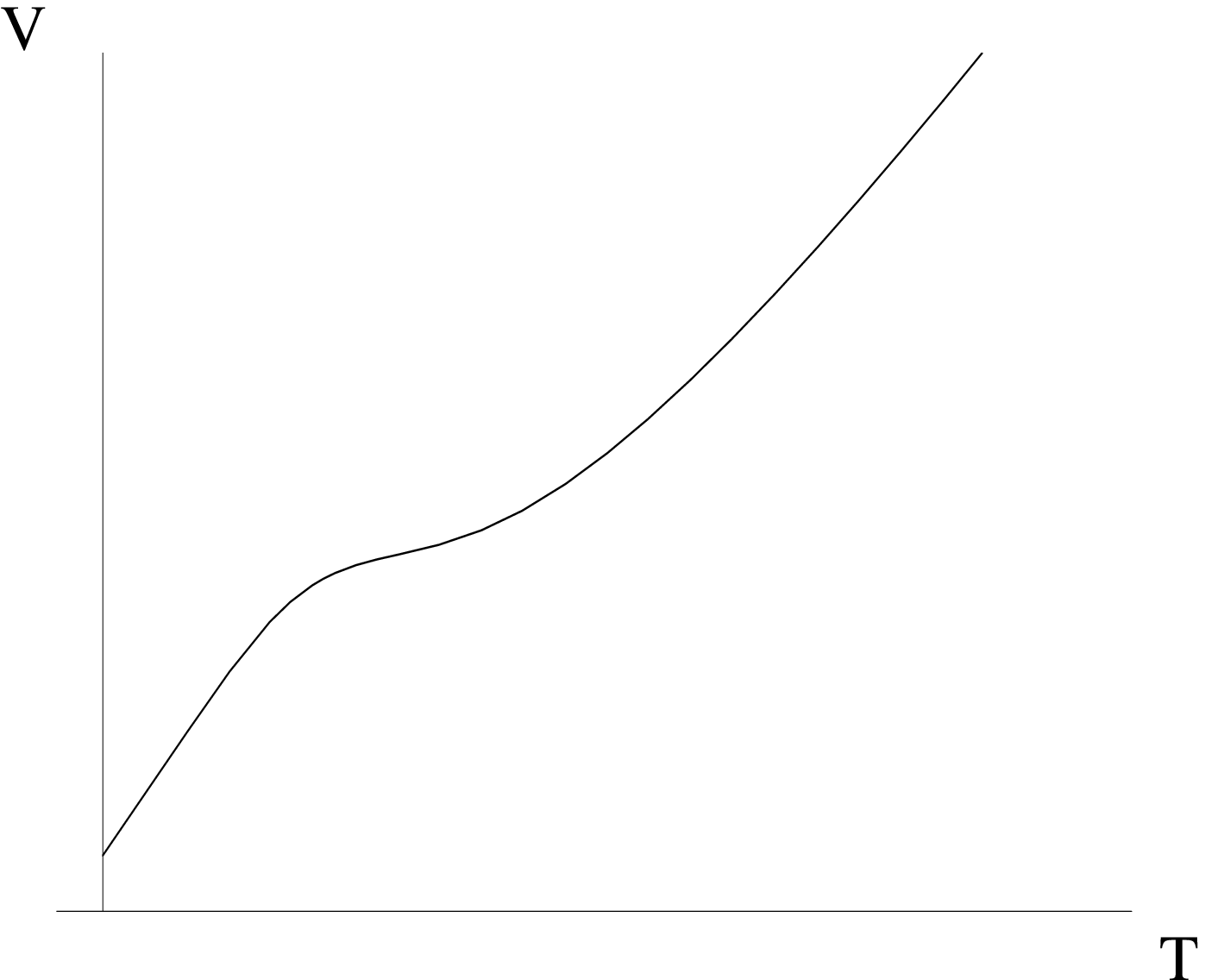,width=0.3\textwidth}\label{Vnomin}}
\caption{$V(T)$ for various choices of $2\pi i\alpha$}
\label{Vexs}
\end{center}
\end{figure}

What we have described above is precisely the phase structure of \cite{Heckman:2007wk}, to which we refer the reader for a more thorough discussion.  The critical point $2\pi i\alpha_{\ast}$ corresponding to a particular choice of $|v_0/|\Delta|$ is determined by the equations
\begin{align}
 \ln\left|\frac{v_0}{\Delta}\right|^4&=\frac{4\left(g_2-3\wp(iT_{\ast}/2)^2\right)}{6\wp(iT_{\ast}/2)\left(3\wp(iT_{\ast}/2)+2\eta_1\right)-g_2}-2L(iT_{\ast}/2,iT_{\ast})-\pi T_{\ast},
\label{tstar}\\
\frac{2\pi i\alpha_{\ast}}{N}&=\frac{4r_0\left(g_2-3\wp(iT_{\ast}/2)^2\right)}{6\wp(iT_{\ast}/2)\left(3\wp(iT_{\ast}/2)+2\eta_1\right)-g_2}.
\label{alphastar}
\end{align}
Unfortunately, these expressions are quite complicated so, even though we can in principle solve for $T_{\ast}$ and $2\pi i\alpha_{\ast}$ in full generality, it is difficult to do so in practice without resorting to numerics.  As noted in \cite{Heckman:2007wk}, though, one should be able to move all of the interesting physics to the regime of large $T$, where \eqref{tstar} and \eqref{alphastar} become more manageable, by tuning $|v_0/\Delta|$ appropriately.  To see this from \eqref{tstar} and \eqref{alphastar}, note that the LHS's of both equations are monotonically increasing functions of $T_{\ast}$ and hence that $T_{\ast}$ and $2\pi i\alpha_{\ast}$ both grow as $|v_0/\Delta|$ is increased.
This means that we can always ensure that $T_{\ast}\gg 1$ by choosing $|v_0/\Delta|$ to be sufficiently large.  Expanding \eqref{tstar} in this regime, we find
\begin{equation}\ln\left|\frac{v_0}{\Delta}\right|^4=\frac{e^{\pi T_{\ast}}}{20}-\pi T_{\ast}
+{\cal{O}}(1)\end{equation}
which, using \eqref{s1s2lt}, is equivalent at large $T_{\ast}$ to the condition for minimizing equation (87) of \cite{Heckman:2007wk}.  From this, we conclude that for $|v_0|\gg |\Delta|$
\begin{equation}e^{\pi T_{\ast}}=
20\ln\left|\frac{v_0}{\Delta}\right|^4+20\ln\left[20\ln\left|\frac{v_0}{\Delta}\right|^4\right]+\ldots
\label{tstarsol}
\end{equation}
Turning now to equation \eqref{alphastar}, we can expand at large $T_{\ast}$
\begin{equation}\frac{2\pi i\alpha_{\ast}}{N_1}=\frac{e^{\pi T_{\ast}}}{20}+{\cal{O}}(1)\end{equation}
and subsequently apply \eqref{tstarsol} to obtain equation (90) of \cite{Heckman:2007wk}
\begin{equation}\frac{2\pi i\alpha_{\ast}}{N_1}=20\left|\frac{v_0}{\Delta}\right|^4\ln\left|\frac{v_0}{\Delta}\right|^4+\ldots
\end{equation}

Another regime in which \eqref{tstar} and \eqref{alphastar} simplify is
that of small $\Im(\tau)$ but physically this is not very interesting
because the description that we have in hand breaks down there.  
As discussed in \cite{Heckman:2007wk}, additional degrees of freedom
that have not been accounted for enter the story.  On the IIB side, D3
and D5 branes wrapping the shrinking compact ${\cal{B}}$ cycle become
important.  On the IIA side, these correspond to D4 and D6 branes with
boundaries on the curved NS5 which wrap the compact $B$-cycle.

Of course, there is much more to say about the phase structure of this system than what we have described here.  For this, we refer the reader to the discussion of \cite{Heckman:2007wk}, which can easily be translated to the IIA/M picture via $T$-duality.

Finally, to address a comment made at the beginning of section
\ref{braneantibraneIIAM}, let us note that from \eqref{tstar} we can
determine numerically that, if we take $|v_0|>\Delta$, then
$T_{\ast}\gtrsim\frac{3}{2}$ and hence all true minima have
$e^{i\pi\tau}\sim (\tilde{\Lambda}/\Delta)^4$ at most of
${\cal{O}}(10^{-2})$.

\section{Discussion}
\label{sec:disc}

In this section, we comment on various lessons that can be drawn from our IIA/M analysis.  First, in section \ref{ss:meaning_gsN} we shall address the possible conclusions one can reach about the nature of supersymmetry breaking in these setups with particular attention paid to when some residual supersymmetric structure remains.  Following this, we turn in section \ref{boundaryconditions} to the implications that observations made in \cite{Bena:2006rg} have on our configurations, how they arise in the $T$-dual type IIB setup, and the general lessons one can draw from them for studies of metastable nonsupersymmetric configurations in local contexts.  Finally, in section \ref{futuredirections} we address some open questions that would be interesting to study further.

\subsection{Brane/antibrane configurations and spontaneous supersymmetry breaking}
\label{ss:meaning_gsN}

Though we have described our type IIA brane/antibrane configurations as
being $T$-dual to the IIB setups studied in \cite{Aganagic:2006ex}, there is an important
point that must be emphasized.  All of our analysis has relied on the
fact that the circle on which we perform $T$-duality is large in the IIA
picture.  On the other hand, the analysis of \cite{Aganagic:2006ex} implicitly assumes
that this circle is large on the IIB side so, in reality, the two
approaches are focusing on quite different regions of parameter space.
In supersymmetric situations, this typically is not important because
BPS arguments guarantee that quantities such as the compact complex
structure moduli in IIB are protected as the radius is varied.  In our
case, though, supersymmetry is broken so our ability to connect the IIB
and IIA pictures at weak coupling indicates that some additional
structure must be present there.

Indeed, the analysis of \cite{Aganagic:2006ex} relies quite heavily on the argument presented therein that, after the geometric transition, supersymmetry is spontaneously broken in these setups, at least at string tree level.  This additional structure is what ensures that the K\"ahler potential determined from special geometry is reliable and can be used to compute the potential \eqref{IIBnonsusypot} that fixes the moduli.  It must also be responsible for our successful comparison of the IIB and IIA pictures.

In a sense, then, our ability to reproduce the results of \cite{Aganagic:2006ex} in weakly coupled IIA despite the fact that we have essentially tuned the IIB radius to zero provides further evidence in support of the assertion made there that the breaking of supersymmetry is spontaneous, at least for some range of parameters.  Moreover, we may be gaining a hint as to what parameters actually control the ``severity'' of supersymmetry breaking because the novel features of our solutions are controlled by the objects $g_sN/\Delta$ and $g_sN/g\Delta\tilde{\Lambda}$.
 Whenever both of these quantities are small, the minimal area curve takes the form \eqref{redcurve} and can be successfully compared to type IIB\@.  On the other hand, when either is large, new phenomena such as the introduction of nonholomorphic bending along $w$ are introduced{\footnote{Note that we can reliably turn on $g_sN/\Delta$ and/or $g_sN/g\Delta\tilde{\Lambda}$ while continuing to use our minimal area curve for a range of parameters at strong coupling $g_s\gg 1$ so they are indeed real and can be studied reliably in the $M$-theory regime}}.

Indeed, the very structure of our curve suggests that supersymmetry is spontaneously broken when $g_sN/\Delta$ and $g_sN/g\Delta\tilde{\Lambda}$ are small because there it factorizes into two pieces which are separately holomorphic but with respect to different complex structures.  In other words, the solution factorizes in this regime into two pieces which are individually supersymmetric but preserve different sets of supercharges.  It is only the simultaneous presence of both which leads to breaking of supersymmetry.  This nice structure strongly suggests to us that supersymmetry thus remains spontaneously broken, and hence our solution remains reliable, even outside of the regimes discussed in Appendix \ref{app:validity} provided $g_sN/\Delta$ and $g_sN/g\Delta\tilde{\Lambda}$ are both small.

A natural question to ask now is whether or not the relevance of $g_sN/\Delta$ and $g_sN/g\Delta\tilde{\Lambda}$ can be understood more directly.  At strong coupling, their importance is rather obvious because, as we have seen, they control the ``backreaction'' of the nonsupersymmetric fluxes, reflected by the nonholomorphicity of $s$, on the $w,v$ geometry.  When they become important, nonholomorphicity enters into $w$ and $v$ and destroys the nice factorized form alluded to above.  Of course, we could have said essentially the same thing about our curved NS5 brane with flux at weak coupling.

A more difficult question, though, is how these parameters appear in the
open string picture with the D4 and \anti{D4} branes.  There, it is
well-known that the presence of the \anti{D4}'s leads, at string tree
level, to the introduction of an FI $D$-term which spontaneously breaks
supersymmetry{\footnote{for a review of this, see for instance
\cite{Brodie:2001pw}.}}.  Because there is no structure to prevent it,
we expect that the breaking becomes more severe when string interactions
are taken into account.  One might therefore expect that the breaking
remains spontaneous provided $g_s\ll 1$.  From our exact solution,
though, we see that it is also necessary for $g_sN$ to be bounded in
some sense.  What, though, is the physical relevance of $g_sN$?

\begin{floatingfigure}{.30\textwidth}
\begin{center}
\epsfig{file=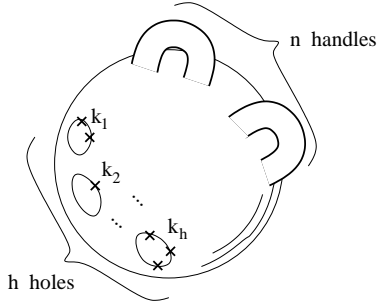,width=0.30\textwidth} \caption{Worldsheet
$\Sigma$ with $n\ge 0$ handles and $h\ge 1$ holes.  ``$\times$''
represents an open string vertex operator.} \label{stringamp}
\vspace*{3ex}
\end{center}
\end{floatingfigure}

For this, consider an open string amplitude, given by
inserting open string vertex operators, corresponding to fields living
on D-branes, on a worldsheet $\Sigma$.  Let the worldsheet be a sphere
with $h\ge 1$ holes and $n\ge 0$ handles added.  There are $h$
boundaries on the worldsheet.  Insert $k_i$ open string vertex operators
on the $i$-th boundary, where $i=1,\dots,h$.  Let the total number of
vertex operators be $k=\sum_{i=1}^h k_i$.
The amplitude then goes like
\begin{align}
 \sim g_o^{k} C_\Sigma \sim g_o^k g_o^{-4+2h+4n}
 = g_o^{k-4+2h+4n}.
\end{align}
Here $g_o\sim g_{s}^{1/2}$ is the normalization constant for open string
vertex operators and $C_\Sigma$ is the amplitude without
insertions{\footnote{Recall that $C_{S^2}\sim g_o^{-4}$, and adding a
hole to a worldsheet gives a factor of $g_o^2$ while adding a handle
gives a factor of $g_o^4$; this is how we got the above power of $g_o$.
}}. This leads to the term in the \emph{five}-dimensional action on the
D4/\anti{D4}-branes:
\begin{align}
 \sim g_o^{k-4+2h+4n} \int d^5 x\,\prod_{i=1}^h \tr(X^{k_i}),
\end{align}
where we denoted the fields collectively by $X$.  $X$ can be the adjoint
or bifundamentals.

To get the conventional normalization where the kinetic terms have
$1/g_o^2$ in front of them, let us rescale fields as $X\to
X/g_o$.
Then we have
\begin{align}
 \sim g_o^{-4+2h+4n} \int d^5x \prod_i \tr(X^{k_i})
 = N^{h}g_o^{-4+2h+4n} \int d^5x\prod_{i=1}^h {\tr(X^{k_i})\over N}.
\end{align}
If we integrate over $x^6$, this becomes
\begin{align}
 \sim N^{h}g_o^{-4+2h+4n} b \int d^4x\prod_{i=1}^h {\tr(X^{k_i})\over N}
 \sim
 {N^2 g_s^{2n}(g_s N)^{h-1}\over \lambda_4}
 \int d^4x\prod_{i=1}^h {\tr(X^{k_i})\over N}
 ,\label{kjnt8Mar07}
\end{align}
where we defined a four-dimensional 't Hooft coupling constant by
\begin{align}
 \lambda_4&\equiv {g_o^2 N\over b}
 \sim {g_s N\over b}.
\end{align}

From \eqref{kjnt8Mar07}, it is clear that $g_s$ counts the number of
handles and $g_s N$ counts the number of boundaries.  Thus, we see that
even if $g_s$ is small we can still get sizeable stringy corrections
from $g_sN$ provided $N$ is large enough.\footnote{ It was noted in
\cite{Berenstein:2002ge} that, in the $g_s\to 0$ limit with fixed $N$,
the gauge theories geometrically engineered by D-branes on CY
singularities receive contributions only from disks and thus have only
single-trace operators.  Here we are refining and generalizing his
argument by introducing parameters $g_s$, $g_sN$, and $\lambda_4$.  }
Determining precisely how large $g_sN$ must be, though, is a more
difficult question which requires a knowledge of how $\Delta$ and
$\tilde{\Lambda}$ appear in the various interaction terms
\eqref{kjnt8Mar07}.  It would be very interesting to pursue this further
and understand the relevance of the specific quantities $g_sN/\Delta$
and $g_sN/g\Delta\tilde{\Lambda}$ from this open string point of view.

\subsection{Metastability and boundary conditions}
\label{boundaryconditions}

We have constructed nonsupersymmetric configurations in type IIA/M theory
by suspending D4's and \anti{D4}'s between curved NS5's.  Because of
their tension, D4's and \anti{D4}'s pull and bend the NS5's
logarithmically in a manner that extends to infinity (Fig.\
\ref{decayfigure}(a)).

D4's and \anti{D4}'s being separated by a potential barrier, this
configuration is classically stable, but in quantum theory they can pair
annihilate by quantum tunneling, as we have already discussed in the
introduction.\footnote{For simplicity of the argument, we are assuming
that the numbers of D4's and \anti{D4}'s are the same,
$N_1=N_2=N$.}$^,$\footnote{A similar analysis of different
configurations appeared in \cite{Giveon:2007fk}.} The time it takes for
this tunneling process to occur can be estimated by a standard analysis,
considering an instanton interpolating the initial and final
configurations \cite{Aganagic:2006ex}.  Note that, for the physics of
this tunneling, the fact that the NS5's are bent logarithmically does
not matter; the D4 and \anti{D4} know only about their vicinity and what
is happening far away is irrelevant for this instantaneous process.
This is very much as the $\beta$-decay of a nucleus in an electric
field; the decay rate, governed by quantum tunneling, is not affected by
the electric field outside the nucleus.
Once all the D4's and \anti{D4}'s have pair annihilated (Fig.\
\ref{decayfigure}(b)), the tension that was holding NS5's is no more and
the NS5's start to straighten (Fig.\ \ref{decayfigure}(c)).  This part
of the decay process proceeds according to the classical equation of
motion, much as the fact that, once the $\beta$-decay has taken place,
the motion of the emitted positron in the electric field can be treated
classically.
The straightening of the NS5's propagates outward (Fig.\
\ref{decayfigure}(c),(d)), but the time it takes for the straightening
to reach the cutoff $v=v_0$ goes to infinity as one takes
$v_0\to\infty$.  Namely, in the $v_0\to\infty$ limit, the system has a
runaway instability.  Note that, the energy released in this process
becomes infinite as $v_0\to \infty$, because not only the mass of the
D4/\anti{D4} but also the tension of the infinitely long bent NS5 gets
converted into energy.
Although here we explained the decay process in the IIA language for the
sake of argument, but the decay process in M-theory is exactly the same.
\begin{figure}[htb]
\begin{center}
\epsfig{file=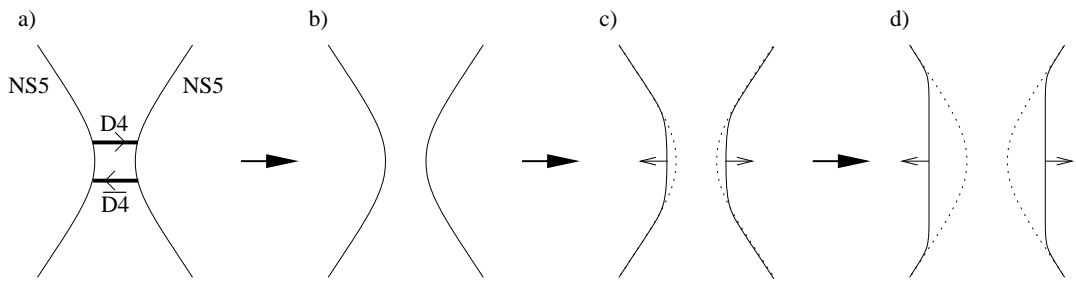,width=0.85\textwidth} \caption{Decay process
of nonsupersymmetric metastable brane/antibrane configuration. The
displayed bending of NS5's is for the $x^6$ direction; the curving in
the $wv$ direction is not shown.  a) Initial metastable
configuration. D4 and \anti{D4} are pulling NS5's inward.  b) D4 and
\anti{D4} have pair annihilated by quantum tunneling. c) Without
tension, NS5's straighten.  d) Straightening propagates to infinity.  }
\label{decayfigure}
\end{center}
\end{figure}

This is in contrast with what happens if there is no logarithmic
bending of NS5's, which is the case if $g_s$ is strictly zero:
$g_s=0$. In that case, there is no ``straightening'' part of the above
decay process, since the NS5's are straight from the beginning (in the
$x^6$ direction; in the $wv$ directions they are holomorphically
curved).  So, the decay process ends in a finite time when the D4's and
\anti{D4}'s have gone through quantum tunneling and pair annihilated.
The final configuration is a supersymmetric configuration of two NS5's
along the holomorphic curves,
\begin{equation}
 {w=\pm W'(v)}.
  \label{susycfg}
\end{equation}

However, for $g_s\neq 0$, which is the case we have been focusing on,
there is logarithmic bending.  Therefore, for $g_s\neq 0$, our
nonsupersymmetric D4/\anti{D4} system and the supersymmetric system
\eqref{susycfg} have different boundary conditions for NS5's at
infinity.  In general, to properly define the quantum theory of
non-compact systems, one must specify the boundary conditions at
infinity.  This means that, the metastable configuration of Fig.\
\ref{decayfigure}(a) and the supersymmetric configuration
\eqref{susycfg} are not different states in the same theory but rather
different states in different theories.  Therefore, the former can never
decay into the latter, which was the emphasis of \cite{Bena:2006rg}.

Note that this does \emph{not} mean that the configuration of Fig.\
\ref{decayfigure}(a) is stable.  As we have seen, it does decay and
shows a runaway instability.
This is not at all in contradiction with \cite{Bena:2006rg} but actually
\cite{Bena:2006rg} gives a nice interpretation of this runaway
instability as follows: although our nonsupersymmetric system wants to
decay into the supersymmetric configuration \eqref{susycfg}, it can only
keep decaying forever toward the latter, because the latter does not lie
in the same theory.  The fact that the energy released in the process is
infinite also implies that this process cannot end in a finite time.

In the above, we used the fact that the D4/\anti{D4} pull and bend the
NS5's to explain that the boundary conditions are different for
nonsupersymmetric and supersymmetric configurations.  An alternative way
to understand this is the following.  A D4-brane ending on an NS5 is a
vortex charge for the 1-form field strength $F_m$ on the NS5 worldvolume
theory. Because D4 and \anti{D4} have opposite vortex charges, the total
charge on each NS5 is $N-N=0$.  This is the same as for the
supersymmetric configuration \eqref{susycfg}, which has no D4/\anti{D4}
and hence no vortex charge.  However, D4-branes ending on an NS5-brane
act also as charges for the $x^6$ scalar field, with respect to which D4
and \anti{D4} have the same charge.  Therefore, the total charge is
$N+N=2N$ and is different from that of the supersymmetric configuration,
which is zero.  So, the two configurations have different charges and
one cannot decay into the other.

The boundary condition being different for nonsupersymmetric and
supersymmetric configurations can be understood in IIB also.
If we wrap D5's on 2-cycles in CY and turn on $B$-field through them, D3
charges are also induced.  A careful study of the system of
$H_3,F_3,F_5$ field strengths shows that in the presence of these
sources, the field $B_2^{NS}$ has to grow logarithmically with $v$, once
we include the leading-order backreaction of the D5's on the
geometry\footnote{The easiest way to see the logarithmic running of
$B_2$ is from the picture after the geometric transition. The running of
$B_2$ is supposed to capture the RG flow running of the gauge
coupling.}. So, while at tree level we can specify the $B_2$ flux
through the 2-cycles to be equal to a constant (which is the $T$-dual
statement of specifying the $x^6$ position of the NS5-brane to be
constant), once we add D5's and consider their backreaction, we see that
$B_2$ grows logarithmically at infinity, necessarily modifying the
boundary conditions of the system.
Now the difference of boundary conditions is easy to see: if we start
with $N$ D5's and $N$ \anti{D5}'s, the logarithmic divergence of $B_2$
at infinity will be proportional to $2N$.  On the other hand in the
supersymmetric configuration, $B_2$ is constant and there is no
logarithmic running. So the boundary conditions at infinity are
different. Also the energy difference between the two configurations
diverges logarithmically if we integrate the energy of the $H$ field
over the entire space.

Finally, as we saw in section 4, the boundary condition difference in M theory 
is more serious. Besides the different logarithmic bending of the $x^6$ coordinate, 
we also have to change the asymptotic behavior on the $w,v$ plane to solve
the minimal area equations.

\subsection{Future directions}
\label{futuredirections}

In the present paper, we studied nonsupersymmetric metastable
configurations and found that they can be analyzed in a reliable way
using the NS5/M5 worldvolume actions in various regimes of type
IIA/M-theory.

We focused on the case of two D4/\anti{D4} stacks between two
quadratically bent NS5-branes in type IIA, or, in the $T$-dual IIB
description, local CY geometry obtained by $A_1$ fibration.  However,
the method of nonholomorphic M5/NS5 curves is applicable to more
complicated configurations.
For example, flavors can be incorporated by including semi-infinite
D4/\anti{D4}-branes attached to NS5's or by including D6-branes
\cite{Witten:1997sc}.  In particular, such systems should give us an
$M$-theory viewpoint to look at the nonsupersymmetric metastable states in
gauge theory which were found in \cite{Intriligator:2006dd} and whose
possible string theory embeddings were pursued in \cite{Ooguri:2006bg,
Bena:2006rg}.
It is also straightforward to generalize our method to $A$-$D$-$E$
quiver gauge theories obtained by stretching D4/\anti{D4}-branes between
multiple NS5's, which is $T$-dual \cite{Dasgupta:2001um, Oh:2001bf} to
the local CY geometries obtained by $A$-$D$-$E$ fibrations
\cite{Cachazo:2001gh, Cachazo:2001sg}.
What is particularly interesting about such more general systems is
that, nonsupersymmetric brane/antibrane configurations can be stable
rather than metastable \cite{Mukhi:2000dn}.  In such cases it may be
possible to take the gauge theory limit and one may be able to study the
``protection'' of holomorphic quantities in gauge theory techniques.

Obtaining exact M5/NS5 curves in such more complicated systems will be
increasingly difficult in practice, if not possible.  However, the fact
that for small $g_s N$ the curve decomposes into the holomorphic part
($wv$) and the harmonic part ($s$) must still hold, and we furthermore
expect that the curve is protected in the same parameter regime because
we conjecture that supersymmetry softly broken there.  This may provide
a tractable window in which one can study the vast nonsupersymmetric
landscape of string/M-theory in a controlled way.
Needless to say, for such a program, it is highly desirable to better
understand the relation between the parameter $g_s N/\Delta$ and the
softness of the supersymmetry breaking, for which we could give only an
indirect argument in subsection \ref{ss:meaning_gsN}.

Nonsupersymmetric configurations such as the ones studied in the current
paper may potentially serve as useful modules or building blocks for
constructing realistic phenomenological and/or cosmological models
\cite{Kawano:2007ru}.  Of course, for that, it is crucial to study if
one can realize such configurations in \emph{compact} CY's as metastable
configurations. For example, in the IIA brane configurations or its dual
IIB local CY geometries, the superpotential $W(v)$ is given ``by hand''
through the boundary condition at the cutoff $v=v_0$.  If we were to
embed these models in compact CY's, those boundary conditions must arise
dynamically from some mechanism in the rest of the CY beyond the cutoff
$v=v_0$.

\section*{Acknowledgments}

We would like to thank D.~Berenstein, J.~de~Boer, J.~Heckman, N.~Iizuka,
T.~Okuda, H.~Ooguri, Y.~Ookouchi, D.~Robbins, J.~Seo, E.~Sokatchev,
C.~Vafa, M.~Van~Raamsdonk and especially A.~Sen for discussions.
J.M. is grateful to the theory groups at the University of Texas, UCSB,
and KITP for their hospitality during the course of this work.  J.M. and
M.S. would also like to thank the theory group at Harvard University for
hospitality while this work was being completed.  
The work of J.M. and M.S. was supported in part by Department of Energy
grant DE-FG03-92ER40701.  J.M. is also supported by a John A McCone
postdoctoral fellowship and M.S. by a Sherman Fairchild Foundation
postdoctoral fellowship.  The work of K.P. is supported by Foundation of
Fundamental Research on Matter (FOM).

\appendix

\section%
[Validity of M5/NS5 descriptions]%
{Validity of M5/NS5 descriptions%
\protect\footnote{We are grateful to Ashoke Sen for a helpful discussion
on this subject.}}  \label{app:validity}

\subsection{Validity of M5/NS5 worldvolume actions}

In $M$ theory in 11D spacetime, which is not compactified or is
compactified on a circle of radius much larger than the 11D Planck
length $\ell_{11}$, the worldvolume dynamics of M5-brane is described by
a Nambu-Goto action $S_{M5}$ at energy scales much lower than
$\ell_{11}^{-1}$.  This description becomes inappropriate if the M5
worldvolume starts to have structures smaller than $\ell_{11}$.  For
example, worldvolume with extrinsic curvature $\CR\sim
\ell_{11}^{-2}$ is bad, and two M5's within distance $\sim\ell_{11}$ is bad.

If we compactify $M$ theory on a small circle of radius $R$, we know
that there is a weakly coupled dual description: type IIA string theory
where F1 string is the lightest object and perturbation theory in
$\gs\ll 1$ is valid \cite{Witten:1995ex}.  The relations between M-theory
and IIA quantities are
\begin{align}
 R&=g_s \ls,\qquad \ell_{11}=g_s^{1/3}\ls.
\end{align}
In type IIA, we have NS5-brane, whose worldvolume action $S_{NS5}$ we
know to be of Nambu-Goto type as given in \eqref{fullns}
\cite{Bandos:2000az}.  This description becomes inappropriate if the NS5
worldvolume starts to have structures smaller than the string length
$\ls$ (note that the 10D Planck length $\ell_{10}\sim\gs^{1/4}\ls$ is
smaller than $\ls$).  Worldvolume fluxes with sufficiently large density
is bad too, which we will discuss later.
If we start from $M$ theory with $\gs\gg 1$ and go to IIA with $\gs\ll
1$, there is transition of these pictures at $\gs\sim 1$ which is highly
nontrivial.  Although we do know that the worldvolume action for M5 is
Nambu-Goto for $\gs\gg 1$, we do not know \textit{a priori} how these
actions get corrected when we go through this ``transition'' at $\gs\sim
1$.  Indeed, when we compactify $M$ theory on a circle of radius $R=\gs
\ls$ and lower $\gs$, the distance between images of M5-branes, which is
$\sim \gs \ls$, will be of the same order as $\ell_{11}=\gs^{1/3}\ls$
when $\gs\sim 1$, and we have no control over dynamics there.
What saves the day is supersymmetry, which fixes the action up to
two-derivative order.  Thanks to this, we can safely say that the action
must be Nambu-Goto all the way through the transition at the lowest
order.
It was logically possible that light objects, which make the M5 action
unreliable at $\gs\gg 1$ at scales $\lesssim \ell_{11}$, become much
lighter than the string mass $M_s=\ls^{-1}$ after the transition and
invalidate the use of NS5 Nambu-Goto action for $\gs\ll 1$.  But we know
\textit{a posteriori} that this actually does not happen, since in IIA
the lightest objects are strings, and as long as we are below the energy
scale $\ls^{-1}$ the Nambu-Goto action for NS5 does not get corrected.
Because of strong coupling, it is in general impossible to follow how
the mass of those light objects changes as one changes $\gs$ in order to
check that the potentially dangerous light objects in $M$ theory indeed
become heavy and safe in IIA after transition.
However, for BPS objects this is possible because their mass does not
get renormalized.  For consistency, the mass of such BPS objects must
continue to be large even after the transition.  One possible object is
the M2-brane stretching between image M5-branes (Fig.\ \ref{M2btM5s}).
To determine the BPS mass, we can safely use the M5/M2 picture at
$\gs\gg 1$, since then the distance between M5's is larger than
$\ell_{11}$.  From a 10D point of view, this looks like a string and its
tension is $T=RT_{M2}\sim R\ell_{11}^{-3}$.  If we decrease $\gs$, we
cannot use the M5/M2 picture any more, but this object continues to
exist and its tension is still given by $T$.  When we get to IIA, where
$\gs\ll 1$, the tension $T$ is $T=(\gs \ls)(1/\gs \ls^3)=1/\ls^2$.  So,
this object has tension of order IIA string tension, and is not dangerous
as far as low energy ($\ll \ls^{-1}$) physics is concerned.
\begin{figure}
\begin{center}
\subfigure[M2 stretching between image M5's]
{\epsfig{file=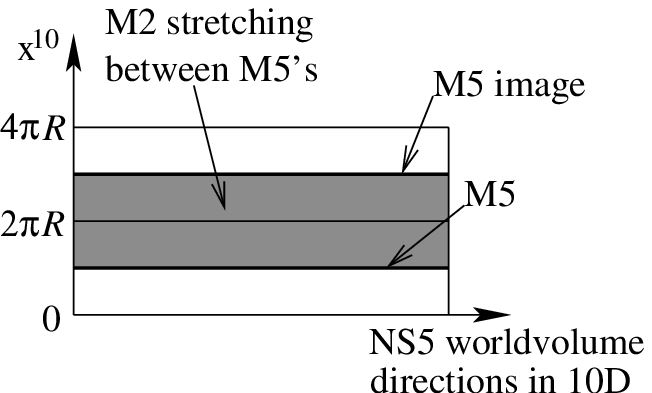,width=0.35\textwidth}
\label{M2btM5s}}
\hspace{0.1\textwidth}
\subfigure[$M$-theory lift of NS5 with flux.]
{\epsfig{file=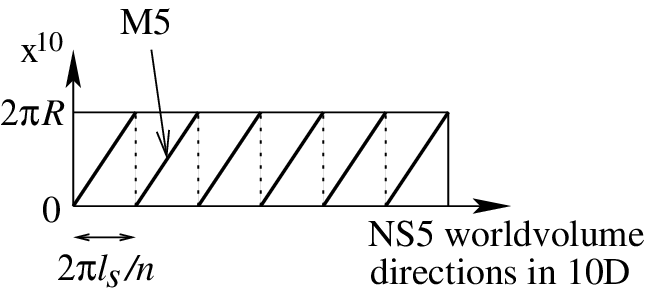,width=0.35\textwidth}\label{tiltedM5}}
\caption{Some features of $M$-theory lifts of NS5 branes with and without flux}
\end{center}
\end{figure}
Now consider the validity of the NS5 worldvolume action \eqref{fullns} in
the presence of flux.  Let there be $n$ fluxes per length $2\pi \ls$
so that the flux density is $n/2\pi \ls$.  The $M$ theory lift is as shown
in Fig.\ \ref{tiltedM5}.  Although tilted, the situation is the same as
in Fig.\ \ref{M2btM5s} in the covering space, except that the distance
between two M5's is now
\begin{align}
 \Delta&={2\pi \gs \ls\over\sqrt{1+(n\gs)^2}}
 \approx
 \begin{cases}
  2\pi \gs \ls & n\gs\ll 1,\\[1ex]
  {2\pi \ls\over n} & n\gs\gg 1.
 \end{cases}
\end{align}
The tension of the BPS object coming from M2 stretching between M5's is
\begin{align}
 T_{M2}\Delta
 \sim
 {T_{F1}\over \sqrt{1+(n\gs)^2}},
\end{align}
where $T_{F1}=1/2\pi\ls^2$ is the string tension.  Because now this is
the lightest object in IIA, its tension must be at least of order of
string tension.  This is the case if $n\gs\ll 1$.  Therefore, the
Nambu-Goto action \eqref{fullns} for NS5 is valid if
\begin{align}
 n\gs\ll 1.\label{smldst}
\end{align}
Note that $\gs\ll 1$ is implied since we are in IIA\@.

\subsection{Validity of M5/NS5 curves}

Let us consider when the curves obtained by using the Nambu-Goto action
to study M5/NS5 setups are reliable.  Consider the simplest case of
$A_1$ with quadratic superpotential studied in \ref{subsec:A1quadIIA}.
Even in more general cases, near each throat the curve can be
approximated by this.
The M5/NS5 curve is, from Eq.\ \eqref{wvcurve},
\begin{align}
 s&=N\log\lambda,\qquad
 v=\Lambda\left(\lambda+\frac{1}{\lambda}\right),\qquad
 w={m}\lambda\left(\lambda-\frac{1}{\lambda}\right).\label{lpwj29Mar07}
\end{align}
Here we redefined $\lambda\to \Lambda\lambda$ in \eqref{wvcurve} and also
shifted $s$ by a constant.  $\Lambda$ is related to $a$ in \eqref{wvcurve} by
$\Lambda^2=a$.  Note that we are setting $\ap=1$.
For the simplicity of the argument, let us set henceforth
\begin{align}
 m&=1.
\end{align}

\subsubsection*{Validity as NS5 curve}

First, consider the validity of \eqref{lpwj29Mar07} as an NS5 curve.
The thinnest part of the throat, where we expect that the flux is
densest and the curvature is largest, is for $|\lambda|=1$.  If we set
$\lambda=e^{i\theta}$,
\begin{align}
 s&=iN\theta,\qquad
 v=2\Lambda\cos\theta,\qquad
 w=-2i \Lambda \sin\theta.
\end{align}
Therefore, as $\theta$ changes from $0$ to $2\pi$, we go once around the
circle of radius $\sim\Lambda$ on the $\Re v$-$\Im w$ plane and $N$ times
along the M-theory circle (Fig.\ \ref{M5throat}).  So, the flux density
is $n\sim N/\Lambda$.  
\begin{figure}
\begin{center}
   \epsfig{file=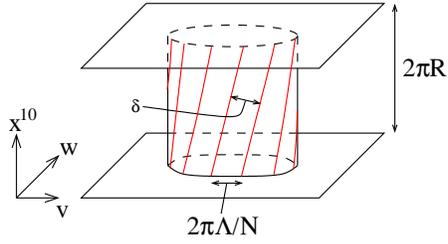,width=0.35\textwidth} \caption{The M5
curve in the throat region, for fixed $x^6$.  As we go once around the
circle in the $v$-$w$ plane, we go $N$ times along the M-theory circle.
$\delta$ is the distance between different ``strands'' of M5.
}
\label{M5throat}
\end{center}
\end{figure}
From \eqref{smldst}, we
obtain the condition for the NS5 worldvolume action to be valid:
\begin{align}
 x\ll 1,\qquad
 x\equiv{g_s N\over \Lambda}.
 \label{ns5cond1}
\end{align}

Another condition comes from requiring that the extrinsic curvature
$\CR$ of the NS5 curve be small in string units.  One can readily
compute the extrinsic curvature of the curve \eqref{lpwj29Mar07}, which
is equal to the curvature of the induced metric on the
$\lambda$-plane:\footnote{This is the extrinsic curvature for $s$, $v$,
and $w$.  Strictly speaking, one must compute the curvature for $x^6$,
$v$ and $w$, but this does not make any qualitative difference.}
\begin{align}
 \CR&=
 {2 |\lambda|^4 \left[x^2(|\lambda|^4+1)+8|\lambda|^2\right]
 \over
 \Lambda^2\left[x^2|\lambda|^2+2(|\lambda|^4+1)\right]^3}.\label{ndtk24Apr07}
\end{align}
It is easy to see that the maximum of this function is as follows:
\begin{equation}
\begin{split}
  x\le 2\sqrt{3}&:~~
 \CR={4\over \Lambda^2(x^2+4)^2}~~\text{at}~~
 |\lambda|^2=1,
 \\
 x\ge 2\sqrt{3}&:~~
 \CR={4x^6\over 27\Lambda^2(x^4-16)^2}~~\text{at}~~
 |\lambda|^2={x^4-48\pm\sqrt{(x^4-48)^2-64x^4} \over 8x^2}.
 \label{Rextr_extrm}
\end{split}
\end{equation}
In the present case \eqref{ns5cond1}, the maximum curvature is
\begin{align}
 \CR&\sim \Lambda^{-2}.
\end{align}
By requiring this to be much smaller than $\ls^{-2}=1$, we obtain another
condition:
\begin{align}
 1\ll\Lambda.
\end{align}

In summary, for the curve \eqref{lpwj29Mar07} to be trustable as an
NS5-brane curve in IIA, the following conditions must be met:
\begin{align}
 g_s\ll 1;\qquad
 1,g_s N\ll \Lambda.\label{nhuj24Apr07}
\end{align}
The first condition is necessary for the IIA string theory is valid in
the first place.    We can write this also as
\begin{align}
 g_s\ll 1,\qquad
 \Lambda\gg1,\qquad x\ll 1.
 \label{validityNS5}
\end{align}

If we restore $m$, one can show that \eqref{nhuj24Apr07} is replaced by
\begin{align}
 g_s\ll 1,\qquad
 g_s N\ll {\Lambda}\min(m,1),\qquad
 1\ll {\Lambda\over m}\min(m^3,1).
\end{align}

\subsubsection*{Validity as M5 curve}

Next, consider the validity of the curve \eqref{lpwj29Mar07} as an M5
curve.  First of all, we need
\begin{align}
 g_s\ll 1
\end{align}
for the M-theory description is valid.

We need that distance $\delta$ between two different ``strands''
of the M5 curve must be smaller than the Plank length
$\ell_{11}=g_s^{1/3}$.  The distance $\delta$ in the throat part of the
curve is given by (see Fig.\ \ref{M5throat})
\begin{align}
 \delta={\Lambda g_s\over\sqrt{N^2g_s^2+\Lambda^2}}
 ={g_s\over\sqrt{x^2+1}}
 \gg g_s^{1/3}
\end{align}
Therefore, we need
\begin{align}
 g_s^{2/3}&\gg \sqrt{x^2+1}.
\end{align}

We also need that the curvature \eqref{ndtk24Apr07} be much smaller than
$\ell_{11}^{-2}=g_s^{-2/3}$ everywhere.  From \eqref{Rextr_extrm}, we
obtain
\begin{equation}
\begin{split}
  x\le 2\sqrt{3}&:~~
 {1\over x^2+4}\sim 1\ll g_s^{-1/3}\Lambda,
 \\
 x\ge 2\sqrt{3}&:~~
 {x^3\over (x^4-16)}\sim {1\over x}\ll g_s^{-1/3}\Lambda
\end{split}
\end{equation}

In summary, for the curve \eqref{lpwj29Mar07} to be trustable as an
M5-brane curve, the following conditions must be met:
\begin{equation}
\begin{split}
  x\lesssim 1&:~~
 g_s^{2/3}\gg 1,\qquad g_s^{1/3}\ll \Lambda
 \\
 x\gtrsim 1&:~~
 g_s^{2/3}\gg x,\qquad g_s^{1/3}\ll \Lambda x=g_sN.
\end{split}\label{validityM5}
\end{equation}
Here we dropped factors such as $2\sqrt{3}$ which are inessential as far
as the order estimates are concerned.  Since $g_s\gg 1$, $N\ge 1$, some
conditions in the above are automatically satisfied.  Therefore, a more
economical way to state the condition is
\begin{equation}
\begin{split}
  x\lesssim 1&:~~ g_s^{1/3}\ll \Lambda,
 \\
 x\gtrsim 1&:~~   g_s^{2/3}\gg x.
\end{split}
\end{equation}

\begin{figure}[htb]
\begin{center}
   \epsfig{file=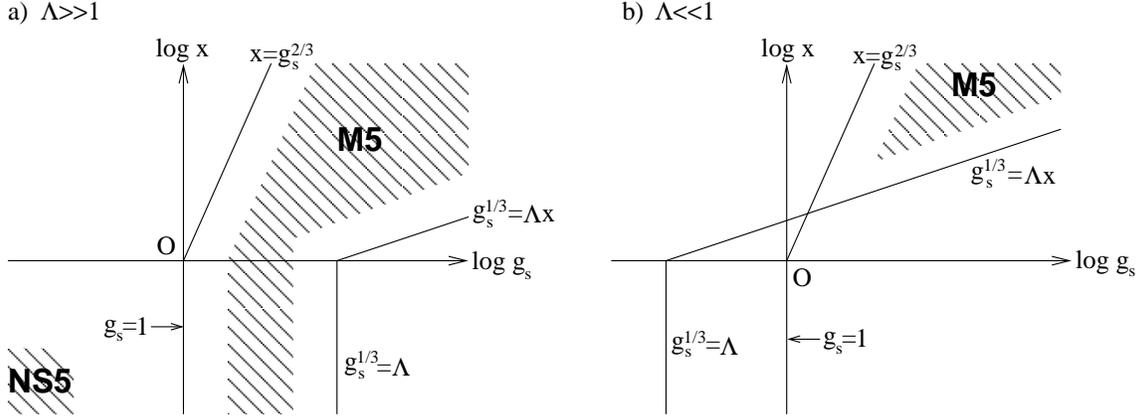,width=.9\textwidth}
\caption{Parameter regimes in which the curve \eqref{lpwj29Mar07} is
trustable as an NS5/M5 curve, for (a) $\Lambda\gg 1$ and (b) $\Lambda\ll
1$. The M5 description is valid in the shadowed regions with ``M5'',
while the NS5 description is valid in the shadowed region with ``NS5''.
The shadowed regions mean that the NS5/M5 descriptions are valid far
away from the surrounding lines, not inside the surrounding lines; see
\eqref{validityNS5} and \eqref{validityM5}.} \label{validregion}
\end{center}
\end{figure}

\subsubsection*{Summary}

Combining the result \eqref{validityNS5} for IIA and the one
\eqref{validityM5} for M-theory, we obtain the parameter regimes of type
IIA/M-theory in which the curve \eqref{lpwj29Mar07} is trustable as an
NS5/M5 curve, as shown in Fig.\ \ref{validregion}.

As discussed in the main text, nonsupersymmetric curves simplify in the
$x\ll 1$ limit and agree with the geometry obtained in type IIB in
\cite{Aganagic:2006ex, Heckman:2007wk}.  Not only IIA but also M-theory
has a parameter regime in which this condition, $x\ll 1$, is satisfied,
although one must take $\Lambda$ to be large too.  Note also that
$\Lambda\ll 1$ is also allowed in M-theory.

\section{T-duality between IIB 2-forms and IIA NS5 position}
\label{app:T-dual_of_B}

In subsection \ref{subsec:T-duality}, we discussed the $T$-duality
between Taub-NUT space in IIB and the NS5-brane configuration in
IIA\@. At the level of the Buscher rule, the NS5-branes are delocalized
(smeared) in the $y$ direction.  However, in string theory, the
NS5-branes are expected to become localized.  Indeed, it is known that
the $y$ position of the NS5-brane is dual to $B$-field through certain
2-cycles (which is sometimes called the ``dyonic coordinate'' in the
literature) in the Taub-NUT geometry \cite{Sen:1997zb, Sen:1997js}.
Although one could study this localization of NS5-branes using
worldsheet CFT techniques \cite{Tong:2002rq, Harvey:2005ab}, here we
present an alternative approach to determine the position of NS5-branes,
which, to our knowledge, is new.  Specifically, we consider wrapping an
imaginary D5-brane around the 2-cycle and follow the $T$-duality.  The
D5-brane gets mapped into a D4-brane stretching between two NS5-branes.
From how much the D4-brane is ``tilted'' in the $y$ direction, one can
read off the $y$ position of NS5-branes on which the D4-brane ends.

In the Taub-NUT geometry \eqref{k-TN_metric} in IIB, there are
nontrivial 2-cycles $c_{pq}$ obtained by fibering the $\yt$ circle
$S^1_\yt$ over a segment connecting $\zv_p$ and $\zv_q$ in $K$.  The
harmonic self-dual 2-form $\Omega_{pq}$ dual to $c_{pq}$ can be locally
written as (see e.g.\ \cite{Cascales:2003ew, Sen:1997js}):
\begin{align}
 \Omega_{pq}
 =\half d(\chi_p-\chi_q),
 \qquad 
 \chi_p=H^{-1}H_p(d\yt+\omega)-\omega_p,\qquad
 d\omega_p=*_3 dH_p.\label{Omega_def}
\end{align}
$\Omega_{pq}$ is localized near the 2-cycle $c_{pq}$, and is normalized
as
\begin{align}
 \int_{c_{pq}}\!\!\!\Omega_{pq}&=2\pi R.\label{Omega_norm}
\end{align}

Let us consider turning on the following NSNS $B$-field:
\begin{align}
 B&= b\,\Omega_{pq},\label{B_Omega}
\end{align}
which is a ``Wilson line'' for $B$ through $c_{pq}$.  From
\eqref{Omega_norm}, one sees that $b$ has the following periodicity:
\begin{align}
 b&\cong b+{2\pi \ap\over R}.\label{b_period}
\end{align}
If we $T$-dualize the IIB metric \eqref{k-TN_metric} in the presence of
the $B$-field \eqref{B_Omega} using the Buscher rule, one still obtains
exactly the same IIA metric as in \eqref{NS5_metric} if we replace the
coordinate $y$ with $y'$ defined by
\begin{align}
 y'&\equiv y+{b\over 2}(\chi_p-\chi_q)_\yt,\label{def_yt'}
\end{align}
where the subscript $\yt$ denotes the $\yt$ component.

Now, consider wrapping an imaginary D5-brane around the 2-cycle $c_{pq}$
(the remaining $3+1$ worldvolume dimensions extend in the Minkowski
space $M_4$). On the D5 worldvolume, the $B$-field \eqref{B_Omega} is
equivalent to the worldvolume gauge field flux
\begin{align}
 2\pi\ap F&\leftrightarrow B=b\,\Omega_{pq}={b\over 2}{d({\chi_p-\chi_q})}.
\end{align}
The corresponding gauge potential 1-form is
\begin{align}
 2\pi\ap A &= {b\over 2}(\chi_p-\chi_q)-b\lambda,\label{A_chi_lambda}
\end{align}
Here $\lambda$ is a closed 1-form which is determined as follows.  Since
the $\yt$ circle $S^1_\yt$ shrinks smoothly to zero at
$\zv=\zv_p,\zv_q$, the Wilson line along the $\yt$-circle,
$\int_{S^1_\yt}A$, must vanish at these points; otherwise there would be
a delta-function like $F$-flux at these points on the D5, which is
unphysical.  On the other hand, from the explicit form of $\chi_p$ in
\eqref{Omega_def}, one sees that $\int_{S^1_\yt}(\chi_p-\chi_q)$ is
nonvanishing even at $\zv=\zv_p,\zv_q$.  The 1-form $\lambda$ must be
chosen to cancel this Wilson line:\footnote{A similar term cannot be
added to \eqref{def_yt'} because that would correspond to a gauge
transformation nonvanishing at infinity, which is a global symmetry
transformation rather than an immaterial coordinate change.}
\begin{align}
 \lambda
 &=d\!\left({H_p-H_q\over 2H}\yt\right)
 ={H_p-H_q\over 2H}d\yt+\yt\,\partial_i\!\left({H_p-H_q\over 2H}\right) dz^i.
\end{align}
If we $T$-dualize along $\yt$, we obtain a D4 stretching between two
NS5-branes at $\zv=\zv_p,\zv_q$, which lies along the curve
\begin{align}
 y(\zv)&=-2\pi \ap A_\yt
 = -{b\over 2}(\chi_p-\chi_q)_\yt+b\lambda_\yt.
\end{align}
In the $y'$ coordinate \eqref{def_yt'}, this curve is
\begin{align}
 y'(\zv)
 &= b\lambda_\yt
 =b{H_p-H_q\over 2H}
 =\begin{cases}
   \phantom{-}b/2 & \zv=\zv_p, \\
   -b/2 & \zv=\zv_q.
  \end{cases}
\end{align}
Therefore, the relative $y'$ position of NS5's, which are the
endpoints of the D4, is given by $b$. The periodicity \eqref{b_period}
is consistent with the radius of the $T$-dual circle, \eqref{y_period}.

To summarize, the $k$-center Taub-NUT geometry \eqref{k-TN_metric} with
$B$-field \eqref{B_Omega} through the 2-cycles $c_{pq}$ in IIB is
$T$-dual to the IIA configuration with $k$ NS5-branes, where the $p$-th
and $q$-th NS5-branes are separated in the $y$ direction by $b$.
Similarly, RR 2-form through the 2-cycle $c_{pq}$,
\begin{align}
 C&=c\,\Omega_{pq},\qquad
 c\cong c+{2\pi\ap\over R},
\end{align}
corresponds to the distance between the NS5-branes in the $x^{10}$
direction, when lifted to M-theory.\footnote{This can be seen by noting
that taking $S$-dual in IIB corresponds to switching the two $S^1$'s of
M-theory on $T^2$.  In the present case, $T^2$ is spanned by $x^{10}$
and $\yt'$.
In the type IIA picture, the $x^{10}$ coordinate appears as a periodic
scalar field living on the NS5 worldvolume.  }  In an equation, the
relation between the 2-forms in IIB and the relative position of
NS5-branes in IIA is
\begin{align}
 \int \left(C_2^{RR}+{i\over g_s^{\rm IIB}}B_2^{NS}\right)
 =-2\pi i\, \Delta s,\label{CB_Delta-s}
\end{align}
where $s=(x^6+ix^{10})/R$ and $R=g_s^{\rm IIA}\sqrt\ap$ is the radius of
the $M$ theory circle.

\section{Properties and Applications of some Elliptic Functions}
\label{app:ellipfcns}

In this appendix, we summarize some of the properties of elliptic functions used throughout the text.
\subsection{Building blocks}

We begin by recalling some basic definitions.  Our primary building blocks for constructing genus one curves are based on the function \eqref{Fdef}
\begin{equation}
F(z)=\ln\theta(z-\tilde{\tau})
\label{appFdef}
\end{equation}
where $\theta$ denotes the Jacobi elliptic function denoted elsewhere by
$\theta_3$ or $\theta_{00}$:
\begin{equation}
\theta(z)=\sum_{n=-\infty}^{\infty}e^{i\pi n^2\tau+2\pi inz},
 \qquad\qquad\tilde{\tau}=\frac{1}{2}(\tau+1).
\label{appthetadef}
\end{equation}
It is easy to see that $\theta(z)$ has a simple zero at $z=-\tilde{\tau}$ and hence that $F(z)\sim\ln z$ near $z=0$.  By taking derivatives, we can introduce functions with poles at the points $a_i$
\begin{equation}
F_i^{(n)}=\left(\frac{\partial}{\partial z}\right)^nF(z-a_i).
\label{Findefd}
\end{equation}
We also typically denote $F_i^{(0)}=F(z-a_i)$ simply by $F_i$.

When building our genus 1 curves, it is important to know the
periodicity properties of the $F_i$.  These are easy to work out from
the periodicities of $\theta(z)$
\begin{equation}
\theta(z+1)=\theta(z),\qquad\qquad\theta(z+\tau)=e^{-i\pi\tau-2\pi iz}\theta(z).
\label{thetapers}
\end{equation}
This leads to the nontrivial periods \eqref{Fmonods}
\begin{equation}
\begin{split}
F_i(z+1)&=F_i(z),\\
F_i(z+\tau)&=F_i(z)-2\pi i(z-a_i)+i\pi,\\
F_i^{(1)}(z+1)&=F_i^{(1)}(z),\\
F_i^{(1)}(z+\tau)&=F_i^{(1)}(z)-2\pi i
\label{appFmonods}
\end{split}
\end{equation}
as well as
\begin{equation}
F_i^{(n)}(z+\tau)=F_i^{(n)}(z+1)=F_i^{(n)}(z)\qquad\text{for }n>1.
\label{appFtmonods}
\end{equation}
The $F_i^{(n)}$ also have nice properties under $z\rightarrow -z$
\begin{equation}
\begin{split}
F(-z) &= F(z)-2\pi i z+i\pi,\\
F^{(1)}(-z)&=-F(z)+2\pi i,\\
F^{(n)}(-z) &= (-1)^nF(z)\qquad n>1.
\label{Fparity}
\end{split}
\end{equation}
It is now a relatively simple matter to expand any elliptic function, say $f(z)$, in terms of the $F_i^{(n)}$.  To do so, we need only identify a linear combination of the $F_i^{(n)}$ with the same pole structure as $f(z)$.  By holomorphy, these two quantities are equivalent up to a constant that is trivially computed.

This permits us to relate our collection of functions, $F_i^{(n)}$, to the more commonly used Weierstrass elliptic functions, whose properties we now review.

\subsection{Weierstrass elliptic functions}

The three primary Weierstrass elliptic functions are defined
by{\footnote{To compare with the usual literature on Weierstrass
functions, note that we have set $\omega_1=\frac{1}{2}$ and
$\omega_3=\frac{\tau}{2}$.  We also explicitly indicate
$\omega_1,\omega_3$-dependence instead of the usual $g_2,g_3$.}}
\begin{equation}
\begin{split}
\sigma(z)&=
 z\!\!\!\!\prod_{\substack{m,n=-\infty\\ (m,n)\ne(0,0)}}^{\infty}\!\!\!\!
 \left(1-\frac{z}{m+n\tau}\right)\exp\left(\frac{z^2}{2(m+n\tau)^2}+\frac{z}{m+n\tau}\right),\\
\zeta(z)&=\frac{1}{z}
 +\!\!\!\!\sum_{\substack{m,n=-\infty\\ (m,n)\ne(0,0)}}^{\infty}\!\!
 \left(\frac{z}{(m+n\tau)^2}+\frac{1}{m+n\tau}+\frac{1}{z-(m+n\tau)}\right),\\
\wp(z)&=\frac{1}{z^2}
  +\!\!\!\!\sum_{\substack{m,n=-\infty\\ (m,n)\ne(0,0)}}^{\infty}\!\!
 \left[\frac{1}{[z-(m+n\tau)]^2}-\frac{1}{(m+n\tau)^2}\right]
\label{weierstrassfcns}
\end{split}\end{equation}
and are related to one another by
\begin{equation}
\zeta(z)=\frac{\partial\ln\sigma(z)}{\partial z},\qquad \wp(z)=-\frac{\partial\zeta(z)}{\partial z}.
\end{equation}
An important fact is that $\zeta(z)$ is odd.  This implies that
$\wp(z)$, as well as its even derivatives, are even while odd
derivatives of $\wp(z)$ are odd.

The analytic structure of the Weierstrass functions on the fundamental
parallelogram is as follows. The function $\sigma(z)$ has a simple zero
at $z=0$
\begin{equation}
\sigma(z)\sim z+\ldots\quad\implies\ln\sigma(z)\sim \ln z+\ldots
\end{equation}
from which we see that
\begin{equation}
\zeta(z)\sim \frac{1}{z}+\ldots,\qquad\wp(z)\sim \frac{1}{z^2}.
\end{equation}
The Weierstrass $\wp$-function is elliptic, having completely trivial periods.  The functions $\sigma$ and $\zeta$, on the other hand, are quasiperiodic
\begin{equation}
\begin{split}
\sigma(z+1)&=-e^{2\eta_1\left(z+\frac{1}{2}\right)}\sigma(z),\\
\sigma(z+\tau)&=-e^{2\eta_3\left(z+\frac{\tau}{2}\right)}\sigma(z),\\
\zeta(z+1)&=\zeta(z)+2\eta_1,\\
\zeta(z+\tau)&=\zeta(z)+2\eta_3,
\end{split}\end{equation}
where $\eta_1$ and $\eta_3$ are the $\zeta$-function half-period values
\begin{equation}
\zeta\left(\frac{1}{2}\right)=\eta_1,\qquad\qquad\zeta\left(\frac{\tau}{2}\right)=\eta_3
\end{equation}
and which satisfy the nontrivial identity
\begin{equation}
\eta_1\tau-\eta_3=i\pi.
\label{etaid}
\end{equation}
By comparing analytic structures and periodicities, it is easy to relate the Weierstrass functions \eqref{weierstrassfcns} to our building blocks $F_i^{(n)}$ \eqref{Findefd}
\begin{equation}
\begin{split}
F(z)&=\ln\sigma(z)-\eta_1z^2+i\pi z +\ln\theta'(\tilde{\tau}),\\
F^{(1)}(z)&=\zeta(z)+i\pi-2\eta_1z,\\
F^{(2)}(z)&=-\wp(z)-2\eta_1,\\
F^{(n)}(z)&=-\left(\frac{\partial}{\partial z}\right)^{n-2}\wp(z)\qquad n>2.
\label{FWeierrels}
\end{split}
\end{equation}
From this we also see that the function $L(a,\tau)$ defined in \eqref{Ldef} is related to $\sigma(z)$ by
\begin{equation}
L(a,\tau)\equiv \ln\left(\frac{12\wp(a)\theta(a-\tilde{\tau})^2}{\theta'(\tilde{\tau})^2}\right)=\ln\left(12\wp(a)\sigma(a)^2\right)-2\eta_1a^2+2\pi i a.
\label{Lsigma}
\end{equation}
The relations \eqref{FWeierrels} and \eqref{Lsigma} prove useful because there are a number of nice identities involving Weierstrass functions that are well-known.  We will list some of these at the end of this appendix.

\subsection{Some simple manipulations}

While we can use analytic structure to relate Weierstrass functions to our $F_i^{(n)}$, it is also useful to work out polynomial relations among more complicated elliptic functions.  Consider, for instance, the elliptic function
\begin{equation}
V(z) = F_1^{(1)}-F_2^{(1)}.
\end{equation}
The function $V(z)^2$ is also elliptic and admits an expansion as a
linear combination of the $F_i^{(n)}$.  To determine which one, we use
the expansion
\begin{equation}
F^{(1)}(z)=\frac{1}{z}+i\pi-2\eta_1z -\frac{g_2z^3}{60}+{\cal{O}}(z^5)
\label{F11exp}
\end{equation}
to study $V(z)^2$ near $a_2$
\begin{equation}
V(z)^2\sim \frac{1}{(z-a_2)^2}-\frac{2\left[F^{(1)}(a)-i\pi\right]}{z-a_2}+\left(\left[F^{(1)}(a)-i\pi\right]^2-2\left[F^{(2)}(a)+2\eta_1\right]\right)+{\cal{O}}(z-a_2)
\label{V2exp}
\end{equation}
where, as in the main text
\begin{equation}
a\equiv a_2-a_1.
\label{appadef}
\end{equation}
From the antisymmetry of $V(z)$, we also know that the even order poles at $a_1$ will have the same coefficients as \eqref{V2exp} while the odd order poles will have opposite signs.  This permits us to write
\begin{equation}
\begin{split}V(z)^2&=\left(F_1^{(1)}-F_2^{(1)}\right)^2\\
&=\left(F_1^{(2)}+F_2^{(2)}\right)+2\left[F^{(1)}(a)-i\pi\right]\left(F_1^{(1)}-F_1^{(2)}\right)+\left(\left[F^{(1)}(a)-i\pi\right]^2-2\left[F^{(2)}(a)+2\eta_1\right]\right).
\label{Vsqexp}
\end{split}
\end{equation}

\subsection{Properties of the curve \eqref{vwparamrep}}

Manipulations such as this are easy to perform even in more complicated situations.  One example of interest is the polynomial relation between $w$ and $v$ of \eqref{vwparamrep}
\begin{equation}
\begin{split}
v&= X\left(F_1^{(1)}-F_2^{(1)}-\left[F^{(1)}(a)-i\pi\right]\right),\\
w&=C\left(F_1^{(2)}-F_2^{(2)}\right).
\label{appvwparamrep}
\end{split}\end{equation}
Because of the antisymmetry of both $w$ and $v$ under $a_1\leftrightarrow a_2$ it is impossible to write $w$ as a quadratic polynomial in $v$.  The simplest possibility then is that $w^2$ may be written as a quartic polynomial in $v$.  Comparing pole structures, one can quite easily deduce the correct relation
\begin{equation}
w^2=\frac{C^2}{X^4}\left[P_2(v)^2+b_1v+b_0\right]
\label{appwsq}
\end{equation}
where
\begin{equation}
\begin{split}
P_2(v)&=v^2+3X^2\left(F^{(2)}(a)+2\eta_1\right),\\
b_1&= -4X^3F^{(3)}(a),\\
b_0&=X^4\left[\frac{g_2}{6}+\frac{5}{3}F^{(4)}(a)-2\left(f^{(2)}(a)+2\eta_1\right)^2\right].
\end{split}\end{equation}
This gives an explicit relation between the deformation parameters $b_0,b_1$ and the quantities $a$ and $\tau$.  It also permits us to relate the variables of our parametric description to the ``physical'' parameters \eqref{physparams} associated with boundary conditions.  In particular, at large $v$ the relation \eqref{appwsq} becomes
\begin{equation}
w\sim \frac{C}{X^2}P_2(v)=\frac{C}{X^2}\left(v^2-3X^2\wp(a)\right)
\end{equation}
where we have used
\begin{equation}F^{(2)}(z)+2\eta_1=-\wp(z).\end{equation}
From this, we can read off the parameters $g$ and $\Delta$ \eqref{physparams}
\begin{equation}
g=\frac{C}{X^2}\qquad\qquad\Delta^2=12X^2\wp(a).
\label{appphysparams}
\end{equation}

Finally, we note that this formalism also makes it quite easy to compute
period integrals.  For instance, the expectation values of the $S_i$ in
type IIB \eqref{SPidef} are mapped to period integrals of the 1-form
\begin{equation}\half w\,dv.\end{equation}
For genus one curves, we have two $S_i$ which can be computed via
\begin{equation}
S_i=\frac{1}{2\pi i}\oint_{A_i}\half w\,dv
\end{equation}
where the $A$ cycles are as in figure \ref{wvparamsgone}.  We can compute these fairly easily by starting from
\begin{equation}
S_i=\frac{1}{4\pi i}\oint_{A_i}w(z)\frac{\partial v(z)}{\partial z}\,dz=\frac{XC}{4\pi i}\left(F_1^{(2)}-F_2^{(2)}\right)^2
\end{equation}
and writing the integrand as a linear combination of the $F_i^{(n)}$ 
\begin{equation}
\left(F_1^{(2)}-F_2^{(2)}\right)^2
 =a\left(F_1^{(4)}+F_2^{(4)}\right)
 +b\left(F_1^{(3)}-F_2^{(3)}\right)
 +c\left(F_1^{(2)}+F_2^{(2)}\right)
 +d\left(F_1^{(1)}-F_2^{(1)}+i\pi\right)+e.
\end{equation}
In particular, once the coefficients $d$ and $e$ are known we can immediately write
\begin{equation}
S_1 = -\frac{g\Delta^3}{4\pi i\left(12\wp(a)\right)^{3/2}}\left(i\pi d+e\right),\qquad S_2 = -\frac{g\Delta^3}{4\pi i\left(12\wp(a)\right)^{3/2}}\left(i\pi d - e\right).
\label{Sexact}
\end{equation}
Actually obtaining the coefficients $a,\dots,e$ in practice is not
difficult and we find
\begin{equation}
\begin{split}
a&=-\frac{1}{6},\qquad b=0,\qquad
c=-2\left(F^{(2)}(a)+2\eta_1\right),\qquad
d=-2F^{(3)}(a),\\
e&=\frac{g_2}{12}-4\eta_1^2+4\eta_1F^{(2)}(a)+3F^{(2)}(a)^2+2F^{(3)}(a)F^{(1)}(a)+\frac{7}{6}F^{(4)}(a).
\end{split}
\end{equation}
In the text, we studied the curve \eqref{appvwparamrep} in the limit of large $\Im(\tau)$.  We can expand $S_1$ and $S_2$ in this regime{\footnote{Note that for $N_1<N_2$, the expansion of $S_2$ is a bit tricky.  One must first demonstrate that all terms of the form $e^{2\pi i N_2\tau n/N}$ cancel so that $e^{2\pi i N_1\tau/N}$ is truly the leading contribution.}} using the first moduli-fixing relation \eqref{ataurel} 
\begin{equation}S_1=g\Delta^3e^{2\pi i N_2\tau/N}+\ldots,\qquad\qquad S_2=-g\Delta^3e^{2\pi i N_1\tau/N}.\end{equation}
To write this in terms of $\Lambda_{UV}$, we must apply the second moduli-fixing relation \eqref{alphataurel}.  At large $\Im(\tau)$, it takes the form
\begin{equation}\Lambda_{UV}^{2N}=v_0^{2N}e^{-2\pi i\alpha(v_0)}=(-1)^N(2\pi X)^{2N}e^{2\pi i N_1N_2\tau/N}.\end{equation}
Noting also that
\begin{equation}\Delta^2\sim -4\pi^2X^2+\ldots\end{equation}
at large $\Im(\tau)$ we obtain \eqref{ldtau}
\begin{equation}
\left(\frac{\Lambda_{UV}}{\Delta}\right)^{2N}=e^{2\pi i N_1N_2\tau/N}+\ldots
\end{equation}
which explicitly demonstrates that large $\Im(\tau)$ corresponds to small $\Lambda_{UV}/\Delta$.  Finally, using all of these results we can express $S_1$ and $S_2$ completely in terms of $g$, $\Delta$, and $\Lambda_{UV}$
\begin{equation}S_1=g\Delta^3\left(\frac{\Lambda_{UV}}{\Delta}\right)^{2N/N_1},\qquad S_2=g\Delta^3\left(\frac{\Lambda_{UV}}{\Delta}\right)^{2N/N_2},\end{equation}
reproducing \eqref{iiati}.

A specific case of interest in the main text corresponds to $N_1=N_2$
and $\tau=2a$.  There, the exact expressions \eqref{Sexact} for the
$S_i$ simplify dramatically
\begin{equation}
S_1=-S_2=-\frac{g\Delta^3}{4\pi i\left(12\wp(\tau/2)\right)^{3/2}}\left(\frac{2g_2}{3}-4\wp(\tau/2)\left[\wp(\tau/2)-2\eta_1\right]\right).
\label{Sexactt2a}
\end{equation}
The first few terms of \eqref{Sexactt2a} at large $\Im(\tau)$ are easily determined
\begin{equation}
S_1=-S_2=g\Delta^3\left(e^{i\pi\tau}-34e^{2\pi i\tau}+984e^{3\pi i\tau}+\ldots\right).
\label{s1s2lt}\end{equation}

\subsection{$q$-series and other useful formulae}

We now list useful formulae, including $q$-series, for some of the quantities that arise throughout our computations.  In what follows, we define
\begin{equation}
q=e^{i\pi\tau}
\end{equation}
as usual.

We start with the differential equation satisfied by $\wp(z)$
\begin{equation}
\left(\frac{\partial\wp(z)}{\partial z}\right)^2=4\wp(z)^3-g_2\wp(z)-g_3
\end{equation}
which also implicitly defines the Weierstrass elliptic invariants $g_2$ and $g_3$.

We now list some $q$-series
\begin{equation}
\begin{split}
\eta_1&=\frac{\pi^2}{6}-4\pi^2\sum_{k=1}^{\infty}\frac{kq^{2k}}{1-q^{2k}}\\
g_2&=20\pi^4\left(\frac{1}{15}+16\sum_{k=1}^{\infty}\frac{k^3q^{2k}}{1-q^{2k}}\right)\\
\wp(\tau/2)&=-\frac{\pi^2}{3}-8\pi^2\sum_{k=1}^{\infty}\frac{kq^{k}}{1+q^{k}}\\
\theta(1/2)&=1+2\sum_{n=1}^{\infty}(-1)^nq^{n^2}\\
\theta'(\tilde{\tau})&=2\pi i\left(1+\sum_{n=1}^{\infty}(2n+1)(-1)^nq^{n(n+1)}\right)
\end{split}
\end{equation}

\begin{equation}
\begin{split}
\ln\sigma(z)&=\ln\left(\frac{\sin(\pi z)}{\pi}\right)+\eta_1z^2+4\sum_{k=1}^{\infty}\frac{q^{2k}}{k(1-q^{2k})}\sin^2(k\pi z)\\
\zeta(z)&=2\eta_1 z+\pi\cot(\pi z)+4\pi\sum_{k=1}^{\infty}\frac{q^{2k}}{1-q^{2k}}\sin(2\pi kz)\\
\wp(z)&=-2\eta_1+\pi^2\csc^2(\pi z)-8\pi^2\sum_{k=1}^{\infty}\frac{kq^{2k}}{1-q^{2k}}\cos(2\pi kz)
\end{split}\end{equation}

We also note that at large $\Im(\tau)$, $L(\tau/2,\tau)$ has the nice expansion
\begin{equation}L(\tau/2,\tau)=20q-262q^2+\ldots\end{equation}

It is also useful to know the $\tau$-derivatives of various quantities
\begin{equation}
\begin{split}
\frac{\partial\ln\sigma(z)}{\partial\tau}&=-\frac{1}{2\pi i}\left[\frac{1}{2}\wp(z)-\frac{1}{2}\zeta(z)^2-\frac{g_2 z^2}{24}+2\eta_1\left(z\zeta(z)-1\right)\right]\\
\frac{\partial\zeta(z)}{\partial\tau}&=-\frac{1}{2\pi i}\left[\frac{1}{2}\wp'(z)+\zeta(z)\wp(z)-\frac{g_2 z}{12}+2\eta_1\left(\zeta(z)-z\wp(z)\right)\right]\\
\frac{\partial\wp(z)}{\partial\tau}&=\frac{1}{2\pi i}\left[2\wp(z)^2+\zeta(z)\wp'(z)-\frac{g_2}{3}-2\eta_1\left(z\wp'(z)+2\wp(z)\right)\right]
\end{split}
\end{equation}

\begin{equation}
\begin{split}
\frac{\partial\eta_1}{\partial\tau}&=-\frac{1}{2\pi i}\left(2\eta_1^2-\frac{g_2}{24}\right)\\
\frac{\partial g_2}{\partial\tau}&=\frac{1}{2\pi i}\left(6g_3-8g_2\eta_1\right)\\
\end{split}\end{equation}

Finally, we list the asymptotic behavior of a few quantities for $\Im(\tau)\ll 1$
\begin{equation}
\begin{split}
\wp(\tau/2)&=\frac{2\pi^2}{3\tau^2}+\ldots\\
g_2&=\frac{4\pi^4}{3\tau^4}+\ldots\\
\eta_1&=\frac{\pi^2}{6\tau^2}+\frac{i\pi}{\tau}+\ldots
\end{split}
\end{equation}

\section{An Exact Solution for $N_{\text{branes}}=N_{\text{antibranes}}$}
\label{app:exactsol}

In this appendix, we describe the derivation of the solution
\eqref{exactsol} as well as some of its properties.

\subsection{Deriving \eqref{exactsol}}

Our goal is to find an exact solution to the minimal area conditions
\eqref{harmonic} and \eqref{virasoro}
\begin{align}
0&=\partial\bar{\partial}v(z)=\partial\bar{\partial}w(z)=\partial\bar{\partial}s(z), \label{appharmonic}\\
0&=g_s^2
 \partial s \,\partial\bar{s} + \partial v\,\partial\bar{v}+\partial w\, \partial\bar{w}\label{appvirasoro}
\end{align}
for $s$ having the appropriate periods
\begin{equation}
\frac{1}{2\pi i}\oint_{A_i}ds = N^i,\qquad\oint_{B_i}ds=-\alpha_i.
\label{appsperiods}
\end{equation}
Recall that $N_i$ denotes the number of D4's or \anti{D4}'s while $N^i$ denotes the $RR$ charges.  Here, we focus on the case of $N_1$ D4's and $N_2=N_1$ \anti{D4}'s so $N_1=N^1=-N^2$.  Recall also that $\alpha_1=\alpha_2=\alpha$.
Solutions to \eqref{appharmonic} are given by $s,v,w$ that are sums of holomorphic and antiholomorphic functions.  We will use this observation to write them as
\begin{equation}
\begin{split}
s(z,\bar{z})&=s_H(z)+\overline{s_A(z)},\\
v(z,\bar{z})&=v_H(z)+\overline{v_A(z)},\\
w(z,\bar{z})&=w_H(z)+\overline{w_A(z)}.\\
\end{split}\end{equation}
In this notation, the second constraint \eqref{appvirasoro} can be written as
\begin{equation}
0=g_s^2\partial s_H\partial s_A+\partial v_H\partial v_A+\partial w_H\partial w_A
\label{appvirasoroHA}
\end{equation}
making manifest that the introduction of an antiholomorphic piece to $s$ necessitates the presence of nonholomorphic terms in $v$ and $w$.

As discussed in the main text, the most general harmonic $s$ with $A$-periods as specified in \eqref{appsperiods} is given by \eqref{sfam} with $N_1=N_2\equiv N$.  We write this here as
\begin{equation}
\begin{split}
s_H(z)&=\gamma(F_1-F_2)+i\pi (N+\delta)z,\\
s_A(z)&=\bar{\gamma}(F_1-F_2)-i\pi (N-\bar{\delta})z.
\label{sgamdel}
\end{split}
\end{equation}
Imposing the compact $B$-period constraint
\begin{equation}
\oint_{B_2-B_1}ds=0
\end{equation}
then leads to
\begin{equation}
2\gamma\Im(a)=\delta\Im(\tau)+iN\Re(\tau).
\label{exbpercons}
\end{equation}
This can be further simplified by noting that switching the sign of $x^{10}$ has the same effect as reversing the fluxes.  In equation form, this means that the curve should be invariant under the operation
\begin{equation}
s\rightarrow\bar{s},\qquad\qquad N\rightarrow -N
\label{ssbarflip}
\end{equation}
and implies that the quantities $\gamma$ and $\delta$ are real.  Combining this with \eqref{exbpercons} we see that
\begin{equation}
\Re(\tau)=0,\qquad\qquad\Im(\tau)=\frac{2\gamma}{\delta}\Im(a).
\label{retauzero}
\end{equation}
Note that there is no condition on $\Re(a)$ here.  This is a peculiarity of the case $N_{\text{branes}}=N_{\text{antibranes}}$.  The final $B$-period constraint will determine the dependence of the moduli on $\alpha$.  We will postpone the discussion of this until after the exact solution \eqref{exactsol} is derived.

For $v$ and $w$, we would like to impose holomorphic boundary conditions
of the form \eqref{holobdryconds} but this is impossible because
\eqref{appvirasoro} implies that at least one of these must have
nontrivial nonholomorphic contributions {\footnote{These contributions
cannot vanish at both infinities without having unwanted poles
elsewhere.}}.  As discussed in the main text, the best we can do is
write a curve whose boundary conditions approach \eqref{appvirasoroHA}
in the limit $g_sN\rightarrow 0$.  To search for a curve of this type,
we start with the expressions for $v$ and $w$ in \eqref{vwparamrep} as a
``seed''
\begin{equation}
\begin{split}
v_H&=X\left(F_1^{(1)}-F_2^{(1)}-\left[F^{(1)}(a)-i\pi\right]\right),\\
w_H&=C\left(F_1^{(2)}-F_2^{(2)}\right)
\end{split}
\end{equation}
and look for holomorphic functions $v_A$ and $w_A$, as well as additions to $v_H$ and $w_H$, that lead to a solution of \eqref{appvirasoro}.  Before proceeding, we should get a better understanding of the contribution to \eqref{appvirasoro} from $s$ \eqref{sgamdel}.  We can write it explicitly as
\begin{equation}
\partial s_H\partial s_A = \left[\gamma\left(F_1^{(1)}-F_2^{(1)}\right)+i\pi(N+\delta)\right]\left[\gamma\left(F_1^{(1)}-F_2^{(1)}\right)-i\pi (N-\delta)\right]
\end{equation}
and use \eqref{Vsqexp} to express it as a linear combination of the $F_i^{(n)}$ \eqref{Findefd}.  The important point to note here is that the analytic structure is such that even (odd) poles are even (odd) under the parity operation $a_1\leftrightarrow a_2$.  As a result, we should only include additional terms in $v_H,v_A$ and $w_H,w_A$ that introduce contributions of this sort into \eqref{appvirasoro}.  This requirement, combined with the fact that $w$ and $v$ be periodic on the torus, leads to the following ansatz
\begin{equation}
\begin{split}
s_H&=\gamma\left(F_1-F_2\right)+i\pi (N+\delta)z,\\
s_A&=\gamma\left(F_1-F_2\right)-i\pi (N-\delta)z,\\
v_H&= X\left(F_1^{(1)}-F_2^{(1)}-\left[F^{(1)}(a)-i\pi\right]\right),\\
v_A &= \alpha \left(F^{(1)}_1-F^{(1)}_2-\left[F^{(1)}(a)-i\pi\right]\right),\\
w_H &= \beta\left(F_1-F_2+\frac{2\pi i a z}{\tau}\right)+\xi\left(F^{(1)}_1+F^{(1)}_2+\frac{2\pi i z}{\tau}\right)+C\left(F_1^{(2)}-F_2^{(2)}\right),\\
w_A &= \bar{\beta}\left(F_1-F_2+\frac{2\pi i az}{\tau}\right)+\frac{2\pi i \bar{\xi}z}{\tau}.
\label{ansatz}
\end{split}
\end{equation}
Note that we had to connect the coefficients of various terms in $w_H$ and $w_A$ in order to make sure that the full harmonic function $w(z,\bar{z})$ is periodic.  We also relied on the fact that $\Re(\tau)=0$ \eqref{retauzero}.  In addition to $C$ and $X$, our ansatz has five parameters $\alpha,\beta,\gamma,\delta,\xi$.

To solve the constraint \eqref{appvirasoro}, we note that the LHS is an elliptic function and hence is completely characterized by its analytic structure.  Indeed, to verify \eqref{appvirasoro} we need only check that the LHS vanishes at $a_1$ or $a_2$.  Cancellation of poles will imply that it is a constant function{\footnote{Because the poles at $a_1$ and $a_2$ are equivalent up to possible minus signs, we need only check that they cancel at one of the two points.}} while actual vanishing at $a_1$ or $a_2$ will guarantee that the value of this constant is zero.

The expressions \eqref{ansatz} contribute terms with poles of degree four and less to \eqref{appvirasoro} so there are five terms, namely the coefficients of the four poles and the constant term in an expansion near $a_1$, that must be set to zero.  This sounds promising because we have introduced five new coefficients.  The system, however, is quite nonlinear and expressions can become quite complicated.

To simplify things a bit further, let us appeal to symmetry yet again.
The original IIA configuration of figure \ref{A1cubTdualD4D4bar} that we
seek to ``lift'' is symmetric under the $\mathbb{Z}_2$ transformation
\begin{equation}
v\rightarrow -v,\qquad w\rightarrow -w,\qquad s\rightarrow -s,\qquad N\rightarrow -N.
\label{parity}
\end{equation}
It is clear how such a transformation should be realized in the $z$-plane as it corresponds to an exchange of $a_1$ and $a_2$ in \eqref{ansatz}.  Requiring that our full $M5$ curve also possess this symmetry{\footnote{Note that we must flip the sign of one of $\delta,\gamma$ as well in accordance with \eqref{retauzero}.  The choice appropriate for realizing the parity flip \eqref{parity} is $\delta\rightarrow -\delta$.}} implies that we must have
\begin{equation}
\xi=0.
\end{equation}
Our final simplification comes from studying the highest order poles of \eqref{appvirasoro}.  In particular, the only remaining contributions to third and fourth order poles are
\begin{equation}
C\bar{\beta}\left(F_1^{(3)}-F_2^{(3)}\right)\left(F_1-F_2+\frac{2\pi ia}{\tau}\right)+X\alpha\left(F_1^{(2)}-F_2^{(2)}\right)^2.
\end{equation}
The fourth order pole can be canceled by choosing
\begin{equation}
\alpha=-\frac{2C\bar{\beta}}{X}
\end{equation}
but this leaves no freedom left for dealing with the third order pole.  This only vanishes if
\begin{equation}
-\frac{2\pi i a}{\tau} = F^{(1)}(a)-i\pi = \zeta(a)-2\eta_1a
\end{equation}
which we can rewrite using \eqref{etaid} as
\begin{equation}
\zeta(a)=\frac{2\eta_3a}{\tau}\implies \frac{\zeta(a)}{a}=\frac{\zeta(\tau/2)}{\tau/2}.
\end{equation}
This admits the obvious solution
\begin{equation}
\tau=2a.
\label{appt2a}
\end{equation}
Using \eqref{retauzero}, we see that this fixes $\Re(a)=0$ as well as requires that the curve parameters $\gamma$ and $\delta$ satisfy
\begin{equation}
\gamma=\delta.
\end{equation}
The condition \eqref{appt2a} is not all that surprising given the symmetry of the problem.  It also leads to tremendous simplifications in the subsequent analysis because
\begin{equation}
F^{(n)}(\tau/2)=0\qquad\text{for $n$ odd}.
\end{equation}
This is easy to see by combining the periodicities \eqref{appFmonods} and parity properties \eqref{Fparity}.

It is now straightforward to check the remaining terms in the expansion of \eqref{appvirasoro} near $a_1$.  There are three such that we need to vanish, corresponding to the two remaining poles and the constant term.  Miraculously, we can achieve a solution by fixing only the two parameters $\beta$ and $\gamma$, leaving $C$ and $X$ arbitrary as before.  This leads to the result
\begin{equation}
\begin{split}
s_H&= Nr_0\cos\theta\left(F_2-F_1-i\pi z\right)+i\pi Nz\\
s_A&=Nr_0\cos\theta\left(F_2-F_1-i\pi z\right)-i\pi Nz\\
v_H&=X\left(F^{(1)}_1-F^{(1)}_2-\left[F^{(1)}(a)-i\pi\right]\right)\\
v_A&=\frac{2\xi (g_sN)^2}{\bar{X}}\left(F^{(1)}_1-F^{(1)}_2-\left[F^{(1)}(a)-i\pi\right]\right)\\
w_H&=g_sNr_0\sin\theta\left(F_2-F_1-i\pi z\right)+\frac{g_sN\xi}{r_0\sin\theta}\left(F_1^{(2)}-F_2^{(2)}\right)\\
w_A&=g_sNr_0\sin\theta\left(F_2-F_1-i\pi z\right)
\end{split}\end{equation}
where
\begin{equation}r_0^2=\frac{3\pi^2\wp(\tau/2)}{3\wp(\tau/2)^2-g_2},\qquad\qquad\xi=\frac{\pi^2}{6\wp(\tau/2)^2-2g_2}.\end{equation}
This is nothing other than \eqref{exactsol}.  Note that we have replaced $C$ by the parameter $\theta$ to make the effective rotation of D4/\anti{D4} stacks illustrated in figure \ref{wbending} manifest.  The final $B$-period constraint from \eqref{appsperiods} now fixes $\tau$ in terms of $\alpha$
\begin{equation}
2\pi i\alpha = r_0N\left[L(a,\tau)+\overline{L(a,\tau)}-i\pi\tau\right]
\label{appalphataurel}
\end{equation}
where $L(a,\tau)$ is defined in \eqref{Ldef} and $a=\tau/2$ \eqref{appt2a}.

As discussed in the text, we can consider this curve in the limit $g_sN\rightarrow 0$ and arrive at an approximate solution with boundary conditions \eqref{holobdryconds}
\begin{equation}
\begin{split}
s&=r_0 N\left[(F_1-F_2+i\pi z)+\text{cc}\right]+i\pi N(z+\bar{z}),\\
v&=X\left(F_1^{(1)}-F_2^{(1)}-\left[F^{(1)}(a)-i\pi\right]\right),\\
w&=C\left(F_1^{(2)}-F_2^{(2)}\right).
\end{split}
\label{appaprxcurv}
\end{equation}

\subsection{Connection with IIB}

Now that we have an approximate curve \eqref{appaprxcurv} for $g_sN\ll 1$ which lifts to an exact solution of \eqref{appharmonic} and \eqref{appvirasoro}, we can compare its moduli to those which extremize the type IIB potential \eqref{IIBnonsusypot}.  The easiest way to do this is to write $V$ directly as a function of $a$ and $\tau$ and simply check whether or not $a$ and $\tau$ satisfying \eqref{retauzero}, \eqref{appt2a}, and \eqref{appalphataurel} correspond to an extremum.  This is straightforward to do using the expression  \eqref{taumat} for the period matrix of our hyperelliptic curve.  The general expressions for derivatives of $V$ are somewhat complicated but if we set $a=\tau/2$ and $\Re(\tau)=0$ they simplify greatly.  For $\partial_aV$, we find in this case that
\begin{equation}
\partial_aV=-\frac{2\pi N(\alpha+\bar{\alpha})(L+\bar{L}-i\pi\tau)-4\pi\alpha\bar{\alpha}\,\partial_aL}{(L+\bar{L}-i\pi\tau)^2}.
\end{equation}
Noting that
\begin{equation}
\partial_aL=0
\end{equation}
we must have $\alpha+\bar{\alpha}=0$.  It is easy to see that \eqref{appalphataurel} automatically has this property so indeed $\partial_aV=0$ .  One might be worried that we just derived a condition on boundary data, which is supposed to be an initial condition for the curve, rather than the moduli.  We only have a small set of solutions in hand, though, corresponding to specific sorts of boundary conditions.  Indeed, we only obtained solutions that are invariant under \eqref{ssbarflip}.  Implicit in this assumption is that we choose boundary conditions consistent with this symmetry.  This, in turn, requires $\alpha+\bar{\alpha}=0$.

We now turn our attention to $\partial_{\tau}V$.  This expression is also complicated but setting $a=\tau/2$, $\Re(\tau)=0$, and $\alpha+\bar{\alpha}=0$ we find that
\begin{equation}\partial_{\tau}V=0\implies \alpha = \frac{N(L+\bar{L}-i\pi\tau)}{2\pi i\sqrt{1-\frac{4}{2\pi i}\partial_{\tau}L}}.\label{ptveqzer}\end{equation}
Noting that
\begin{equation}
\partial_{\tau}L=\frac{1}{2\pi i\wp(\tau/2)}\left(\wp(\tau/2)^2-\pi^2\wp(\tau/2)-\frac{g_2}{3}\right)
\end{equation}
we see that \eqref{ptveqzer} is identical to \eqref{appalphataurel} and
hence the moduli of the reduced curve \eqref{appaprxcurv} are an exact
extremum of the IIB potential \eqref{IIBnonsusypot}.

\end{document}